\documentclass[lettersize,journal]{IEEEtran}
\usepackage{amsmath,amsfonts}
\usepackage{algorithmic}
\usepackage{algorithm}
\usepackage{array}
\usepackage[caption=false,font=normalsize,labelfont=sf,textfont=sf]{subfig}
\usepackage{textcomp}
\usepackage{stfloats}
\usepackage{url}
\usepackage{breakurl}
\usepackage{verbatim}
\usepackage{graphicx}
\usepackage{cite}
\usepackage{enumitem}
\usepackage{xcolor}
\usepackage{soul}
\usepackage{multirow}
\sethlcolor{yellow}
\usepackage{graphicx}
\usepackage{booktabs}
\usepackage{subfig}
\usepackage{caption}
\usepackage{acronym}
\usepackage{lscape}
\usepackage{rotating} 
\usepackage{array} 

\newcolumntype{C}[1]{>{\centering\arraybackslash}m{#1}}
\begin{document}

\title{Toward 6G Optical Fronthaul: A Survey on Enabling Technologies and Research Perspectives}

\author{Abdulhalim Fayad,  \IEEEmembership{Graduate Student Member,} \IEEEmembership{IEEE}, Tibor Cinkler, and Jacek Rak, \IEEEmembership{Senior Member,} \IEEEmembership{IEEE}
\thanks{ This work was supported by CHIST-ERA Sustainable and Adaptive Ultra-High-Capacity Micro Base Stations (SAMBAS) Grant
 funded by Research Foundation–Flanders (FWO), Agence nationale de la Recherche (ANR), National Research, Development and
 Innovation Office (NKFIH), and UK Research and Innovation (UKRI), under Grant CHIST-ERA-20-SICT-003. The work of Jacek Rak was funded in part by the Passau International Centre for Advanced Interdisciplinary Studies (PICAIS) of the University of Passau, Germany.
}
\thanks{Abdulhalim Fayad is with the Department of Telecommunications and Media Informatics, Budapest University of Technology and Economics, 1111 Budapest, Hungary.}%
\thanks{Tibor Cinkler is with the Department of Telecommunications and Media Informatics, Budapest University of Technology and Economics, 1111 Budapest, Hungary, HUN-REN-BME Cloud Applications Research Group, 1111 Budapest, Hungary, and the Department of Computer Communications, Gda\'nsk University of Technology, G. Narutowicza 11/12, 80-233 Gdansk, Poland.}%
\thanks{Jacek Rak is with the Department of Computer Communications, Gda\'nsk University of Technology, G. Narutowicza 11/12, 80-233 Gda\'nsk, Poland. He is also the PICAIS Visiting Research Fellow with the Chair of Computer Networks and Computer Communications at the University of Passau, Germany.}
\thanks{Manuscript received October 7, 2023.}}

\markboth{IEEE COMMUNICATION SURVEYS \& TUTORIALS,~Vol.~14, No.~8, August~2021}%
{Shell \MakeLowercase{\textit{et al.}}: A Sample Article Using IEEEtran.cls for IEEE Journals}


\maketitle

\begin{abstract}
The anticipated launch of the Sixth Generation (6G) of mobile technology by 2030 will mark a significant milestone in the evolution of wireless communication, ushering in a new era with advancements in technology and applications. 6G is expected to deliver ultra-high data rates and almost instantaneous communications, with three-dimensional coverage for everything, everywhere, and at any time.
	In the 6G Radio Access Networks (RANs) architecture, the  Fronthaul connects geographically distributed Remote Units (RUs) to Distributed/Digital Units (DUs) pool. Among all possible solutions for implementing 6G fronthaul, optical technologies will remain crucial in supporting the 6G fronthaul, as they offer high-speed, low-latency, and reliable transmission capabilities to meet the 6G strict requirements.
	This survey provides an explanation of the 5G and future 6G optical fronthaul concept and presents a comprehensive overview of the current state of the art and future research directions in 6G optical fronthaul, highlighting the key technologies and research perspectives fundamental in designing fronthaul networks for 5G and future 6G. 
	Additionally, it examines the benefits and drawbacks of each optical technology and its potential applications in 6G fronthaul networks.
	This paper aims to serve as a comprehensive resource for researchers and industry professionals about the current state and future prospects of 6G optical fronthaul technologies, facilitating the development of robust and efficient wireless networks of the future.
\end{abstract}

\section*{Acronyms}
\begin{acronym}
	\acro{1G}{First Generation}
	\acro{2G}{Second Generation}
	\acro{3G}{Third Generation}
	\acro{4G}{Fourth Generation}
	\acro{5G}{Fifth Generation}
	\acro{6G}{Sixth Generation}
	\acro{3GPP}{3rd Generation Partnership Project}
	\acro{AI}{Artificial Intelligence}
	\acro{AR}{Augmented Reality}
	\acro{ARQ}{Automatic Repeat Request}
	\acro{ATM}{ Asynchronous Transfer Mode}
	\acro{BBU}{BaseBand Unit}
	\acro{BS}{ Base Station}
	\acro{CA}{Carrier Aggregation}
	\acro{CDMA}{Code Division Multiple Access}
	\acro{CoE}{ CPRI over Ethernet}
	\acro{CoMP}{Coordinated Multi-Point}
	\acro{COTS}{Commercial Off-The-Shelf}
	\acro{CPRI}{ Common Public Radio Interface}
	\acro{C-RAN}{Centralized/Cloud Radio Access Network}
	\acro{CU}{Central Unit}
	\acro{CWDM}{Coarse Wavelength Division Multiplexing}
	\acro{D-RAN}{Distributed RAN}
	\acro{DA-RAN}{DisAggregated-RAN}
	\acro{DAS}{Distributed Antenna Systems}
	\acro{DBA}{Dynamic Bandwidth Allocation}
	\acro{DFT-S}{Discrete Fourier Transform Spread}
	\acro{DL}{DownLink}
	\acro{DT}{Digital Twin}
	\acro{DU}{Distributed Unit}
	\acro{DWBA}{Dynamic Wavelength and Bandwidth Allocation}
	\acro{DWDM}{Dense Wavelength Division Multiplexing}
	\acro{EB}{ Exabytes}
	\acro{eICIC}{enhanced Inter-Cell Interference Cancellation}
	\acro{eCPRI}{ethernet Common Public Radio Interface}
	\acro{EDR}{ Enhanced Data Rate }
	\acro{eMBB}{enhanced Mobile BroadBand}
	\acro{FDMA}{Frequency Division Multiple Access}
	\acro{FiWi}{Fiber-Wireless }
	\acro{F-RAN}{Fog Radio Access Network}
	\acro{FSO}{Free Space Optics}
	\acro{FTTH}{Fiber To The Home}
	\acro{GSM}{Global System for Mobile communication}
	\acro{HAP}{High Altitude Platform}
	\acro{H-CRAN}{Heterogeneous C-RAN }
	\acro{HDR}{High Data Rate}
	\acro{HetNet}{Heterogeneous Network}
	\acro{HPC}{High-Performance Computing}
	\acro{HSP}{High Speed PON}
	\acro{IA}{Interference Alignment}
	\acro{I/Q}{In-phase Signal/Quadrature Waveform}
	\acro{IEEE}{ Institute of Electrical and Electronics Engineers}
	\acro{IoT}{Internet of Things}
	\acro{LAN}{Local Area Network}
	\acro{LoS}{Line-of-Sight}
	\acro{LTE}{Long-Term Evolution}
	\acro{M2M}{Machine-to-Machine}
	\acro{MAC}{Media Access Control}
	\acro{MIMO}{Multi-Input-Multi-Output}
	\acro{mMTC}{massive Machine-To-Machine Communications}
	\acro{mmWave}{ millimeter Wave}
	\acro{MNO}{ Mobile Network Operator}
	\acro{NDR}{Next Data Rate}
	\acro{NFV}{Network Function Virtualization}
	\acro{NG-OAN}{Next-Generation Optical Access Network}
	\acro{NG-PON2}{Next Generation Passive Optical Networks stage~2}
	\acro{OBSAI}{Open Base Station Architecture Initiative}
	\acro{OCDM}{Optical Code Division Multiplexing}
	\acro{ODN}{Optical Distribution Network}
	\acro{OFDM}{Orthogonal Frequency Division Multiplexing}
	\acro{OLT}{Optical Line Terminal}
	\acro{OMA}{Orthogonal Multiple Access}
	\acro{ONU}{Optical Network Unit}
	\acro{OPM}{Optical Performance Monitoring}
	\acro{O-RAN}{Open Radio Access Network}
	\acro{ORI}{Open Radio Interface}
	\acro{P2MP}{Point-to-MultiPoint}
	\acro{P2P}{Point-to-Point}
	\acro{PAM-4}{Four-level Pulsed Amplitude Modulation}
	\acro{PDCP}{Packet Data Convergence Protocol}
	\acro{PDM-PON}{Polarization Division Multiplexing-PON}
	\acro{PDU}{Protocol Data Unit}
	\acro{PHY}{Physical Layer}
	\acro{PON}{Passive Optical Network}
	\acro{QAM}{Quadrature Amplitude Modulation}
	\acro{QDR}{Quad Data Rate}
	\acro{QoS}{Quality of Service}
	\acro{QSFP}{Quad Small Form-factor Pluggable}
	\acro{RAN}{Radio Access Network}
	\acro{RCA}{Root Cause Analysis}
	\acro{RE}{ Radio Equipment}
	\acro{REC}{Radio Network Controller}
	\acro{RF}{Radio Frequency}
	\acro{RLC}{Radio Link Control}
	\acro{RNC}{Radio Network Controller}
	\acro{RoF}{Radio over Fiber }
	\acro{RRC}{Radio Resource Control}
	\acro{RRH}{Remote Radio Head}
	\acro{RSP}{Radio Signal Processing}
	\acro{RU}{Radio Unit}
	\acro{SAN}{Storage Area Network}
	\acro{SDM}{Space Division Multiplexing}
	\acro{SM-MIMO}{Sparse Massive-Multiple Input Multiple Output}
	\acro{SDN}{ Software Defined Network}
	\acro{SFP}{Small Form-factor Pluggable}
	\acro{SIC}{Successive Interference Cancellation}
	\acro{SMS}{Short Messaging Service}
	\acro{SSII}{Self-Induced Intermodulation Interference}
	\acro{TDM}{Time Division Multiplexing}
	\acro{TDMA}{Time Division Multiple Access}
	\acro{TWDM}{Time and Wavelength Division Multiplexing }
	\acro{UAV}{Unmanned Aerial Vehicle}
	\acro{UMTS}{Universal Mobile Telecommunications System}
	\acro{URLLC}{Ultra-Reliable and Low-Latency Communications}
	\acro{V2X}{Vehicle-to-everything}
	\acro{vBBU}{Virtual BaseBand Unit}
	\acro{VR}{Virtual Reality }
	\acro{vRAN}{Virtualized Radio Access Network}
	\acro{WDM}{Wavelength Division Multiplexing}
	\acro{XDR}{eXtreme Data Rate}
	\acro{XR}{Extended Reality}
	\acro{ZB}{Zetabytes}
\end{acronym}

\section{Introduction}\label{sec1}
\begin{figure}[htbp]
	\centering
	\includegraphics[width=0.5\textwidth]{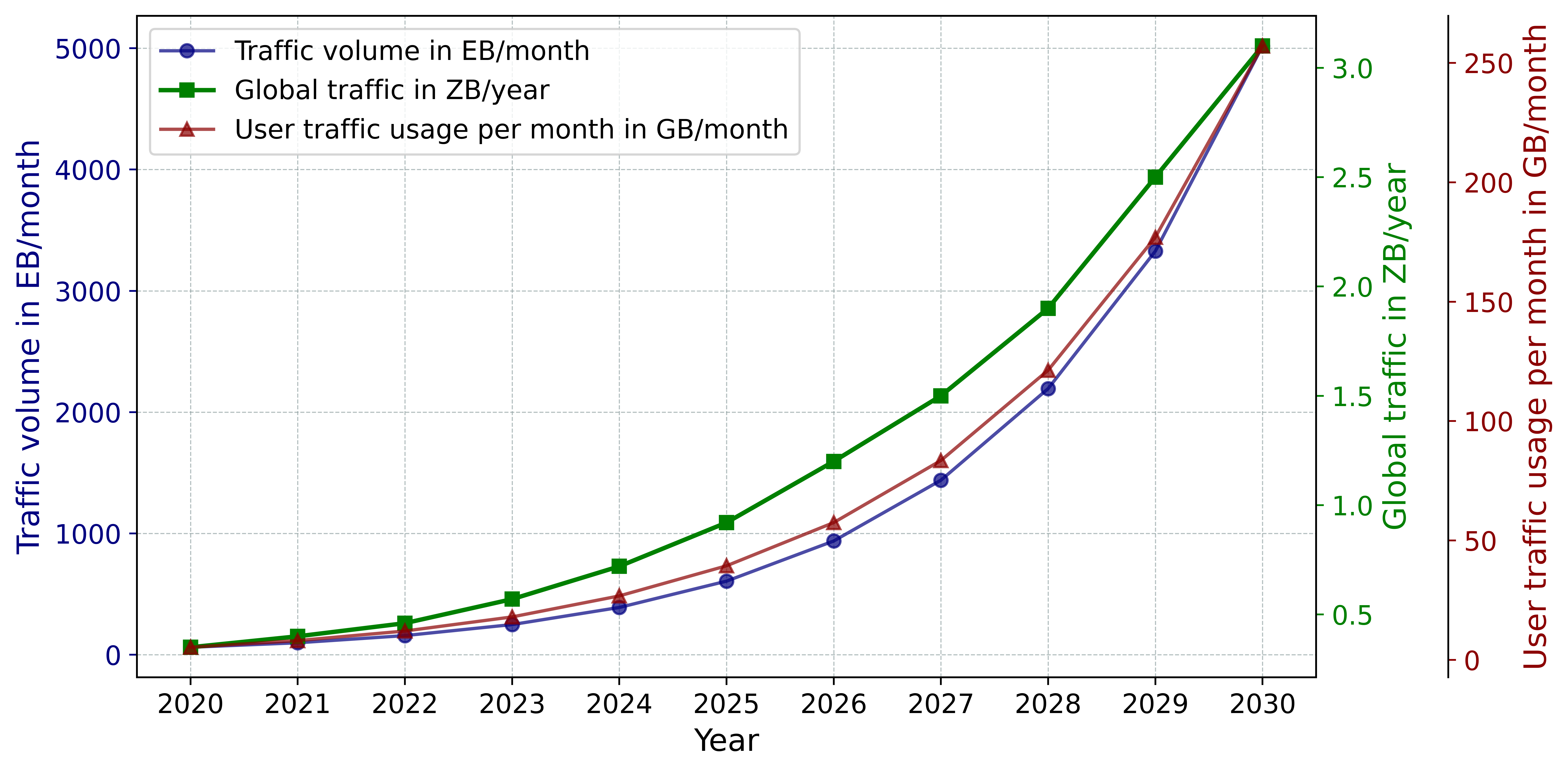}
	\caption{Estimated global mobile traffic and user data traffic from 2020 to 2030 based on ITU-R Report M.2370-0 \cite{itu}.}
	\label{fig:traffic}
\end{figure}

\IEEEPARstart{R}{ecently}, particularly in the aftermath of the \mbox{COVID-19} pandemic, communication networks have become increasingly important in all aspects of our lives (i.e., healthcare, smart manufacturing, distance learning, smart cities, etc). For that, the number of devices connected to the Internet is expected to grow significantly each year, leading to a substantial increase in traffic volumes served by mobile networks. By 2030, it is projected that these networks will handle at least 5000 Exabytes of data per month, while global data traffic is forecasted to exceed 3 Zetabytes. Additionally, the user mobile data usage will rise from 5 GB in 2020 to more than 250 GB per month in 2030 as shown in Fig.~\ref{fig:traffic} \cite{itu}.
Over the past four decades, communication technologies have consistently transformed to address the changing demands of users. The evolution of wireless mobile networks has been particularly notable, advancing through successive generations. Each new generation has introduced substantial improvements, reflecting the relentless pace of innovation in this field. Starting from the First Generation (1G) and advancing to the current Fifth Generation (5G). This evolution has been driven by the increasing demand for low-latency and high-capacity wireless communication services \cite{5g1}. Recent technological advancements towards the Sixth Generation (6G), such as network softwarization, virtualization, massive Multiple-Input-Multiple-Output (MIMO), ultra-densification of devices, and the addition of new frequency bands, have resulted in substantial benefits across various sectors. These advancements have brought advantages to commercial providers, academic research institutions, standards organizations, and, most importantly, end-users.

5G represents a major step forward from previous mobile network generations, as it is able to meet recent communication needs of a wide range of use cases in three main application areas:
\begin{itemize}
	\item \emph{enhanced Mobile BroadBand (eMBB)}, which supports applications including ultra-high-definition video, 3D video, cloud work, and Augmented/Virtual Reality (AR/VR).
	\item \emph{ultra-Reliable and Low-Latency Communications (uRLLC)} enabling applications in areas such as autonomous vehicles, drones, industrial automation, telemedicine, and other critical missions.
	\item  \emph{mMTC} enabling \emph{massive Machine Type Communication} and providing services for smart homes, buildings, cities, and Internet of Things (IoT).
\end{itemize}

While 5G has undeniably revolutionized connectivity, the surge of interest in 6G mobile technology stems from the relentless pursuit of even greater technological advancements. The insatiable demand for faster, more reliable, and ubiquitous connectivity has propelled researchers, industries, and policymakers to set their sights on the next frontier. 6G is envisioned to surpass the capabilities of its predecessor, offering unprecedented data speeds, ultra-low latency, and the ability to seamlessly integrate with emerging technologies such as Artificial Intelligence (AI), Extended Reality (XR), and the Internet of Everything (IoE), holographic communications, that can not be adequately served by 5G. This paradigm shift is not merely an incremental upgrade but represents a quantum leap forward in the realm of wireless communication. As society becomes increasingly dependent on digital connectivity, the allure of 6G lies in its potential to redefine the boundaries of what is possible, unlocking new realms of innovation and transforming the way we live, work, and interact with the world \cite{6g1, 6g2}.

\begin{figure}[htbp]
	\centering
	\includegraphics[width=0.4\textwidth]{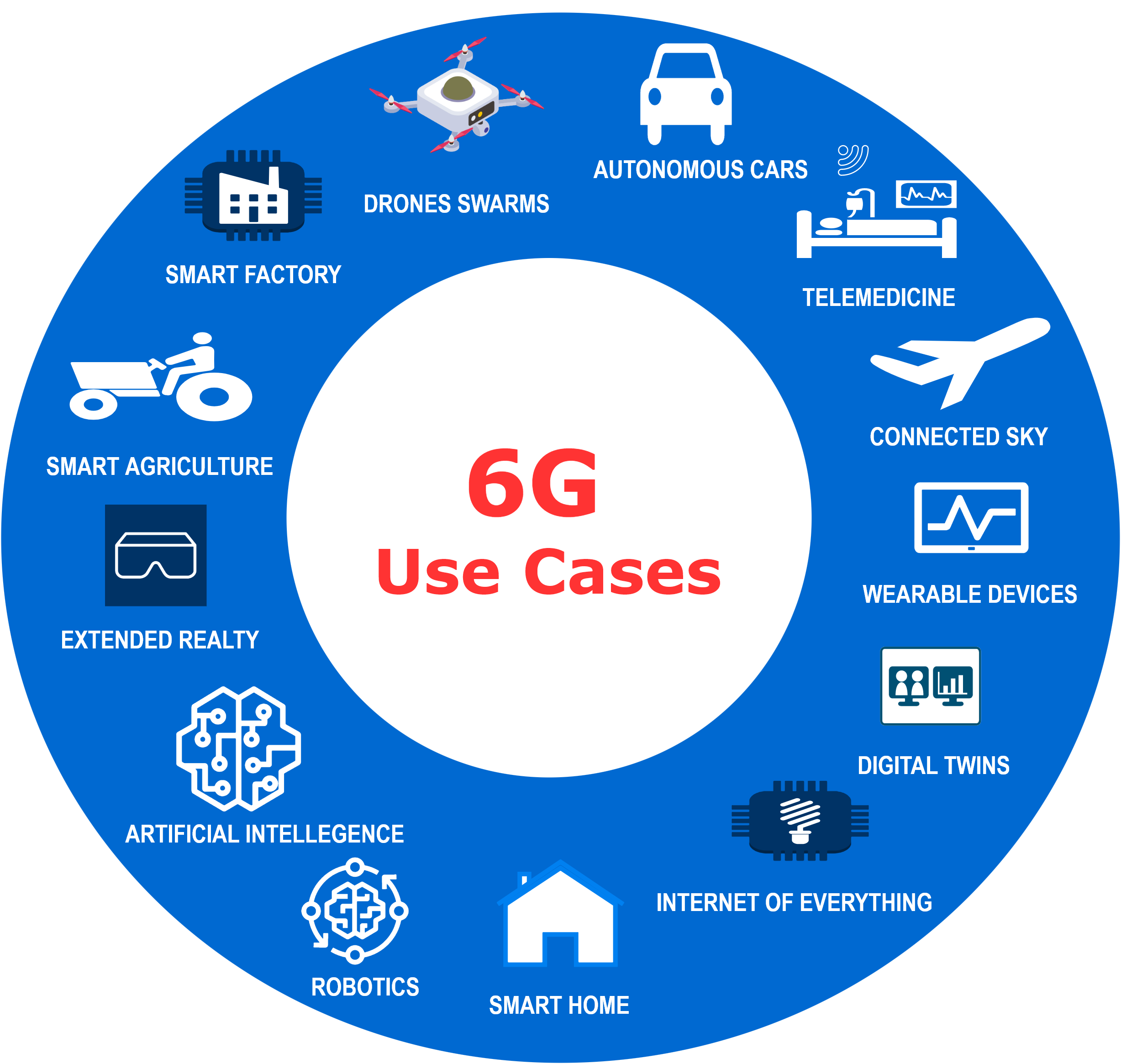}
	\caption{The expected 6G use cases.}
	\label{fig:6g}
\end{figure}

Figure~\ref{fig:6g} illustrates the anticipated primary use cases for 6G, encompassing the following range of applications.
\begin{itemize}
\item \emph{Autonomous Cars:} Enabling uRLLC is crucial for the real-time data exchange necessary for autonomous vehicles to navigate safely and efficiently.
\item \emph{Telemedicine:} Providing the necessary bandwidth and reliability for high-definition video streaming and real-time data transmission, revolutionizing healthcare delivery.
\item \emph{Connected Sky:} From unmanned aerial vehicles (UAVs) to air traffic control systems, 6G is anticipated to enhance connectivity and communication reliability in the airspace, facilitating safer and more efficient air transportation.
\item \emph{Wearable Devices:} The proliferation of wearable technology is expected to benefit from high data rates and low latency, enabling seamless integration into daily life and enhancing user experiences.
\item \emph{Digital Twins (DTs):} Enabling the creation and management of digital replicas of physical objects and systems, allowing for real-time monitoring, analysis, and optimization across various industries, from manufacturing to urban planning.
\item \emph{IoE:} Supporting seamless integration and communication of billions of devices, leading to more intelligent and interconnected ecosystems.
\item \emph{Smart Home:} Enhancing connectivity within households, enabling more efficient energy management, enhanced security, and personalized user experiences.
\item \emph{Robotics:} With advancements in artificial intelligence and robotics, 6G can enable robots to communicate and collaborate more effectively, leading to breakthroughs in industrial automation, healthcare, and disaster response.
\item \emph{Drone Swarms:} Providing the high throughput and the low latency that are anticipated to support the coordination and communication of large fleets of drones, unlocking new possibilities for applications such as aerial surveillance, disaster relief, and package delivery.
\item \emph{Artificial Intelligence (AI):}  Accelerating the development and deployment of AI applications by providing the necessary computational resources and connectivity for real-time data processing and analysis.
\item \emph{XR:} From virtual reality to augmented reality, 6G promises to deliver immersive and interactive experiences with ultra-high-definition visuals and seamless connectivity, revolutionizing entertainment, education, and communication.
\item \emph{Smart Agriculture:} By enabling precision agriculture techniques, such as remote sensing and autonomous machinery, 6G can enhance productivity, sustainability, and resource efficiency in the agricultural sector.
\item \emph{Smart Factories:} With the advent of Industry 4.0, 6G can facilitate the integration of cyber-physical systems, IoT devices, and AI-powered analytics, leading to more flexible, efficient, and interconnected manufacturing processes.
\end{itemize}

 Major telecommunication companies, including, e.g., Ericsson \cite{ericsson}, Nokia Bell Labs~\cite{nokia}, Samsung Research~\cite{samsung}, and Huawei \cite{huawei}, have recently presented their perspectives on the future of 6G. Several ongoing projects are aimed at developing long-term strategic plans for 6G \cite{projects}.

Moreover, this shift is complemented by the evolution of Radio Access Networks (RAN), culminating in the rise of Open Radio Access Networks (O-RAN). O-RAN, a paradigm gaining momentum in the 5G and 6G era, embodies openness, accessibility, and interoperability—qualities that transcend vendor restrictions and proprietary technologies. This multi-vendor network solution not only enhances performance and efficiency but also drives cost reduction, fostering an environment conducive to continuous innovation \cite{oran1,oran2, oran22}.

In the dynamic evolution of 5G and 6G networks, a pivotal challenge arises in the form of implementing an efficient fronthaul within the framework of O-RAN. \emph{Fronthaul}, denoting the high-speed links connecting baseband processing functions in the Digital Unit (DU) or BaseBand unit (BBU) to remote Radio Units (RUs) at the cell site, stands as a critical element in this architectural paradigm. Realizing an optimized fronthaul is paramount to unlocking the full potential of O-RAN in the context of 5G and 6G networks.
The fronthaul segment can be implemented using either wireless (microwave and millimeter Wave (mmWave)) or optical (optical fiber and Free Space Optics (FSO)) technologies. Wireless communications, prevalent in earlier generations (up to Fourth Generation (4G)), face significant constraints such as limited capacity to meet the demands of future 5G/6G high-bandwidth applications (such as splitting options 7.1 and 8 as will be illustrated in Section IV), restricted spectrum, high interference, and stringent control \cite{wireless}. In contrast, optical technologies are recognized as the most sustainable for handling 5G and beyond. Optical fiber, known for its high bandwidth, reliability, security, scalability, and cost-effectiveness \cite{wireless, optical, pon2}, emerges as a particularly suitable choice for 6G fronthaul. It not only offers high reliability and security but also supports long-distance communications. On the other hand, FSO, utilizing lasers for data transmission over the air, presents a cost-efficient alternative to optical fiber in specific scenarios such as remote areas, temporary installations, areas with obstacles, etc. Together, these optical technologies are expected to play a vital role in enhancing wireless communication systems to meet the demands of future high-bandwidth applications in 5G and 6G.
 
\subsection{Scope of This Survey}
In this survey, we aim to review 6G optical fronthaul and its enabling technologies and research perspectives. We start by reviewing the related survey and highlighting the main contributions of our survey. Progressing forward, we present the evolution of wireless mobile networks from 1G to 6G, as well as the development of RAN architectures toward the O-RAN paradigm. After, we discuss the concept of the fronthaul interface and the various splitting options. Additionally, we examine the benefits and drawbacks of different optical technologies for 6G networks. Furthermore, we review the current research efforts on the optical fronthaul of 5G and 6G networks. Finally, it discusses the main research directions and challenges for the optical fronthaul of 6G networks.
\subsection{Motivation}
We have chosen this survey topic due to the importance of the fronthaul segment in the future wireless communications architecture. Moreover, in the existing works, there is a lack of comparative works covering all available perspectives on optical fronthaul for 6G networks. While previous surveys \cite{related10, related1, related2, related7, related8, related9, related3, related5, related6, related4, xhaul, rofsurvey} have focused on 5G wireless backhauling or the general usage of optical communications for 5G and beyond, unfortunately they fail to capture the latest achievements and advancements in the field of 5G/6G fronthauling (detailed in Section~\ref{sec2}). Also, none of them has dedicated substantial attention to the area of optical fronthaul for 6G networks. This gap in the literature prompted us to select this subject as the focal point of our paper, as we believe that the optical fronthaul will play a pivotal role in the upcoming advancements of network technologies.
Moreover, the research area of designing optical fronthaul for mobile networks is rapidly expanding, with new technologies and standards being introduced. Thus, there is a clear need for an up-to-date survey paper that provides an up-to-date and comprehensive review of the state-of-the-art optical fronthaul for 6G networks.
	
Our survey aims to fill this gap by providing a detailed analysis of various optical fronthaul technologies and architectures for beyond 5G networks, covering their pros, cons, and major use cases. Additionally, it outlines the key challenges in the field and suggests the related potential solutions. 
We believe this survey can serve as a valuable resource for researchers and practitioners, enabling them to stay informed about recent developments regarding optical fronthaul technologies.

\subsection{Structure of the Paper}
\begin{figure}[htbp]
	\centering
	\includegraphics[width=1\columnwidth]{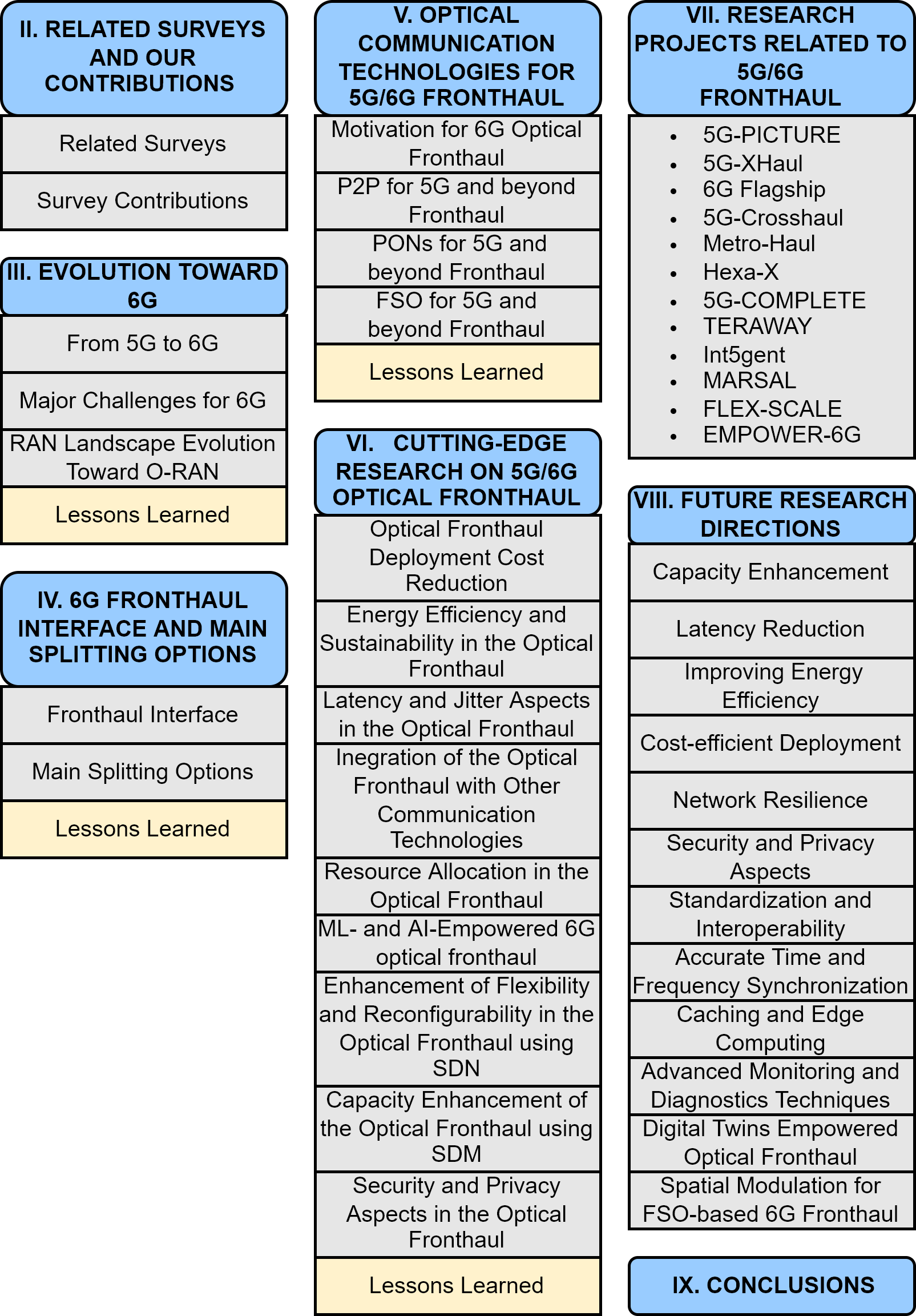}
	\caption{The structure of this  survey.}
	\label{fig:outlines}
\end{figure}

In Section~\ref{sec2}, we highlight the main contributions of this article and provide an overview of the related surveys.
In Section~\ref{sec3}, we provide an overview of the evolution of wireless networks toward 6G, discuss the main challenges for future 6G networks, as well as the evolution of the radio access networks landscape.
Section~\ref{sec4} provides a brief overview of the fronthaul interface and the main splitting options currently available. In Section~\ref{sec5}, we explain how different optical technologies including Point-to-Point (P2P) optical fiber, Passive Optical Networks (PON), and FSO) can meet the requirements of 5G/6G fronthaul.
Section~\ref{sec6} analyzes in detail the recent research efforts on the optical fronthaul of 5G and 6G networks. We explore the different approaches to address the challenges in this field and discuss the potential of various optical technologies. In Section~\ref{sec7}, we briefly discuss the objectives of the recent projects related to 5G/6G optical fronthaul. 
Section~\ref{sec8} highlights the main research directions for optical fronthaul of 6G networks. It also discusses the related emerging trends and technologies.
Finally, Section~\ref{sec9} concludes the paper. The general structure of the paper is also shown in Fig.~\ref{fig:outlines}.

\section{Related Surveys and Our Contributions}\label{sec2}
\subsection{Related Surveys}
In recent years, several survey studies have focused on 5G and future 6G network fronthauling. For instance, the study in \cite{related10} provided a comprehensive review of C-RAN architecture, encompassing its backhaul and fronthaul segments. It explored various fronthaul transport options supporting increased wireless bandwidth demand and small cell deployment. Additionally, it discussed emerging research directions, opportunities, and challenges in optical networking for mobile fronthaul networks.

In \cite{related1}, existing and emerging backhaul technologies were reviewed with a focus on wireless technologies, highlighting their benefits and drawbacks. The importance of combining radio access and backhaul was underlined. Key drivers, lessons learned, and challenges were also discussed, providing insights into addressing the backhauling challenge beyond 2020.

Reference \cite{related2} provided a thorough tutorial on \mbox{C-RAN} optical fronthaul technology, architectures, and problems for 5G networks.  It presented a review of optical fiber and FSO fronthaul technologies and ideas for decreasing complexity, cost, bandwidth needs, latency, and power consumption.

The work in \cite{related7} discussed the challenges and opportunities of providing Internet access to rural communities. Several rural connectivity solutions are evaluated, focusing on access/fronthaul and backhaul techniques.

In \cite{related8}, the authors highlighted the challenges of the fronthaul interface, such as strict latency and bandwidth requirements.  The main focus is to study the suitability of the PON solution in meeting the requirements of 5G C-RAN architecture.

Reference \cite{related9} reviewed optical and wireless backhaul technologies crucial for meeting the demanding requirements of 5G and beyond networks. Furthermore, the discussion extended to the challenges and prospects of 6G technologies.

The authors of \cite{related3} provided a review of the Next Generation-Optical Access Networks (NG-OANs) for 6G RANs with a primary focus on PON technology. An optical access solution of integrated Fiber-Wireless (FiWi) access networks is introduced, harnessing the synergies of NG-OANs and 6G RANs to support super-broadband services in diverse scenarios. Furthermore, the paper identified key challenges and open issues in developing NG-OANs and 6G RANs.

Reference \cite{related5} discussed how wireless networks evolve to meet higher capacity demands. The paper focused on the wireless fronthaul interface of 5G and beyond, highlighting technologies like fronthaul compression and Line-of-Sight Multiple-Input Multiple-Output (LoS-MIMO) solutions. Additionally, they mentioned challenges and future research directions to improve the performance of 5G and beyond wireless fronthaul.

The main focus of  \cite{related6} is on analyzing the issues, challenges, opportunities, and applications of wireless backhaul in the context of 5G networks, similar to \cite{related5}. Furthermore, the paper highlighted the importance of wireless backhaul in beyond 5G networks, including cell-free networking, ultra-massive MIMO, and highly dense networks.

The authors of \cite{related4} discussed the rising interest in developing 6G mobile communication systems requiring higher throughput, lower latency, and massive connections. Moreover, they reviewed the convergence of wireless and optical technologies as a fundamental step toward developing the 6G network infrastructure.

In \cite{xhaul}, the importance of optical \mbox{x-haul} in the 6G ecosystem was emphasized. The work examined the challenges associated with integrating optical \mbox{x-haul} with wireless technologies. Additionally, they reviewed research on optical \mbox{x-haul} networks, covering O-RAN architecture and coordination functionalities in radio-over-fiber networks. Additionally, they highlighted future research challenges in 6G optical \mbox{x-haul}.

The authors of \cite{rofsurvey} discussed the challenges and opportunities of using Radio over Fiber (RoF) technology to increase the capacity of 5G fronthaul systems. They focused on the C-RAN architecture, PON design, optical millimeter-wave generation, and resource allocation schemes. They also summarized the key performance metrics and evaluation methods for RoF systems, such as spectral efficiency, energy efficiency, latency, and optical performance monitoring. Additionally, they provided future research directions for RoF-based 5G fronthaul.

\begin{table*}
	\centering
	\caption{Related surveys}
	\label{tab:differentsurvey}
	\begin{tabular}{|m{0.4 cm}|m{0.35 cm}|m{3.2 cm}|m{6 cm}|m{6 cm}|}
		\hline
		\textbf{Ref} & \multicolumn{1}{c|}{\textbf{Year}} &\textbf{Focus} & \textbf{Main Contributions}& \textbf{Limitations} \\ \hline
		\multicolumn{1}{|c|}{\cite{related10}}& \multicolumn{1}{c|}{2016}& 5G backhaul technologies&Advantages and limitations of 5G wireless backhauling technologies; Challenges and future research directions of  5G backhaul networks.& Outdated (2016); Does not cover optical fronthaul solutions for 6G.\\ \hline

		\multicolumn{1}{|c|}{\cite{related1}}&\multicolumn{1}{c|}{2018}&Fronthaul options for \mbox{C-RAN}& Different fronthaul transport options and emerging research directions in \mbox{C-RAN}; Challenges and requirements of fronthaul networks.& Does not provide detailed study of 6G RAN architectures; Does not compare various fronthaul transport options; Does not cover FSO technology.\\ \hline
		
		\multicolumn{1}{|c|}{\cite{related2}}&\multicolumn{1}{c|}{2018}&Optical fronthaul for \mbox{C-RAN}& Technologies, requirements, challenges, and solutions of optical fronthaul for C-RANs: Solutions for improving efficiency of optical fronthaul.& Does not provide a study about 6G RAN architectures and main challenges towards efficient optical fronthaul as in our paper.\\ \hline
		
		\multicolumn{1}{|c|}{\cite{related7}}&\multicolumn{1}{c|}{2020}& Rural connectivity in 6G era&Challenges in rural connectivity, focusing on fronthaul and backhaul solutions; Potential of emerging technologies such as low-earth orbit satellites and 5G networks.& General study about fronthaul and backhaul solutions; Does not cover all the recent works, projects and future directions about optical fronthauling.\\ \hline
		
		\multicolumn{1}{|c|}{\cite{related8}}&\multicolumn{1}{c|}{2020}& PON-based 5G fronthaul&Suitability of PON for 5G fronthaul.& Limited to PON for 5G fronthaul; Does not coverage of various optical technologies for 6G fronthaul.\\ \hline
		
		\multicolumn{1}{|c|}{\cite{related9}}&\multicolumn{1}{c|}{2020}&Optical and wireless technologies for 5G/6G&Review of optical and wireless backhaul network technologies; Analysis of key backhaul network requirements for 5G and 6G.& Limited to backhaul solutions; No detailed explanation of the fronthaul interface; Does not cover recent studies about optical fronthaul.\\ \hline 
		
		\multicolumn{1}{|c|}{\cite{related3}}&\multicolumn{1}{c|}{2022}& Optical access networks in 6G&Architectures, enabling technologies, applications, and challenges of NG-OAN (i.e., PON) for 6G; Comparison of 6G RANs and NG-OANs.& Focuses only on PON technology; Does not cover of other optical fronthaul technologies like P2P, and FSO; Does not provide detailed explanation of the fronthaul interface; Does not highlight the possible research directions in the field.\\ \hline
		
		\multicolumn{1}{|c|}{\cite{related5}}&\multicolumn{1}{c|}{2022}& Wireless fronthaul for 5G and future RANs&Challenges of implementing wireless fronthaul in 5G networks; Proposal of wireless fronthaul architecture integrating FSO and wireless technologies.& Does not focus on optical technologies; No discussion on main challenges and future research directions.\\ \hline
		
		\multicolumn{1}{|c|}{\cite{related6}}&\multicolumn{1}{c|}{2022}& Wireless backhaul for 5G and beyond&Focused on wireless technologies for 5G and beyond; Covers main challenges and future research directions.& Does not focus on the fronthaul interface; Does not cover optical technologies for 6G fronthaul. \\ \hline
		
		\multicolumn{1}{|c|}{\cite{related4}}&\multicolumn{1}{c|}{2023}& Wireless and optical technologies for 6G access&Use cases, requirements, wireless and optical convergence, and technologies to enable 6G networks; Role of ML in 6G network physical layers.& Does not focus on fronthaul interface, current solutions, projects, and future research directions.\\ \hline
		
		\multicolumn{1}{|c|}{\cite{xhaul}}&\multicolumn{1}{c|}{2023}&Optical x-haul network design for 5G, 6G&Discussion about 6G optical x-haul deployment, \mbox{O-RAN} architecture, and software solutions.& Does not cover different P2P and PON architectures, usage of FSO, and current solutions in optical 6G fronthaul.\\ \hline
		
		\multicolumn{1}{|c|}{\cite{rofsurvey}}&\multicolumn{1}{c|}{2023}&5G fronthaul systems using RoF&Review of 5G system capacity enhancement using RoF;  Various architectures,  applications and technologies for 5G mobile networks.& Does not study different optical fronthaul technologies; Does not cover various optical technologies and their evolution; Does not cover the recent studies, projects, and future research directions related to 5G and beyond fronthaul.\\ \hline
	\end{tabular}
\end{table*}

\subsection{Survey Contributions}
Compared to the previous studies, the main contributions of our survey can be listed as follows:
\begin{enumerate}
	\item Conducting a comprehensive overview and analysis of the latest advancements in the field of 5G/6G optical fronthaul studies, as well as providing a thorough examination of up-to-date research.
	\item Presenting an insightful overview of the development of mobile communications toward 6G, emphasizing the anticipated challenges for 6G,  highlighting the evolving landscape of RANs, as well as the importance of the fronthaul.
	\item An in-depth discussion of the main splitting options for the 6G fronthaul interface and the analysis of capacity and latency requirements for the efficient operation of 6G networks.
	
	\item Explaining the critical role of optical communications in meeting the demands of 5G and the future 6G fronthaul, enhanced by a compelling motivation for leveraging optical technologies to address the evolving needs of mobile communications networks. 
	
	\item Offering a comprehensive overview of the main optical technologies considered for the 6G fronthaul use cases, including P2P, PON and FSO (in particular, their suitability in various 6G fronthaul scenarios).
	\item Discussing the recent studies on optical fronthaul for 5G and beyond, additionally mentioning recent related projects to demonstrate the current state of the field.
	\item Highlighting potential future research directions aimed at advancing the efficiency of optical fronthaul for 6G mobile communications as well as exploring promising areas where further investigation and innovation are required to optimize the performance and capabilities of 6G optical fronthaul systems.
\end{enumerate}
\section{Evolution Toward 6G}\label{sec3}
\subsection{From 5G to 6G}
In this section, we provide a brief overview of the evolution of characteristics of wireless mobile networks starting from 1G up to the future 6G networks.
First Generation (1G) was introduced in the 1980s, provided basic analog voice communication but had limitations in capacity and data support. 

Second Generation (2G), in the 1990s, improved voice quality and introduced SMS and basic internet browsing \cite{1g1, 1g2}. Third Generation (3G) marked a pivotal moment with mobile internet, video calls, and multimedia messaging \cite{3g1, 3g2}. Fourth Generation (4G), commonly referred to as Long-Term Evolution Advanced (LTE-A), offered higher data rates, lower latency, and improved support for multimedia services \cite{4g1, 3g2}.
Fifth Generation (5G) mobile technology is the latest cellular wireless standard, promising improvements over its predecessor, 4G mobile technology, in different aspects (higher capacity, lower latency, etc.).

5G has a long history of development dating back to the early 2010s. Compared to the previous generations, 5G represents a significant leap forward, capable of meeting diverse communication requirements across three primary categories \cite{5g1, 5g2}, uRLLC supports mission-critical applications such as self-driving vehicles, unmanned aerial vehicles, industrial automation, and remote medical care. uRLLC ensures seamless and instantaneous data transmission with unparalleled reliability and minimal latency. eMBB offers support for bandwidth-demanding applications like high-definition video, three-dimensional video, cloud-based tasks, and AR/VR. mMTC facilitates extensive device connectivity and serves applications such as smart homes, buildings, cities, and IoT. mMTC enables a vast network of interconnected devices, fostering smart environments and efficient communication between machines. The technologies used for 5G include the latest advancements in the RAN, core network technologies, and new spectrum bands in the sub-6~GHz and mmWave range. 

The data rates for 5G can reach up to 20 Gbps for stationary users and multi-Gbps for mobile users, with the end-to-end transmission latency as low as 1 millisecond. The frequencies used for 5G include both traditional sub-6 GHz and higher mmWave frequencies with bandwidth in the range of hundreds of MHz. 

The use cases for 5G include high-speed mobile Internet access, VR/AR, autonomous vehicles, smart cities, and the IoT. However, as mobile network capacity and the number of connected devices grows, as do new traffic patterns and service diversification, there will be a large gap between 5G capabilities and market requirements projected
after 2030. As a result, academia and industry are already focusing on the next big evolution in the telecommunications industry, known as 6G mobile technology. New 6G applications will surely imply even more stringent Quality of Service (QoS), compared to 5G networks.

 As the 6G is in the early stage, it is mentioned in the literature as 5G+ and Beyond 5G (B5G). 6th generation systems shall provide extensive 3D coverage, spanning terrestrial, aerial, space, and sea domains, reaching remote and underserved areas and allowing anything to communicate anywhere and anytime. Moreover, they are expected to bring new capabilities such as providing data rates higher than  \mbox{1 Tbps} (50 times higher than 5G), latency lower than \mbox{100 $\mu$s} (more than 10 times lower than 5G), high reliability up to 7 nines (5 nines in 5G), and massive 3D coverage, among other benefits, as shown in Table~\ref{tab:wirelessg}. It is expected to play a critical role in shaping the future of wireless communications and enabling new possibilities in areas such as AI, IoE, and XR. Additionally, new specifications and technologies are planned to be introduced, such as new man-machine interfaces, universal computation distribution, data fusion, mixed reality, precision sensing, and controlling, as well as massive and scalable connectivity.\cite{6g1, 6g2, 6g3, 6g4,6g4, 6g6, hexa}. 
Table~\ref{tab:wirelessg} presents a comparison of the main characteristics for all consecutive cellular generations, i.e. ranging from 1G to 6G.
\begin{table*}
	\caption{A comparison of the main features of wireless network generations from 1G to 6G.}
	\label{tab:wirelessg}
	\begin{tabular}{|p{1.8 cm}|p{1.4 cm}|p{1.4 cm}|p{1.6 cm}|p{1.6 cm}|p{2.5 cm}|p{4.5 cm}|}
		\hline
		\textbf{Generation}     & \textbf{1G} & \textbf{2G} &\textbf{ 3G} & \textbf{4G} &\textbf{ 5G} & \textbf{6G}  \\ \hline
		 Beginning of operation &  1980&  1990 &  2000& 2010& 2020& 2030 (expected)  \\ \hline
		Data rate & 2.4 kbps& 64 kbps & 2 Mbps&100 Mbps& 20 Gbps    & $\ge$100 Gbps    \\ \hline
	 Bandwidth  & 30 kHz&200 kHz&5 MHz&1.25-20 MHz&0.25-1 GHz &Up to 3 THz \\ \hline
	Access &FDMA&TDMA, FDMA&WCDMA&OFDMA, MIMO& OFDMA, NOMA, Massive MIMO & AI, ML, SM-MIMO \\ \hline
	Min Latency & NA&$\ge$500 ms& 120 ms&1-5 ms& 1 ms& 100 $\mu$s \\ \hline
	Max Reliability &90\%&99\%&99.9\%&99.99\%&99.999\%&99.99999 \% \\ \hline
	Max Density &NA&NA&NA &10$^{5}$ devices/km$^{2}$&10$^{6}$ devices/km$^{2}$& 10$^{7}$ devices/km$^{2}$\\ \hline
	Max Mobility& 56 km/h&120 km/h& 250 km/h  &350km/h&500 km/h   &1000 km/h\\ \hline
	Energy and Spectrum Efficiency&Very low&Low&Medium&High&Very high&  Extreme\\ \hline
	Frequency&800 MHz&900 and 1800 MHz&from 1.6 to 2.1 GHz&1-2.5 GHz&3-300 GHz&   73-140 GHz, 1-10 THz\\ \hline
	Use cases& voice&voice, data&voice, data, multimedia&voice, data, video streaming, HD video, mobile TV&IoT, AR/VR, UHD video streaming, Vehicle-to-everything (V2X), smart life, smart healthcare, wearable devices &XR, tactile Internet, IoE, autonomous cars, holographic-type communication, digital sensing, digital Twins, space connectivity.     \\ \hline
	\end{tabular}
\end{table*}

\subsection{Major Challenges for 6G}
Moving toward 6G will bring several challenges for academia and industry in the radio and transport layers, particularly in the following aspects.  In the \emph{radio layer}, to meet the demands required by different bandwidth-hungry applications,  there is a need for utilizing new frequencies that have never been used before, such as the 7 GHz-20 GHz range, millimeter wave bands, sub-THz range of \mbox{100 GHz–300 GHz} for short-range applications, and the Terahertz frequency band ranging from 100 GHz to 10 THz  \cite{thz, thz2, thz3, thz4}. Moving deeper to the \emph{transport layer} (including fronthaul, midhaul, and backhaul), there is a need for \emph{resilient}, \emph{ultra-high-capacity} and \emph{ultra-low latency} transport networks from the cell site to the core network \cite{capacity, latency}. 

Additionally, the increased number of network connections may drive the adoption of shared connections instead of the current P2P links to lower both the \emph{cost of deployment and ownership} of 6G networks \cite{Integrated, xhaul}. Further, as the 6G is expected to provide connectivity at the 3D level, the \emph{Space-Earth integration} network is needed, which would use the stratosphere platform base stations and low-earth orbit satellites to provide full network coverage in remote areas \cite{ozger20236g}. This solution may open up new possibilities for various emerging transport technologies such as FSO \cite{sky}. 

Moreover, in both the radio and transport layers, the infusion of \emph{AI/ Machine Learning (ML)} is imperative for the realization of end-to-end intelligent 6G networks \cite{AI}.
The question of how to overcome these 6G challenges will manifest themselves remains a concern and will be a significant driver for innovation and research during the next decade.
	In our study, we place emphasis on the transport layer, specifically focusing on the pivotal role played by the fronthaul segment in advancing the development of future 6G networks and beyond.

\subsection{Lessons Learned}
From this section we highlight the following learning points:
\begin{itemize}
	\item Exploring new frequency ranges, including the 7 GHz-20 GHz range, sub-THz, and Terahertz bands, is crucial for addressing the bandwidth requirements of 6G. This entails continuous spectrum exploration, ensuring the adaptation of 6G to diverse needs in bandwidth-intensive applications and evolving communication demands.
	
	\item The transport layer's resilience, ultra-high capacity, and ultra-low latency are pivotal considerations for overcoming the challenges posed by 6G. Advances in transport networks, covering fronthaul, midhaul, and backhaul, are essential to support the varied requirements of emerging 6G services.
	
	\item Achieving 3D connectivity in 6G involves the integration of space and Earth-based solutions, leveraging stratosphere platform base stations and low-earth orbit satellites. Embracing space-based solutions opens new avenues for transport technologies like FSO, introducing innovative approaches to ensure seamless connectivity in 6G.
	\item Intelligent systems powered by AI/ML are imperative for addressing the complex challenges in 6G, emphasizing the role of adaptive algorithms and smart decision-making in network optimization. Integrating AI/ML technologies enhances the efficiency, adaptability, and self-optimization capabilities of 6G networks, laying the foundation for intelligent and autonomous communication systems.
\end{itemize}

\subsection{RAN Landscape Evolution Toward O-RAN}
The evolution of RAN has been a continuous process aimed at improving performance, increasing its coverage and capacity, and reducing the latency and deployment cost in each generation of mobile technologies, i.e., 1G to 6G \cite{ran1, ran2, ran3, ran4}.

Starting with the \emph{Distributed RAN (D-RAN)}, that was introduced as the first generation of RAN architectures, where the BBU was collocated with RRHs at the cell site. RRHs handle the wireless signal transmission to the end user, while BBUs manage baseband processing. The architecture relies on a high-speed backhaul network connecting the BBU to the core network \cite{ran1}. 
 D-RAN architecture has primarily been utilized in wireless generations up to 4G. However, it is not cost-efficient and lacks the flexibility and scalability required for advanced features such as network slicing and edge computing. For that, it is not suitable for 5G and future 6G.

\begin{figure}[ht]
	\centering
	\includegraphics[width=0.4\textwidth]{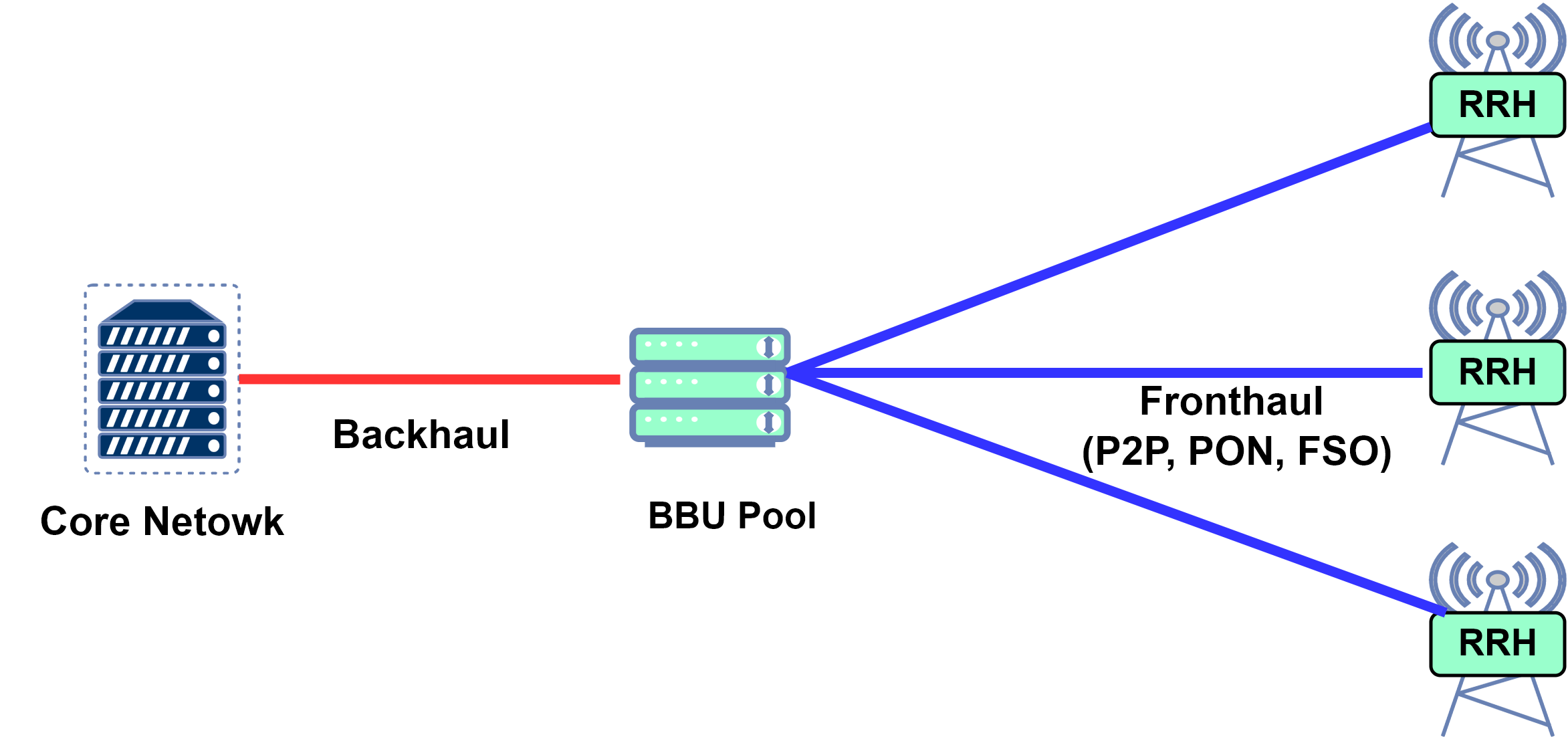}
	\caption{C-RAN architecture.}
	\label{fig:CRAN}
\end{figure}

Moving forward, \emph{Centralized RAN (C-RAN)} was introduced as a solution to address the limitations of D-RAN. In C-RAN, the BBUs are consolidated into a single centralized BBU pool. This BBU pool is connected to RRHs distributed throughout the coverage area via a high-capacity fronthaul network and to the core network via a backhaul link \cite{cran}, as depicted in Fig~\ref{fig:CRAN}.
C-RAN architecture supports advanced Radio Signal Processing (RSP) techniques, such as Coordinated Multi-Point (CoMP), Carrier Aggregation (CA), and Enhanced Inter-Cell Interference Cancellation (eICIC). These techniques effectively mitigate intra-tier and inter-tier interference, improving the overall network performance and enhancing user experience. Additionally, it reduces the costs associated with deploying and maintaining multiple RAN sites. This consolidation also enables efficient resource allocation, optimizing network capacity and spectrum utilization..
However, implementing C-RAN requires high-speed fronthaul connectivity to ensure seamless communication between the BBU pool and RRHs which presents challenges in terms of cost and infrastructure requirements. Nonetheless, the benefits offered by C-RAN, make it suitable architecture for 5G and beyond cellular networks as discussed in \cite{ran5, ran6}.

\emph{Heterogeneous C-RAN (HC-RAN)} emerged to meet the demands of 5G dense networks, integrating small BSs with macro BSs to create a heterogeneous environment.
The macro BSs play a~crucial role in providing network control, mobility management, and overall system performance enhancement. 
The smaller BSs and RRHs are strategically deployed throughout the network, complementing the macro BSs. By doing so, they effectively improve the system capacity while reducing the power consumption required for transmission.
By leveraging cloud computing, HC-RAN can efficiently distribute computational tasks and manage network resources dynamically and flexibly \cite{hran}. 
Moreover, it improves the network capacity and reduces the power consumption The authors of \cite{hcran6g} showed that the HC-RAN architecture can achieve up to 56\% improvement in average throughput. Additionally, the authors of \cite{hcran6g1} showed that, compared to C-RAN, the HC-CRAN can achieve up to 40\% improvement in network throughput, 30\% improvement in energy efficiency, and 20\% improvement in user satisfaction ratio.

\emph{Fog-RAN (F-RAN)} was presented as a solution to meet the increased demand for latency reduction in 5G networks by decentralizing some of the network functions closer to the end users \cite{fran}. F-RAN takes advantage of the fog computing proximity to the RRHs (see \cite{fogcompution}), thereby minimizing data transmission delay and enhancing user experience.
The distributed architecture of F-RAN optimizes data processing at the network edge, reducing congestion by offloading computational tasks from central servers to fog nodes. This approach enhances overall network efficiency, particularly in 5G networks, addressing latency demands for applications like autonomous vehicles and IoT. F-RAN's decentralized execution of critical functions, as demonstrated in \cite{fran}, improves resource utilization and eases the load on the fronthaul network. Notably, experiments in \cite{frantrail} showcase F-RAN's advantages, including a 70\% reduction in end-to-end latency, up to 50\% throughput improvement, and a 40\% reduction in energy consumption compared to C-RANs, affirming its potential for 6G success. 
In response to the growing demand for the softwarization of RAN architectures, the concept of \emph{virtualized RAN (v-RAN)} has recently emerged. In v-RAN, conventional hardware-based RAN functions are ingeniously supplanted by software-based solutions seamlessly executed on commercially available Commercial Off-The-Shelf (COTS) servers, often referred to as virtual BBU (vBBUs). This shift eliminates reliance on proprietary hardware, leading to cost reductions and improved scalability. 

Additionally, v-RAN offers high flexibility and agility in responding to evolving network requirements and cost savings. By leveraging software-based solutions, operators can swiftly adapt and deploy new services, features, and updates across the network. The decoupling of hardware and software allows for easier upgrades and scalability, as operators can add new virtual instances of BBUs without needing additional physical equipment \cite{vran, vRANValue}.
v-RAN architecture finds applications across multiple wireless generations, including 4G, 5G, and beyond. Its flexibility and scalability make it suitable for evolving networks that require efficient resource allocation, rapid service deployment, and cost-effective scaling. As stated in \cite{vran1}, v-RAN architecture can achieve power consumption reduction of up to 46.1\% and 84.1\% compared to C-RAN and D-RAN, respectively, as well as network throughput improvement up to 25\%.

O-RAN was recently introduced as an industry initiative aiming to revolutionize the RAN for future mobile networks by creating an open, interoperable, intelligent network infrastructure \cite{oran1}. The primary objective of O-RAN is to break the traditional barriers of proprietary hardware and software solutions, fostering a more competitive and innovative RAN market. Established in 2018, the \mbox{O-RAN} Alliance is the driving force behind this initiative, bringing together telecommunications companies, system vendors, and service providers to collaboratively define open interfaces and develop common specifications for various RAN components.
O-RAN introduces a new level of flexibility and cost-effectiveness in the deployment and operation of RAN, empowering network operators to adapt swiftly to changing market conditions and meet evolving customer demands.

\begin{figure}[ht]
	\centering
	\includegraphics[width=0.4\textwidth]{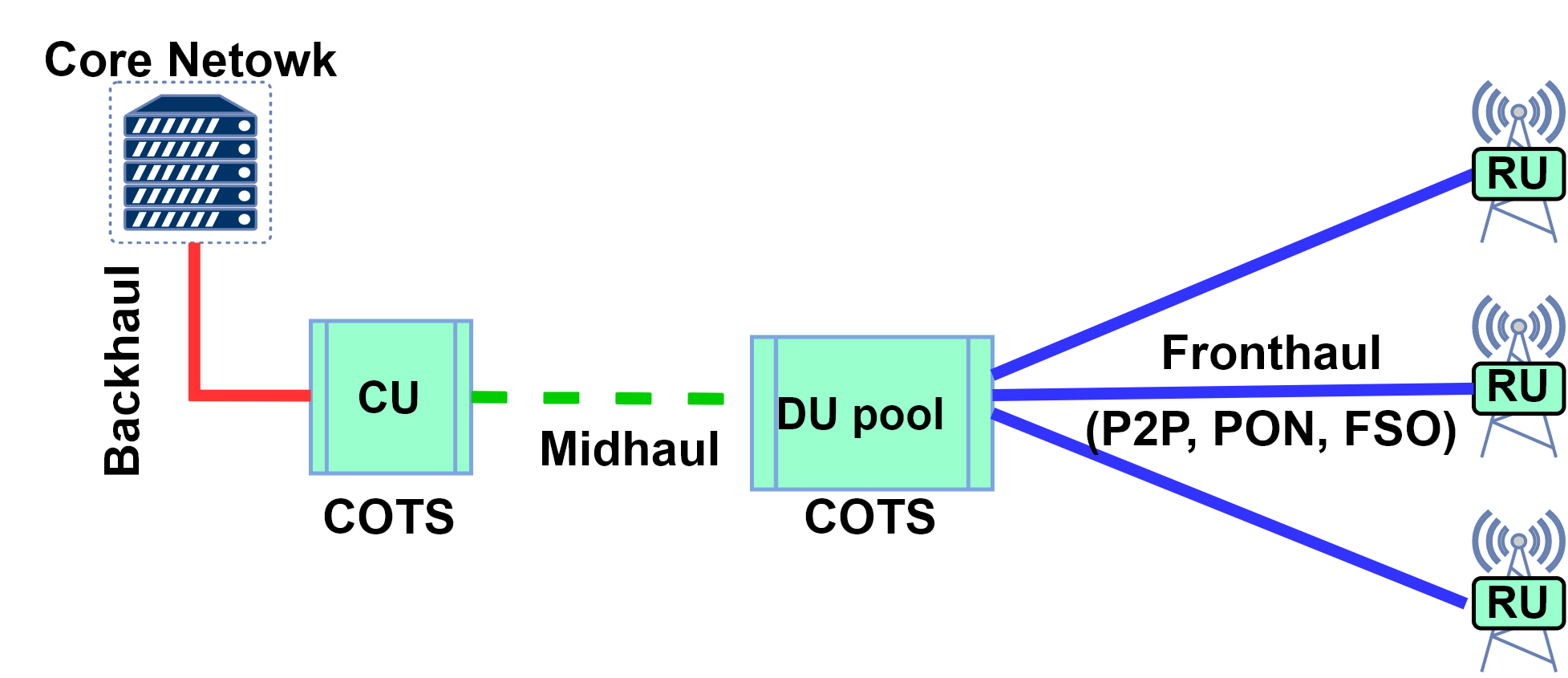}
	\caption{O-RAN architecture.}
	\label{fig:ORAN}
\end{figure}

Incorporating open interfaces within O-RAN allows for the seamless integration of diverse RAN components from multiple vendors, reducing vendor lock-in and promoting the smooth deployment of new services and applications. This interoperability enables network operators to mix and match components according to their specific requirements, fostering a more dynamic and efficient network environment.
One of the significant advantages of O-RAN lies in its open architecture, which provides an ideal platform for implementing AI and ML algorithms. Network operators can leverage these advanced technologies to optimize network performance, efficiently allocate network resources, and enhance the overall user experience by dynamically adapting to changing traffic patterns and user behavior, \mbox{O-RAN} facilitates intelligent network management, ensuring optimal service quality and network efficiency \cite{oran1, oran2, oran22, ran3, oran3}. 
Figure~\ref{fig:ORAN} illustrates the architecture of O-RAN.

The development of RAN architecture has been driven by the need to improve network performance and efficiency, as well as to reduce costs. Each evolution has brought new advancements and capabilities to the network, making it more flexible and scalable to meet the increasing demands of users.
Since C-RAN has been proposed as a proper solution for 4G and 5G, the performance characteristics of its successors discussed in this section show their high potential of being applied in beyond 5G and future 6G architectures.  
In the context of 6G, v-RAN and O-RAN tend to be favored because of the need for network slicing, agility, dynamic resource allocation, and rapid deployment. However, despite their promising potential, implementing an efficient fronthaul poses a hurdle to practically implementing these architectures in real-world applications \cite{related2}. 
The primary obstacle in implementing an efficient fronthaul lies in the complex and diverse nature of the devices. In the future 6G networks, the number of devices connected to the network is expected to dramatically increase,  including smartphones, tablets, wearable devices, IoT devices, and autonomous vehicles, each with unique requirements and characteristics.

To ensure optimal performance and seamless connectivity for this type of device, the fronthaul infrastructure must be capable of handling the diverse traffic demands, bandwidth requirements, and latency constraints of each device. This necessitates the development of sophisticated techniques and protocols for efficient data transmission, synchronization, and management across the fronthaul network.

In addition to the technical challenges, the cost of implementing fronthaul is another significant factor that limits the practical implementation of RAN architectures in real-world applications. Deploying and maintaining a robust fronthaul infrastructure can financially burden network operators and service providers.

Furthermore, the scalability of the fronthaul infrastructure adds to the cost consideration. As the number of devices and network traffic increases, other RUs must be deployed to expand the network capacity. This implies additional fronthaul connections, which can contribute to the overall cost. Upgrading or modifying the existing fronthaul infrastructure to accommodate new technologies and higher bandwidth requirements can incur significant expenses. The benefits and challenges of each RAN architecture are summarized in Table~\ref{tab:ran}.

\begin{table*}[htbp]
	\centering
	\caption{ Comparison of RAN architectures.}
	\begin{tabular}{|m{0.75 cm}|p{7.5 cm}|p{7.5 cm}|}
		\hline
		\multicolumn{1}{|c|}{\textbf{RAN architecture}} & \textbf{Benefits} & \textbf{Challenges}\\
		\hline
\multicolumn{1}{|c|}{D-RAN} & Reduced latency; improved network performance through the distribution of baseband processing; simple architecture. & Higher operational costs due to a distributed nature; limited scalability; increased complexity in network management and deployment; unfair resource allocation and energy inefficiency; vendor lock.\\\hline
	\multicolumn{1}{|c|}{C-RAN} & Improved network scalability and efficiency through centralization of network functions; improved network flexibility through the use of cloud technologies; reduced Capital expenditures (Capex) and Operating expenses (Opex); spectral and energy efficiency.& The need for high capacity; low latency fronthaul connection;  BBU complexity;  fronthaul deployment cost; vendor dependency.
\\\hline
	
	\multicolumn{1}{|c|}{HC-RAN} & Less frotnhaul constraints compared to C-RAN; combines benefits of D-RAN \& C-RAN; improved coverage and capacity;  supports heterogeneous networks.&	Increased complexity in network management and deployment; requires advanced coordination and interference management; fronthaul deployment cost; vendor dependency.\\\hline
	
	\multicolumn{1}{|c|}{F-RAN} & Improved network performance; enhanced latency and reliability; lower BBU complexity; lower fronthaul constraints; distributed processing at edge nodes; supports network slicing.&Increased complexity in network resource management and deployment; inter-tier interference; fronthaul deployment cost; vendor dependency.\\\hline
 	
	\multicolumn{1}{|c|}{v-RAN} & Faster deployment of new services and applications; cost reduction; network functions virtualization; improved scalability; improved network flexibility.& High computational requirements; requires robust and reliable virtualization platforms; fronthaul deployment cost.\\\hline
	\multicolumn{1}{|c|}{O-RAN} & Cost reduction; open interfaces for multi-vendor interoperability; encourages innovation and competition; simplifies network management; improved flexibility. & Increased complexity in network management and deployment; interoperability issues; standardization and compatibility challenges; security concerns due to open interfaces; integration with current technologies challenges; fronthaul deployment cost.\\\hline
		 
	\end{tabular}
	\label{tab:ran}
\end{table*}
\subsection{Lessons Learned}
We can conclude the following lessons from RAN evolution:
\begin{itemize}
	\item The evolution of RAN architectures underscores the industry's commitment to iterative improvement, emphasizing the need for continual adaptation to meet evolving communication demands.
	
	\item While each RAN architecture brings innovations, balancing benefits and challenges is crucial. The industry must carefully consider factors such as deployment costs, scalability, and technical complexities to ensure practical and effective implementation.
	
	\item Recognizing the challenges in efficient fronthaul implementation, including diverse device nature, scalability issues, and high deployment costs, highlights the need for addressing these challenges to enable the practical application of advanced RAN architectures, especially in the context of 6G networks.
\end{itemize}

\section{6G Fronthaul Interface and Main Splitting Options}\label{sec4}
\subsection{Fronthaul Interface}
The fronthaul typical interfaces include Common Public Radio Interface (CPRI), ethernet CPRI (eCPRI), Open Radio Interface (ORI), the Open Base Station Architecture Initiative (OBSAI), and Open-RAN 7.2x \cite{related2, fronthaul1, fronthaul2, fronthaul3}. 
 As illustrated in Fig.~\ref{fig:2}, there are several functional layers and detailed as follows:

\begin{figure*}[htbp]
	\centering
  \includegraphics[width=0.9\textwidth]{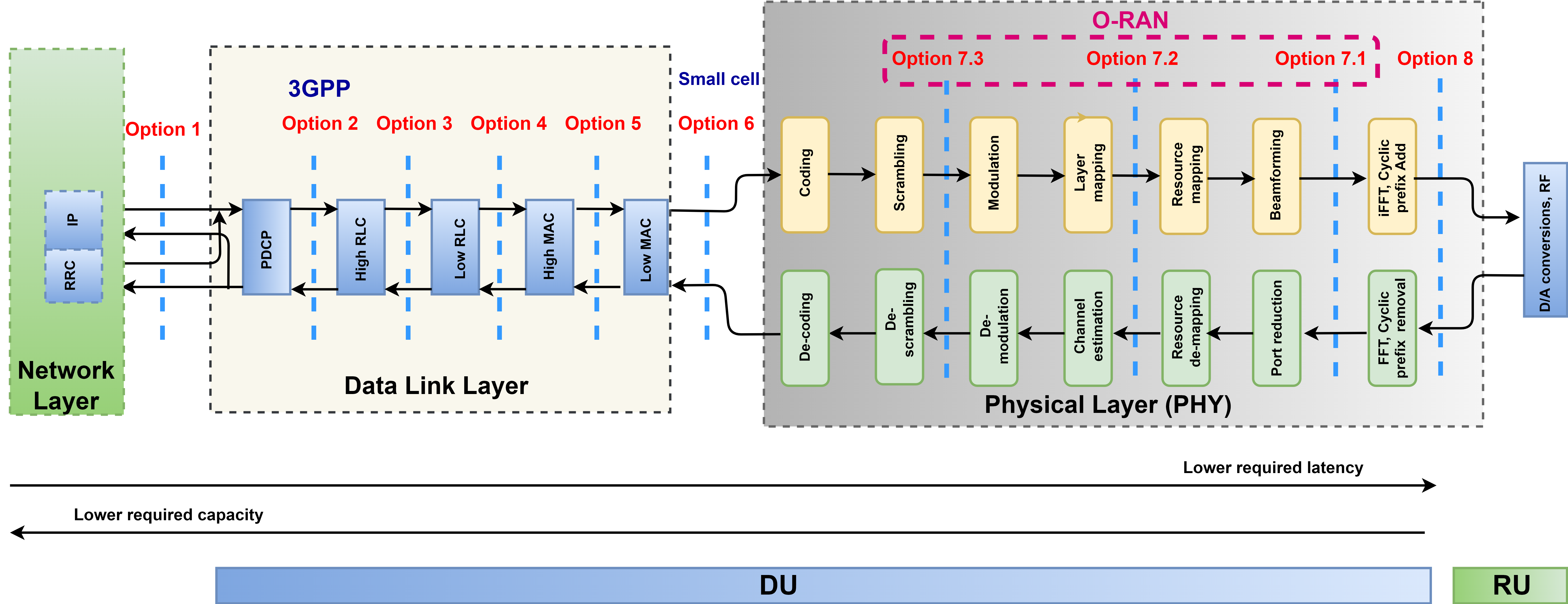}
  \caption{RAN functional splitting options.}
  \label{fig:2}
\end{figure*}

 \begin{itemize}
 	\item \textbf{Radio Resource Control (RRC):} This layer plays a~crucial role in transmitting control plane signals between the base station and the end user. It handles the necessary signaling procedures for establishing and maintaining the radio connection.
 	\item \textbf{Packet Data Convergence Protocol (PDCP):} This layer focuses on two main tasks. First, it compresses the header of the data packets to optimize bandwidth usage. Second, it ensures the confidentiality and integrity of the data through encryption and data integrity checks.
 	\item \textbf{Radio Link Control (RLC):} This layer is responsible for ensuring reliable and ordered data transmission in both the downlink and uplink directions. It handles tasks such as segmenting and reassembling data packets, error detection, and retransmissions, if needed.
 	\item \textbf{Medium Access Control (MAC):} This layer oversees the Hybrid Automated Repeat Request (HARQ) mechanism, allowing error correction and retransmission of data packets. The MAC layer also handles the multiplexing and prioritization of logical channels, ensuring efficient use of the available resources. Additionally, it schedules the transmissions, determining when and how data is sent over the wireless network.
 	\item \textbf{Physical layer (PHY):} The role of this layer encompasses various tasks related to the physical transmission of data. It handles the modulation and demodulation of signals, converts data into electromagnetic waves for transmission over the air interface, and performs channel coding and decoding. Additionally, the PHY layer is responsible for managing power control, channel estimation, and adapting the transmission parameters to ensure efficient and reliable communication. It also plays a crucial role in implementing advanced technologies such as MIMO and beamforming for enhancing spectral efficiency and overall system performance.
 \end{itemize}

 The 3rd Generation Partnership Project (3GPP) has defined several options for splitting RAN functions as  trade-offs between gains following from a centralized solution and fronthaul bandwidth needs \cite{split}.
 \subsection{Main Splitting Options}
  Eight functional splits have been introduced to address different fronthaul split options, ranging from (higher-layer split) Option 1 to (lower-layer split) Option 8 (RF/PHY), as illustrated in Fig.~\ref{fig:2}. These options offer various trade-offs in terms of complexity, fronthaul capacity, latency, and jitter, catering to different deployment scenarios and performance requirements \cite{sp1}. 
  Those splitting options starting from Option~8 and ending with Option 1 are as follows:
  \begin{itemize}  	
\item \textbf{Option 8 (Lower-Layer Split, RF/PHY)}: This split represents  a fully centralized architecture, where all functions, including the RF and PHY processing, are moved from the RU to the DU. This split significantly reduces the complexity of the RU structure, as it only requires basic analog and digital processing. However, it comes with a trade-off: the required fronthaul capacity is high, especially as the number of antennas increases. Additionally, this split is highly latency- and jitter-constrained due to the centralized processing in the DU.
  
 \item \textbf{Option 7 (Low PHY)}  This split moves the PHY function to the RU, while other higher-layer functions remain in the DU. This split offers a reduction in the required fronthaul bitrate compared to Option 8. However, it lacks symmetry between the uplink and downlink characteristics, meaning that the required bitrate and bandwidth scales with the number of used antennas. This split is suitable for scenarios where the reduction in fronthaul capacity outweighs the asymmetry in uplink and downlink requirements.
 
  \item \textbf{Option 6 (MAC–PHY split)} This option separates the physical layer (PHY) and the data link layer (MAC-PHY). The RU handles PHY processing, while the higher-layer MAC processing resides in the DU. This split achieves a lower fronthaul capacity requirement compared to the previous options. It is particularly well-suited for small cell deployments, where the lower complexity of the RU can be advantageous
  
 \item \textbf{Option 5 (Intra MAC)} In this split, all time-critical processing operations are performed in the DU. This includes MAC processing, scheduling, and control functions. By centralizing these functions, the split reduces the delay requirements for time-sensitive operations. However, it also leads to a more complex fronthaul interface to support the increased coordination and signaling between the DU and RU. 
  
  \item \textbf{Option 4 (RLC/MAC)} This split  separates the functions related to the Radio Link Control (RLC) and Medium Access Control (MAC) layers. The RU handles the transmission of MAC Service Data Units (SDUs) in the uplink direction and receives RLC and Protocol Data Units (PDUs) in the downlink direction. This split requires a relatively low fronthaul data rate. By centralizing the Automatic Repeat Request (ARQ) function in the DU, this split can enhance resilience to non-ideal transmission conditions and mobility scenarios.
  
  \item \textbf{Option 3 (Intra RLC)} This split divides the RLC functionality into high RLC and low RLC. The DU performs asynchronous RLC processing and PDCP processing, while other RLC functions, including synchronous RLC network operations, remain in the RU. This split is particularly sensitive to latency in some cases, as it involves distributed processing and coordination between the DU and RU. 
  
  \item \textbf{Option 2 (RLC/PDCP)} establishes a split between the RLC and the Packet Data Convergence Protocol (PDCP) functions. In this split, all real-time components remain in the RU, resulting in lower data rate and latency requirements compared to previous options. The centralization of real-time processing in the RU makes it suitable for scenarios where lower complexity and lower fronthaul capacity are desired. 
  
  \item \textbf{Option 1 (PDCP/RRC)} This split places the RRC function in the DU, while the PDCP, RLC, MAC, PHY, and RF functions are located in the RU. Option 1 relaxes the required latency and bit rate compared to other splits. However, it comes with a higher level of complexity in the RU structure due to the inclusion of multiple functions.
\end{itemize}

\vspace{4 mm}
The O-RAN is a new mobile network, has recently come to light as a platform for fully interoperable, open, intelligent, virtualized RAN systems. The O-RAN framework facilitates compatibility between different vendors' equipment by adopting open interfaces and standardized protocols, promoting vendor diversity and avoiding vendor lock-in \cite{oran1, oran2}. 

At the core of the O-RAN architecture lies the division of the traditional base station into three distinct components, each playing a crucial role in the network ecosystem. The first component is the Open-Radio Unit (O-RU), which handles the radio frequency transmission and reception, enabling wireless communication with user devices. The second component is the Open-Distributed Unit (O-DU), which is responsible for processing and routing data between the O-RU and the core network. Lastly, the Open-Central Unit (O-CU) is the central intelligence hub, managing the overall network orchestration and optimization. One of the key elements defining the \mbox{O-RAN} architecture is the concept of Open Fronthaul, which refers to the connection between the O-RU and the O-DU.

In the O-RAN, which is the the candidate architecture for future 6G, new functional splits have been introduced, enhancing flexibility and adaptability. The previously established Splitting Option 7 has been further divided into three sub-splits: 7.1, 7.2, and 7.3. Collectively referred to as the 7.x splitting family, these sub-splits enable network operators to tailor their network deployments to specific use cases, optimizing performance and resource allocation \cite{7x}.

These new splitting options offer enhanced functionality by dividing the PHY between the RU and the DU, providing greater flexibility and efficiency in wireless communication systems. The new splitting options can be summarized as follows:
\begin{itemize}
	\item \textbf{Option 7.1:} In this split, the In-phase and Quadrature (I/Q) symbols are transmitted in the frequency domain. This split effectively avoids the overhead caused by frequency-to-time conversion, resulting in improved performance. However, it is essential to note that this option requires high bandwidth and low latency as the bitrate scales with the number of MIMO layers. The more MIMO layers used, the higher the data rate.
	\item \textbf{Option 7.2:} This split is similar to Option 7.1, as it also transmits I/Q signals in the frequency domain. However, in this case, the signals from multiple antenna ports are combined, which reduces the required bandwidth compared to Option 7.1. This combination of signals allows for efficient data transmission while maintaining a high level of performance. 
	\item \textbf{Option 7.3:} This split takes the concept of split functionality even further by achieving a significant reduction in fronthaul bit rate, compared to Options 7.1 and 7.2. It accomplishes this by allocating additional functionality, such as demodulation and modulation, near the antennas at the RU. This localization of functionality enables more efficient use of resources, reducing the required capacity for fronthaul transmission.
	
	For example, considering a cell bandwidth of 10 MHz or 20 MHz, 2x2 MIMO, 4 antenna ports, and 16-QAM modulation, the required capacity for the 7.x splitting options would be as follows: 134.4 Mbps for Option~7.3, 1075.2 Mbps for Option 7.2, and 4300.8 Mbps for Option 7.1 ~\cite{7x}. These figures demonstrate the increased efficiency achieved by Option 7.3 in reducing the required bit rate for fronthaul transmission.
	
	However, it is important to note that Option 7.3 is characterized by a  trade-off of increased complexity in the RU structure compared to Options 7.1 and 7.2. This complexity arises due to the allocation of additional functionality, such as demodulation and modulation at the RU. Despite the increased complexity, Option 7.3 offers substantial benefits regarding the fronthaul bit rate reduction, making it an attractive choice in specific scenarios.

\end{itemize}

Realizing an efficient fronthaul is a critical and complex task that poses significant challenges for mobile network operators in designing 5G and beyond networks. The fronthaul must address the demanding requirements of high data rate and low latency communications inherent in these advanced networks~\cite{Integrated}. To illustrate the capacity and latency demands, let's consider the example of splitting option 8. In this scenario, a~5G system is required to support the average data rate of 1.5 Gbps at an RU. The RU operates with \mbox{64 QAM} modulation, an 8x12 MIMO antenna array, 200 MHz bandwidth, and 96 antenna ports. This particular configuration results in the fronthaul capacity need exceeding 800 Gbps while maintaining a latency requirement of less than \mbox{250 $\mu$s \cite{bitrate}}.  

Moreover, the significance of fronthaul extends beyond just meeting the demands of data rate and latency. It plays a crucial role in deploying macro and small cells in future networks, facilitates distributed antenna systems (DAS), and supports the evolution toward RAN openness and virtualization. Fronthaul is a vital link enabling seamless integration and coordination of diverse network elements.

To achieve an efficient fronthaul, various optical communication technologies can be employed, each suitable for different splitting options such as P2P, PON, and FSO. The following subsection discusses these technologies and their applicability to fronthaul solutions. Table~\ref{tab:split} shows the bandwidth (transport capacity) and latency requirements for different fronthaul options and the associated optical technologies.

\begin{table}[htbp]
	\centering
	\caption{Requirements on different fronthaul split options based on \cite{split, split1, skubic2017optical} (DL-- Downlink capacity in Gbps; UL-- Uplink capacity in Gbps).}
	\begin{tabular}{ |c|c|c|c|c|}
		\hline
		\textbf{Split option}  & \textbf{DL}& \textbf{UL} & \textbf{Latency}  & \textbf{Possible technology}\\
		\hline
		1& 1&1 & 10 ms & \multirow{7}{*}{P2P, PON, FSO} \\\cline{1-4}
		2& 1&1&1.5–10 ms & \\\cline{1-4}
		3& 1&1 & 100  $\mu$s–10 ms &\\\cline{1-4}
		4& 1& 1&10–200  $\mu$s &\\\cline{1-4}
		5& 1&1& 10–200  $\mu$s&\\\cline{1-4}
		6& 1.2& 1.2& 10–200  $\mu$s&\\\cline{1-4}
		7.3&3.2 &3.2 & 10–200  $\mu$s&\\\cline{1-4}
		7.2&6&2& 10–200  $\mu$s&\\\hline
		7.1& 323& 323&10–200  $\mu$s& \multirow{2}{*}{P2P}\\\cline{1-4}
		8&885 &885 &10–200  $\mu$s&\\\hline
	\end{tabular}
	\label{tab:split}
\end{table}

\subsection{Lessons Learned}
We highlight the following learning points from this section:
\begin{itemize}
	\item The 6G fronthaul interface involves multiple functional layers such as RRC, PDCP, RLC, MAC, and PHY. It is crucial to understand their roles and splitting options, ranging from fully centralized (Option 8) to distributed (Option 1). Each option presents trade-offs regarding fronthaul capacity, latency, and complexity, addressing diverse deployment scenarios and performance requirements.
	
	\item The emergence of O-RAN introduces a paradigm shift in mobile networks, promoting interoperability, openness, and intelligence. The O-RAN architecture divides the traditional base station into O-RU, O-DU, and O-CU, enhancing flexibility. New splitting options (7.1, 7.2, 7.3) provide tailored solutions for specific use cases, optimizing performance and resource allocation. The trade-offs involve bandwidth, latency, and RU complexity.
	
\end{itemize}
\section{Optical Communications Technologies for 5G/6G Fronthaul}\label{sec5}
\subsection{Motivation for 6G Optical Fronthaul}
Based on the ITU-T report, by 2030, the 6G technology peak data rate will be $\geq$ 1 Tbps or at least 50 times faster than that in 5G, with a deliver latency $\leq$ 0.1 ms or at least 10 times lower than in 5G  under all conditions \cite{forcast2030}. The "Mobile Backhaul and Fronthaul Market" Research Report for 2023 projects significant growth in the global market. As of 2022, the market size was valued at USD 15,460 million, and it is expected to witness substantial expansion, reaching USD 56,340 million by 2030, with a remarkable Compound Annual Growth Rate (CAGR) of 18.7\% during the forecast period of 2023-2028 \cite{opticalforcast}. Moreover, as we mentioned in the introduction, the amount of global data traffic and the number of devices connected to the network is anticipated to grow exponentially during the next decade. Alongside this, there is an increase in the number of high bandwidth-low latency-demanding applications. That will significantly impact the future 6G fronthaul, a critical segment of O-RAN that provides high throughput, low delay, high reliability, and secure communications.

One possibility is to leverage various optical communication technologies to fulfill these requirements concerning the fronthaul for 5G and beyond. P2P optical fiber and Point-to-MultiPoint (P2MP) technologies like PON are notably recognized for their exceptional ability to provide high-capacity and low-latency data transmission. However, despite their advantages, these technologies often come with high costs and limited deployment options, potentially limiting their widespread implementation. To overcome these challenges, FSO has emerged as a viable alternative to wired optical fibers, especially in scenarios where physical or cost constraints exist. 
In this section, we provide a comprehensive understanding of these architectural approaches, outlining their distinct features and benefits for 6G fronthaul implementation. To achieve this, we address the following key questions:
\begin{enumerate}[label=\Alph*]
	\item Is the architecture suitable for 5G/6G fronthaul?
	\item What splitting configurations can the respective architecture accommodate?
	\item What benefits and drawbacks does each architecture offer?
\end{enumerate}

\subsection{Point-to-Point Optical Fiber (P2P) for 5G and Beyond Fronthaul}
\begin{figure}[htbp]
	\centering
		\includegraphics[width=0.4\textwidth]{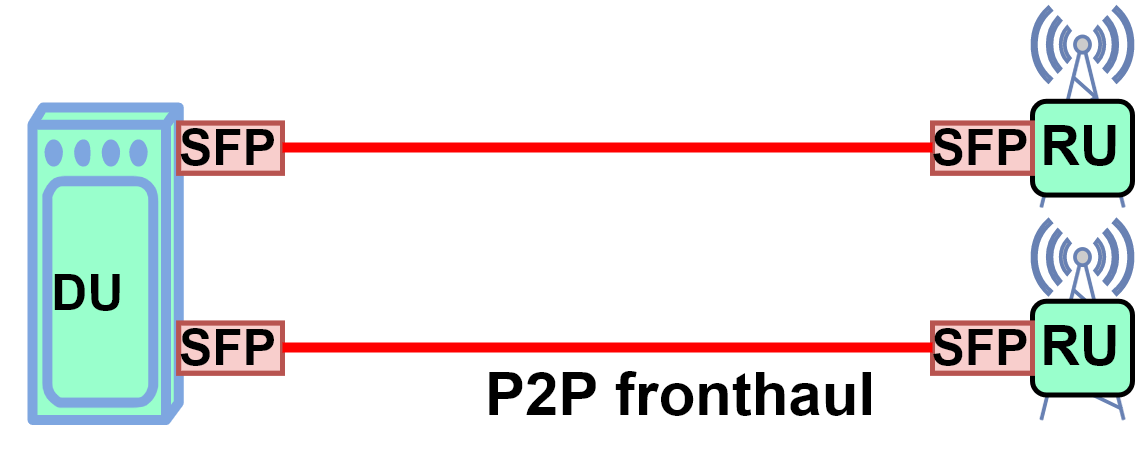}
		\caption{P2P for 5G and beyond fronthaul.}
		\label{fig:p2p}
	\end{figure}

This architecture provides dedicated, high-capacity optical links or Wavelength Division Multiplexing (WDM) links between individual RUs and the DU pool without intermediate networking equipment. Figure~\ref{fig:p2p} illustrates an example of a P2P configuration designed for 5G and beyond fronthaul applications. P2P fronthaul has several advantages, including \emph{reduced latency} (dedicated links reduce latency by minimizing contention and packet delays inherent in shared fronthaul architectures), \emph{secure data transmission}, \emph{simplified troubleshooting}, \emph{increased capacity}, and \emph{better network management}. It also allows for more flexible deployment and higher scalability of the network. This architecture is particularly beneficial in areas rich with dark fiber resources.
P2P architecture is needed in several use cases, such as (1) highly available and secure connectivity requirements and (2) ultra-high bandwidth demands \cite{p2p2, p2p3}. Also, this architecture may improve transmission availability, as any branch failure will not influence the others \cite{p2p}.
	
Despite its advantages, P2P fronthaul has its drawbacks, notably the need for extensive fiber resources, which can lead to cost inefficiencies due to the high expense of deploying optical fibers. An additional drawback is the lack of flexibility within the P2P architecture. In this configuration, each endpoint allocates resources based on peak traffic demand. Moreover, given the dynamic nature of real-time traffic, fixed configurations and independent P2P transmissions create a challenge wherein idle endpoints are unable to share resources with busy ones. This limitation hinders the achievement of efficient overall resource utilization in the P2P architecture.

To overcome the limitations of P2P architecture and usher in a new era of network economics. It is particularly relevant as operators face new challenges with the advent of 5G, fiber-deep, and hyper-scale cloud connectivity. \emph{XR optics} utilizes coherent optical subcarrier aggregation to present a novel, pluggable, and software-driven framework that aims to lower the cost of deploying and managing optical networks substantially. Additionally, XR optics offers enhanced flexibility in deployment, allowing a single coherent pluggable to be effortlessly reprogrammed through software to function in  P2P or P2MP configurations, catering to diverse networking requirements \cite{Infinera2022XROptics}. 

In 5G and beyond fronthaul, P2P is a viable option when there is a demand for high capacities, such as in splitting options 7.1 and 8, where the required fronthaul capacity per RU will exceed 300 Gbps. As highlighted in \cite{p2p1}, the capacity of 100 Tbps can be achieved using advanced optical fiber techniques such as a  modulation format called Multi Dimensional Optimized Constellation (MDOC). Moreover, to ensure the low latency of the fronthaul, which is set at \mbox{100 $\mu$s} or less, the length of the optical fiber fronthaul can not exceed 10 km given that the one-way propagation delay of approximately 5  $\mu$s/km in optical fiber. Table~\ref{tab:sfp} provides an overview of various optical transceivers suitable for P2P architecture, along with their corresponding applications. The main applications can be listed as follows.

	\begin{itemize}
		\item \textbf{Fibre Channel} being a rapid data transfer protocol primarily employed in Storage Area Networks (SANs), enabling dependable and expandable communication between servers and storage devices (ensuring reliable data transfer,  while also being able to adapt to changing network demands). It facilitates the swift transfer of data at high transmission rate, reaching several Tbps \cite{fiberchannel}. 
		\item \textbf{Gigabit Ethernet} referring to a commonly adopted Ethernet-based networking standard that facilitates data transmission at a speed of 1 Gbps. It is used in Local Area Networks (LANs) in residential and commercial environments. By enabling swifter data transfer between computers and network devices, such as routers and switches, Gigabit Ethernet surpasses slower Ethernet variants like Fast Ethernet (100~Mbps) and Ethernet (10~Mbps) \cite{ethernet}.
		\item \textbf{InfiniBand} being  a high-performance interconnection technology that offers low-latency and high-speed data transfer between nodes in High Performance Computing (HPC) systems and data centers. It aims to establish a fast and responsive connection between servers, storage systems, and various devices in a data center or HPC cluster \cite{edr}. The different InfiniBand types are as follows:
		\begin{enumerate}
			\item \textit{Quad Data Rate (QDR):} This type of InfiniBand technology provides a data rate of 40 Gbps (4x10~Gbps).
			\item \textit{Enhanced Data Rate (EDR):} EDR InfiniBand supporting a data rate of 100 Gbps (4x25 Gbps).
			\item \textit{High Data Rate (HDR):} HDR InfiniBand offering a~data rate of 200 Gbps (4x50 Gbps)
			\item \textit{Next Data Rate (NDR):} HDR InfiniBand offering a~data rate of 200 Gbps (4x50 Gbps)
			\item \textit{Extreme Data Rate (XDR): } XDR InfiniBand supporting data rates of 800 Gbps (4x200 Gbps).
			
		\end{enumerate}
	\end{itemize}

\begin{table}[htbp]
	\centering
	\caption{ Different optical transceiver types for P2P configuration based on \cite{p2p4, fronthaul4, p2p5, p2p6}.}
	\begin{tabular}{|c|p{1 cm}|c|}
		\hline
		\multicolumn{1}{|c|}{\textbf{Type}} & \textbf{Capacity (Gbps)} & \textbf{Application or Use-Case} \\
		\hline
		SFP & \multicolumn{1}{|c|}{0.1-1}  & Gigabit Ethernet, Fibre Channel \\\hline
		SFP+ & \multicolumn{1}{|c|}{10}  & 10 Gigabit Ethernet, 8/16G Fibre Channel \\\hline
		SFP28 &\multicolumn{1}{|c|}{ 25} & 25 Gigabit Ethernet, 32G Fibre Channel \\\hline
		SFP56 & \multicolumn{1}{|c|}{50} & 50 Gigabit Ethernet, 64G Fibre Channel \\\hline
		SFP-DD & \multicolumn{1}{|c|}{100} & 100 Gigabit Ethernet, 128G Fibre Channel \\\hline
		QSFP &\multicolumn{1}{|c|}{ 4} & 4x1 Gigabit Ethernet \\\hline
		QSFP+ & \multicolumn{1}{|c|}{40} & 40 Gigabit Ethernet, InfiniBand QDR \\\hline
		QSFP28 & \multicolumn{1}{|c|}{50-100} & 100 Gigabit Ethernet, InfiniBand EDR \\\hline
		QSFP56 & \multicolumn{1}{|c|}{200} & 200 Gigabit Ethernet, InfiniBand HDR \\\hline
		QSFP-DD & \multicolumn{1}{|c|}{400} & 400 Gigabit Ethernet, InfiniBand NDR \\\hline
		QSFP‐DD800& \multicolumn{1}{|c|}{800} & 800 Gigabit Ethernet, InfiniBand XDR \\
		\hline
	\end{tabular}
	\label{tab:sfp}
\end{table}

\subsection{Passive Optical Networks (PONs) for 5G and Beyond Fronthaul} 
\begin{figure}[htbp]
	\centering
	\includegraphics[width=0.4\textwidth]{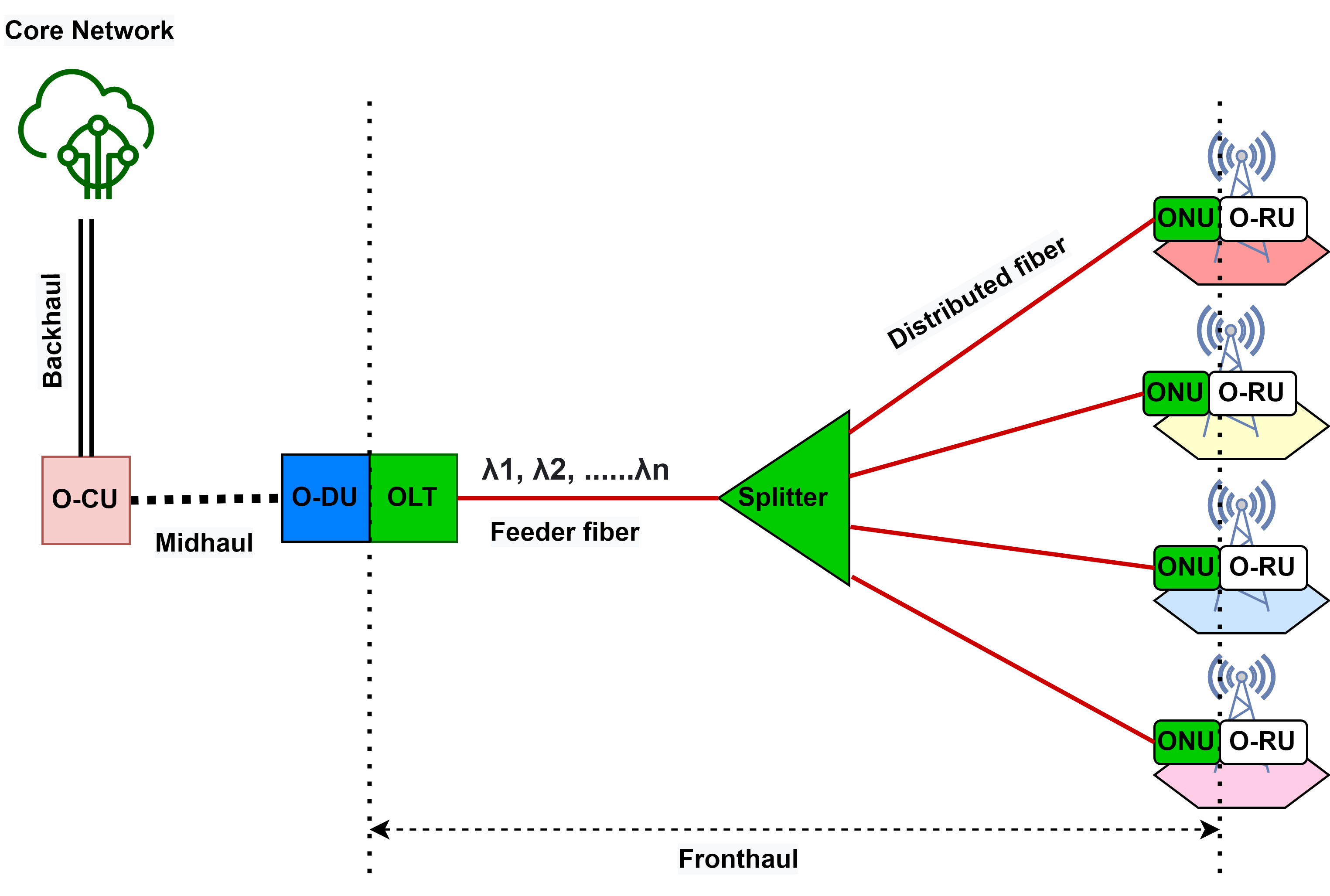}
	\caption{PON-based 5G and beyond fronthaul.}
	\label{fig:pon}
\end{figure}

PONs have recently become popular and promising solutions for 5G and beyond for providing fronthauling/backhauling services for large numbers of RUs. PON architecture has a P2MP configuration by using a single fiber to serve several RUs based using power splitter \cite{pon1}, \cite{pon3}, \cite{ocdm2}. Additionally, it offers a cost-efficient solution due to fiber resource sharing, low-cost aggregation, and a lower number of optical interfaces \cite{pon}. The illustrative Fig.~\ref{fig:pon} depicts an example of employing PON for 5G and beyond fronthaul.

Three leading organizations are responsible for developing standards for PON architectures, including the ITU Telecommunication Standardization Sector (ITU-T) Question 2/Study Group 15 (Q2/15), the Full Service Access Network (FSAN) Group, and the IEEE 802.3 Ethernet Working Group.
\begin{figure}[htbp]
	\centering
	\includegraphics[width=0.47\textwidth]{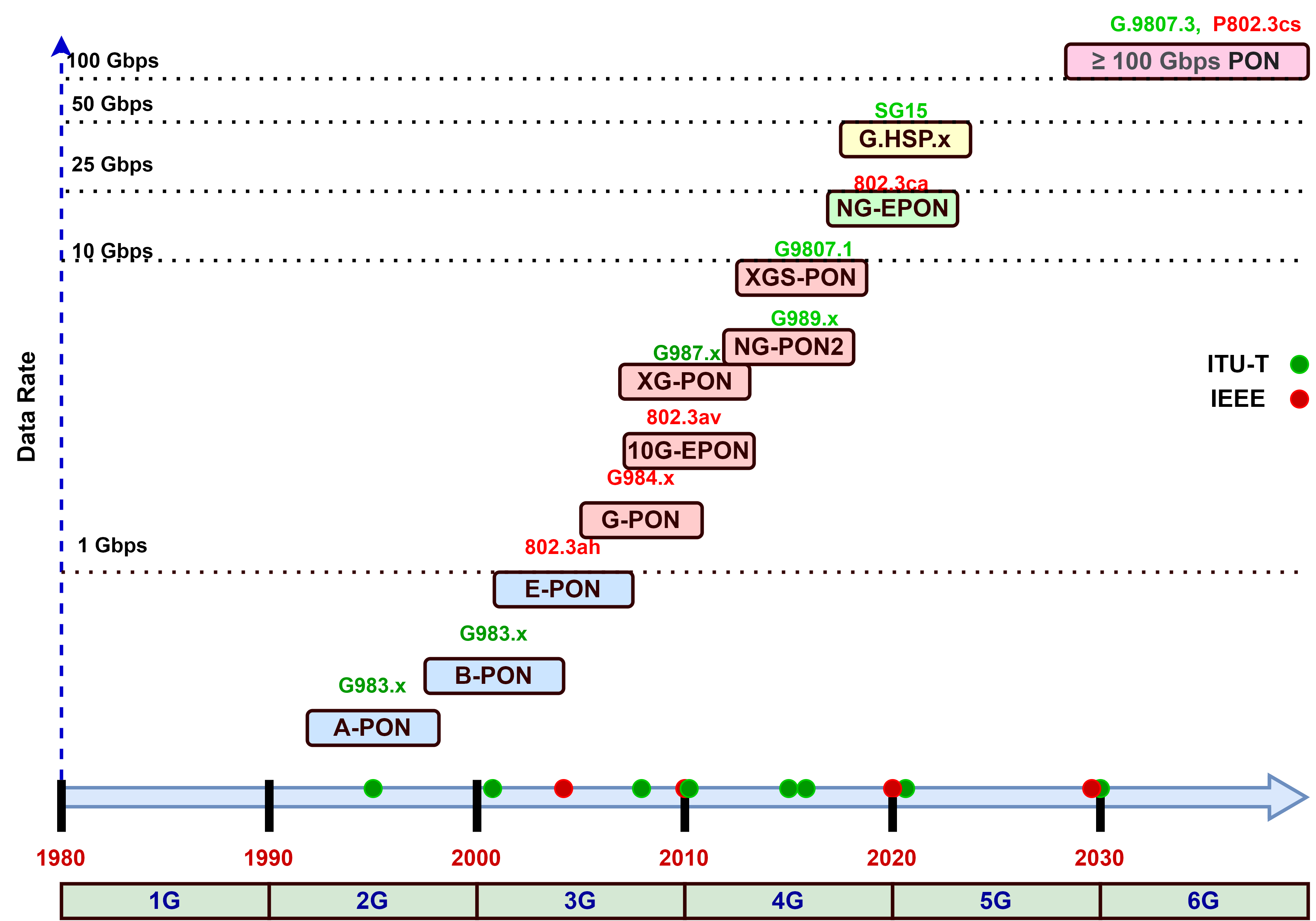}
	\caption{PON standards based on IEEE and ITU-T.}
	\label{fig:ponstd}
\end{figure} 
 
The roadmap of PONs went through several developments over the last two decades (see Fig.~\ref{fig:ponstd}) \cite{pon0}. Starting from \mbox{ITU-T} G.983 standard in 1998 that specified an Asynchronous Transfer Mode (ATM) protocol for PONs  (A-PON) as in Broadband-PON \mbox{(B-PON)} with data rates of 622 Mbps for downlink and 155~Mbps for uplink. The PON was developed toward  TDM-PON, standardized in 2003 as ITU-T G.984 Gigabit PON (GPON), which provides data rates of 2.5 and 1.25~Gbps for downlink and uplink, respectively. In addition to  IEEE 802.3ah Gigabit Ethernet PON (EPON) in 2004, which provides downlink/uplink capacities of 1/1 Gbps.

To facilitate future bandwidth expansion on existing Optical Distribution Networks (ODNs), 10 Gbps PON standards were established in 2008 by both the IEEE (IEEE 802.3av 10 GE-PON) providing a symmetrical 10 Gbps capacity and the ITU-T (ITU-T G.987 10G-PON, XG-PON) providing 10 and 2.5 Gbps for downlink and uplink. The ITU-T G.989 Next-Generation PON 2 (NG-PON2) standard, established in 2013, uses Time and Wavelength Division Multiplexing (TWDM) to achieve a link capacity of 40 Gbps by utilizing four symmetrical 10 Gbps wavelength pairs. TWDM-PON is recommended to support 5G fronthaul splitting options ranging from 1 to 6, 7.2, and 7.3  \cite{skubic2017optical, ponsplit}. 
 
 However, NG-PON2 was not widely used because of the high cost of ONUs that required the usage of tunable lasers and filters. To address this issue, the XGS-PON (10G Symmetric PON) standard was next created, which provides a symmetrical 10 Gbps PON. XGS-PON was established in 2016 to allow 10-Gigabit-capable symmetric PON for residential, business, and 5G backhaul and fronthaul solutions \cite{xgpon, xgpon1}. 

The letter "X" represents the number ten in the XGS, whereas the letter "S" is standard for symmetry. Furthermore, the IEEE 802.3ca 25G/50G-EPON standard was defined in 2020 to provide large capacity for future applications \cite{25pon}.
Similarly, the ITU-T G.9804 High Speed PON (HSP) standardized in 2021, defining a set of ITU-T guidelines to support data rates of up to 50 Gbps \cite{itu25pon}. Moreover, operators are getting engaged in mobile transport over PON, and a new operator-initiated project, "G.sup.5GBH," is being studied in ITU-T SG15/Q2 to investigate the usage of TDM PONs for 5G transport \cite{transport}. Extensive research is underway to explore the possibilities of implementing PON technology, aiming to achieve data rates exceeding 100 Gbps (super-PON), i.e., ITU-T G.9807.3 and IEEE P802.3cs. Building on these advancements, PON Overlay Networks are designed to unlock the revenue potential of existing PON access infrastructure by bringing high-capacity 25G-200G services \cite{InfineraPON2024}. Moreover, for more capacities, XR optics are proposed to transmit coherent capacity levels over an existing PON infrastructure \cite{InfineraBlog2024}.
This evolution is driven by the need to fulfill the demanding specifications of 6G fronthaul after the year 2030 \cite{coherentpon, coherentpon1, coherent2, coherent3, coherent4, coherent5}.

Table~\ref{tab:PON} overviews different standards of PON with their capacities.

\begin{table*}[htbp]
	\centering
	\caption{PON standards based on \cite{pon, pon2, B-PON, E-PON, G-PON, 10G-EPON, XG-PON, NG-PON2, XGS-PON, NG-EPON, hpon1, hpon2}.}
	\begin{tabular}{|l |c|c|c|c|c|c|c|}
		\hline
		\textbf{PON type}  &\textbf{Year}& \textbf{Capacity (DL)}& \textbf{Capacity (UL)} & \textbf{Reach} & \textbf{Max Split Ratio} & \textbf{Wavelengths (DL/UL)} & \textbf{Standard} \\
		\hline
		A-PON/B-PON&1995/2001&622 Mbps &155 Mbps & 20 km & 1:32 & 1490 nm/1310 nm & ITU-T G983.x\\\hline
		E-PON&2004& 1 Gbps &1 Gbps & 20 km & 1:32 & 1490 nm/1310 nm & IEEE 802.3ah\\\hline
		G-PON&2008& 2.5 Gbps &1.25 Gbps & 20 km & 1:64 & 1490 nm/1310 nm & ITU-T G984.x\\\hline
		10G-EPON&2010&10 Gbps &10 Gbps & 20 km & 1:32 & 1577 nm/1270 nm & IEEE 802.3av \\\hline
		XG-PON&2010& 10 Gbps &2.5 Gbps  & 20 km & 1:64 & 1577 nm/1270 nm & ITU-T G987.x\\\hline
		NG-PON2&2015&10/2.5 Gbps &10/2.5 Gbps & 40 km & 1:64 & 1525-1544 nm/1596-1603 nm & ITU-T G989.x\\\hline
		XGS-PON&2016& 10 Gbps &10 Gbps & 20 km & 1:64 & 1577 nm/1270 nm & ITU-T G9807.1\\\hline
		NG-EPON&2020& 25 Gbps& 25 Gbps & 25 km & 1:64 & 1525-1544 nm/1310 nm & IEEE 802.3ca\\\hline
		G.HSP.x &2021&50 Gbps&50 Gbps& 20 km & 1:64 & 1530-1565 nm/1260-1280 nm & ITU-T SG15\\
		\hline
	\end{tabular}
	\label{tab:PON}
\end{table*}

\vspace{2 mm}
\noindent \textbf{\textit{PON Architectures and Classifications}}\\
Several PON architectures and classifications have been developed by academia and industry for different use cases, such as high-speed Time Division Multiplexing-PON (TDM-PON), WDM-PON, TWDM-PON, Orthogonal Frequency Division Multiplexing-PON (OFDM-PON), Optical Code Division Multiplexing-PON (OCDM-PON), Non-Orthogonal Multiple Access-PON (NOMA-PON), and Polarization Division Multiplexing-PON (PDM-PON) \cite{pon0, optical}. In the following, we detail these architectures.
\begin{itemize}
	\item \textbf{TDM-PON:}
This architecture enables multiple users to share a single optical fiber bandwidth. This is accomplished by allocating a certain time slot for each user to transmit data upstream, which is controlled by the central office, known as the Optical Line Terminal (OLT). This means that each user is given a specific time window to communicate their data, ensuring they do not interfere with other users \cite{tdm}. It was deployed for Fiber-To-The-Home (FTTH) applications. Additionally, TDM-PON is an suitable solution for providing optical connectivity in a cost-effective manner due to the simple architecture, as multiple remote nodes with simple optical power splitters are able to provide us with easy-to-use optical connections anywhere in the Optical Distributed Unit (ODN). As a result, TDM-PONs are currently expanding their applications in sectors other than residential FTTH, such as mobile communication networks to handle 5G and 6G \cite{tdm1}.
	 However, TDM-PON technology is currently incompatible with 5G and 6G fronthaul due to various challenges, including high
capacity, low latency, virtualization, and slicing \cite{tdm1}. The upstream cycle in Time Division Multiple Access (TDMA) is 125  $\mu$s and cannot be lowered, which is higher than the stringent latency requirements of 5G and future 6G. For example, for 5G, there is a one-way fronthaul latency requirement of 100  $\mu$s and around 2 ms for backhaul end-to-end latency. The latency constraints for 6G are expected to be even higher (ie., 10  $\mu$s) \cite{fronthaul4}. Due to the time needed to process the Dynamic Bandwidth Allocation (DBA) in the OLT, ordinary TDM-PON technology with DBA has increased latency. As DBA processing time has a direct impact on network latency, it is critical to avoid sophisticated approaches that can further increase delay \cite{tdm3}. A low-latency DBA approach has been proposed in \cite{tdm2} to reduce the DBA cycle or delay while improving bandwidth efficiency, where DBA cycle length varies depending on the traffic load.  
	\item \textbf{WDM-PON:}
	This architecture offers various distinct benefits for 5G fronthaul applications, including high capacity, low latency (it does not require DBA), cost savings, and ease of implementation. Every ONU in WDM-PON is allocated a particular wavelength for transmission. Passive wavelength splitters (in ODN) or wavelength filters can be used to adjust the wavelengths (in ONUs). While it is not ideal for highly stat-muxable residential traffic, it is suitable for mobile fronthaul \cite{wdm, optical}.  
	
	The ITU-T standardization sector  recommended using 
	WDM-PON for 5G fronthaul \cite{wdmfh,wdmfh1 }. To meet 5G fronthaul requirements, each wavelength must offer the capacity of at least 25 Gbps \cite{wdm}. The trial done by the authors of \cite{capacity6} concerning using WDM-PON for 5G fronthaul shows that the WDM-PON system can achieve the 25 Gbps data rate, the one-way system delay is approximately 56.2 ms with 10~km fronthaul transport, and the power budget is 24.97~dB. The WDM-PON technology can be categorized into DWDM and CWDM (Coarse WDM) based on the number of available wavelengths.
	However, up to now, there have been almost no deployments of the technology due to high costs associated with optics, a lack of business demand, and the absence of standardized protocols \cite{Integrated}.
\item \textbf{TWDM-PON:}	
This architecture represents an advanced and highly efficient architecture that combines the benefits of WDM-PON and TDM-PON with the added capability of utilizing both time and wavelength for signal transmission. This innovative approach has attracted significant attention and has been recommended to standardize the NG-PON2 \cite{twdm, twdm1}.

In TWDM-PON, each channel within the system has the flexibility to operate at a different data rate, enabling the network to cater to varying bandwidth requirements efficiently. This inherent adaptability makes TWDM-PON a mature technology that seamlessly supports legacy and new technologies, positioning it as a flexible and suitable solution for the demanding fronthaul requirements of 5G networks.

One of the significant advantages of TWDM-PON is its ability to coexist harmoniously with existing Gigabit Passive Optical Network (GPON) and Ethernet Passive Optical Network (EPON) networks. This coexistence capability allows for a gradual deployment of TWDM-PON without disrupting the functioning of the already established infrastructure. This graceful migration path to TWDM-PON ensures a smooth transition to enhanced network capabilities while safeguarding the investments made in the current network infrastructure.

Regarding performance, TWDM-PON stands out by supporting a high data rate of 40 Gbps \cite{pon}. Moreover, the trial done by the authors of \cite{twdmpon} shows the development of a 100 Gbps per wavelength PON architecture. Their experiment demonstrated effectiveness over distances up to 100 km and was adaptable for Terabit-capable WDM-PON and TWDM-PON. This substantial capacity allows for an efficient transmission of large volumes of data, meeting the high-bandwidth demands of modern applications and services in the 6G era.

TWDM-PON implements a Dynamic Wavelength and Bandwidth Allocation (DWBA) mechanism to make the most of the bandwidth available. This intelligent allocation scheme optimizes the utilization of the network's resources by dynamically assigning wavelengths and bandwidth to different channels based on real-time traffic demands. This flexibility ensures efficient resource allocation, effective load balancing, and the ability to adapt to changing network conditions, resulting in improved network performance and enhanced user experience.

Based on that, TWDM-PON can be a promising solution for beyond 5G and 6G fronthaul in some cases, but it is not yet mature enough. That is because there are still some challenges and open issues that need to be addressed, such as how to efficiently allocate the optical and wireless resources, how to optimize the functional split of baseband processing, and how to increase the capacity of the network to reach the Tbps levels.
	\item \textbf{OFDM-PON:}
It is an advanced optical fiber communication technology that utilizes the OFDM modulation technique to achieve high-speed data transmission capabilities \cite{ofdm0}. In OFDM-PON, the subcarriers employed for data transmission are carefully designed to be orthogonal to each other, ensuring minimal interference and maximizing the overall spectral efficiency. Additionally, OFDM-PON utilizes two distinct wavelengths for downstream and upstream data transmission, further optimizing the system performance \cite{ofdm1}. 
One of the key advantages of OFDM-PON lies in its ability to significantly enhance bandwidth utilization over longer distances compared to other PON technologies, such as TWDM-PON. This characteristic makes OFDM-PON a suitable solution for various cutting-edge applications, particularly those related to 5G and 6G fronthaul, especially when at least Tbps capacity is needed \cite{ofdm2}. By leveraging its improved bandwidth utilization, OFDM-PON facilitates the seamless transmission of large volumes of data, enabling faster and more efficient communication in demanding scenarios.

However, it's worth noting that adopting OFDM-PON can be limited in specific applications due to the requirement for coherent detection. Coherent detection refers to the need for specialized equipment that can accurately recover transmitted signals, thereby ensuring reliable and error-free data reception. The cost associated with this coherent detection equipment can be relatively higher than non-coherent solutions, which may pose challenges to the widespread adoption of OFDM-PON in specific contexts. Nonetheless, ongoing technological advancements and increasing demand for high-speed connectivity drive continuous improvements and cost reductions in coherent detection equipment, which may gradually overcome this limitation.
	\item \textbf{OCDM-PON:}
It is an advanced technology that enhances the efficiency of optical signal transmission. By utilizing a unique code, data is encoded onto the optical signal, enabling multiple users to share the same wavelength without causing interference or degradation of the signal quality \cite{ocdm1}. Another advantage of OCDM-PON is its ability to support asynchronous transmission. This means that different users can transmit and receive data independently without the need for strict synchronization. As a result, the system can handle numerous users simultaneously, enabling high-capacity and scalable network architectures. Moreover, the implementation of OCDM-PON enhances network security. The unique code used in the encoding process adds a layer of security, making it difficult for unauthorized users to access or intercept the transmitted data \cite{ocdm0}.
The key benefit of OCDM-PON are highly efficient bandwidth utilization, good correlation performance, asynchronous transmission, the ability to handle numerous users simultaneously, minimal signal processing delay, and improved network security \cite{ocdm0}. 
All of the mentioned advantages make the OCDM-PON suitable for 5G and beyond applications.

OCDM can be categorized into two main categories: coherent systems and incoherent systems. In coherent systems, OCDM uses a bipolar technique that relies on carrier phase information. On the other hand, incoherent systems utilize a unipolar technique, which does not require phase synchronization detection. The simplicity and reduced complexity of the incoherent system have made it the preferred detection strategy in OCDM-PON deployments. 

\item \textbf{NOMA-PON:} NOMA is a technique that allows multiple users to share the same frequency or time resource by assigning different power levels to each user. NOMA can improve the spectral efficiency and user fairness of wireless communication systems. Recently, NOMA has been applied to PONs, addressing challenges such as limited bandwidth, high power consumption, and complex network management.
	
	NOMA-PON is a novel scheme that combines NOMA with OFDM to overcome these challenges. OFDM is a modulation technique that divides the available bandwidth into multiple orthogonal subcarriers, each carrying a fraction of the data. OFDM can achieve high data rates and robustness against chromatic dispersion and inter-symbol interference in optical transmission. NOMA-PON uses power-domain NOMA to multiplex multiple ONUs on the same subcarrier with different power levels. This way, NOMA-PON can support more users and higher data rates than conventional OFDM or Discrete Fourier Transform-Spread (DFT-S) OFDM PONs, which use Orthogonal Multiple Access (OMA) to allocate different subcarriers to different ONUs.
	
	NOMA-PON also employs adaptive bit and power-loading algorithms to optimize the performance of each ONU according to its channel conditions and power loss. The bit and power loading algorithms determine the optimal number of bits and power level for each subcarrier and each ONU, respectively, to maximize the throughput and minimize the bit error rate. NOMA-PON has been experimentally demonstrated to outperform OFDM and DFT-S OFDM PONs in terms of throughput and power budget \cite{noma-pon1}.
	
	However, NOMA-PON also faces some challenges, such as the mitigation of Self-Induced Intermodulation Interference (SSII) (which is a nonlinear distortion caused by the superposition of multiple signals with different power levels) and the design of efficient multiple access control protocols. SSII is a nonlinear distortion that occurs when considerable ONUs transmit on the same subcarrier with different power levels. SSII can degrade the signal quality and reduce the achievable data rate. NOMA-PON can mitigate SSII by using appropriate power allocation schemes, such as Successive Interference Cancellation (SIC) or Interference Alignment (IA) \cite{noma-pon2}. Multiple access control protocols are needed to coordinate the transmission and reception of multiple ONUs in NOMA-PON. These protocols should consider the trade-off between complexity and performance and the fairness and quality of service requirements of different ONUs \cite{noma-pon3}.
	
	NOMA-PON is a promising technique for future 6G fronthaul applications that can provide high spectral efficiency and power budget and support flexible and scalable network architectures. However, NOMA-PON is not a mature technology and is still an active research topic that requires further investigation and optimization to achieve its full potential.
	
		\item \textbf{PDM-PON:}
	It is a type of PON that uses PDM to increase the data rate and spectral efficiency of the optical signals. PDM-PON can achieve 100 Gb/s or beyond per wavelength by modulating the signal with different polarization states and using coherent detection at the receiver. One of the modulation formats that can be used for PDM-PON is PDM-PAM-4, which is a four-level pulse amplitude modulation that encodes two bits per symbol. PDM-PAM-4 PON can offer high receiver sensitivity, colorless frequency selectivity, and linear detection, enabling channel impairment compensation in the digital domain \cite{pdm-pon1, pdm-pon2}. PDM-PON can also use low-cost and low-complexity components, such as intensity modulators, DFB lasers, and Kramers-Kronig receivers, to reduce the cost and power consumption of the system \cite{pdm-pon1, pdm-pon2}. 
	
	PDM-PON is a relatively mature technology for optical access networks, as it has been demonstrated and tested in various experiments and scenarios \cite{pdm-pon1, pdm-pon2, pdm-pon4}. However, it still faces some challenges and limitations, such as the complexity of polarization management, the trade-off between bandwidth and reach, and the compatibility with legacy PON systems. Therefore, PDM-PON may need further optimization and innovation to meet the requirements of 6G fronthaul, such as high capacity, low latency, and flexible resource allocation. Some possible solutions include using advanced modulation formats, coherent detection techniques, and joint resource allocation schemes \cite{pdm-pon3}.
\end{itemize}
\subsection{FSO for 5G and Beyond Fronthaul}
FSO is a transmission system that uses modulated light beams to send data over the atmosphere wirelessly between two fixed nodes. It has seen rapid growth in the recent decade \cite{fso0}.	Due to different functional splits and relaxing the fronthaul capacity requirements, FSO becomes an attractive alternative for microwave and optical fiber in handling 5G and beyond fronthaul data traffic \cite{fso, fso1, fso4}.

FSO is characterized by a large optical bandwidth, which can lead to a much higher fronthaul capacity of 100 Gbps~\cite{fso3} over a distance of few kilometers and immunity to electromagnetic interference. FSO can support splitting options ranging from 1 to 6 and 7.3. Line-of-sight (LoS) conditions are required for  FSO, which operates worldwide in the unlicensed  800-1700~(nm) wavelength range (above 300 GHz) \cite{fso2}. FSO links can also be easily installed, making it a viable fronthaul solution for ``on-demand'' capacity deployments. 
 
Several RUs need to be deployed to upgrade existing networks during a specific event for a brief period of time. Furthermore, in certain situations, it may be difficult or impossible to lay a fiber optic cable, such as when crossing a surface of water or when the RU is located in a remote or hard-to-reach location, such as on a mountain. In these cases, FSO could potentially be used as an alternative. There are several benefits to using FSO for fronthaul in 5G and future 6G networks. FSO is a wireless technology, which means that it does not require the physical infrastructure (such as cables) that traditional wired communication methods do. This can make it easier and less expensive to deploy \cite{fso5}.

Additionally, FSO can transmit data over long distances of several kilometers with minimal signal loss, making it well-suited for use in areas where fiber optic cables are impractical. However, there are also some challenges to using FSO for fronthaul in 5G and 6G networks. One major challenge is that FSO is susceptible to interference from weather conditions such as heavy clouds, intense fog, snow, and dust storms, which can attenuate the light beam and disrupt the signal \cite{fso6}.

However, hybrid FSO/mmWave configuration can be used to increase availability up to 99.999 percent because it is resilient in rainy or foggy conditions \cite{fso55}. FSO can be further integrated with optical fiber in order to provide cost-efficient, high-capacity, and low-latency connection \cite{fiber-fso1, fiber-fso} or with PON to provide flexible connections \cite{pon-fso, pon-fso1}.

FSO can be employed in \emph{terrestrial} settings for 5G and beyond fronthauling/backhauling by deploying FSO devices on top of towers in the cell sites to ensure connectivity between different RUs and the DU pool as point-to-point links core network, similar to traditional microwave links.

Additionally, FSO can be used for ``vertical'' (\emph{non-terrestrial}) communications, where it facilitates data transfer between different airborne entities such as High Altitude Platforms (HAPs), Unmanned Aerial Vehicles (UAVs), balloons, and satellites, as well as between these entities and ground stations as point-to-multipoint links. This unique deployment allows FSO to support various applications, including military, emergency response, and commercial activities such as broadcasting, data centers, and telecommunications \cite{fso7, fso9, fso10}.

\begin{figure*}[htbp]
	\centering
	\includegraphics[width=0.8\textwidth]{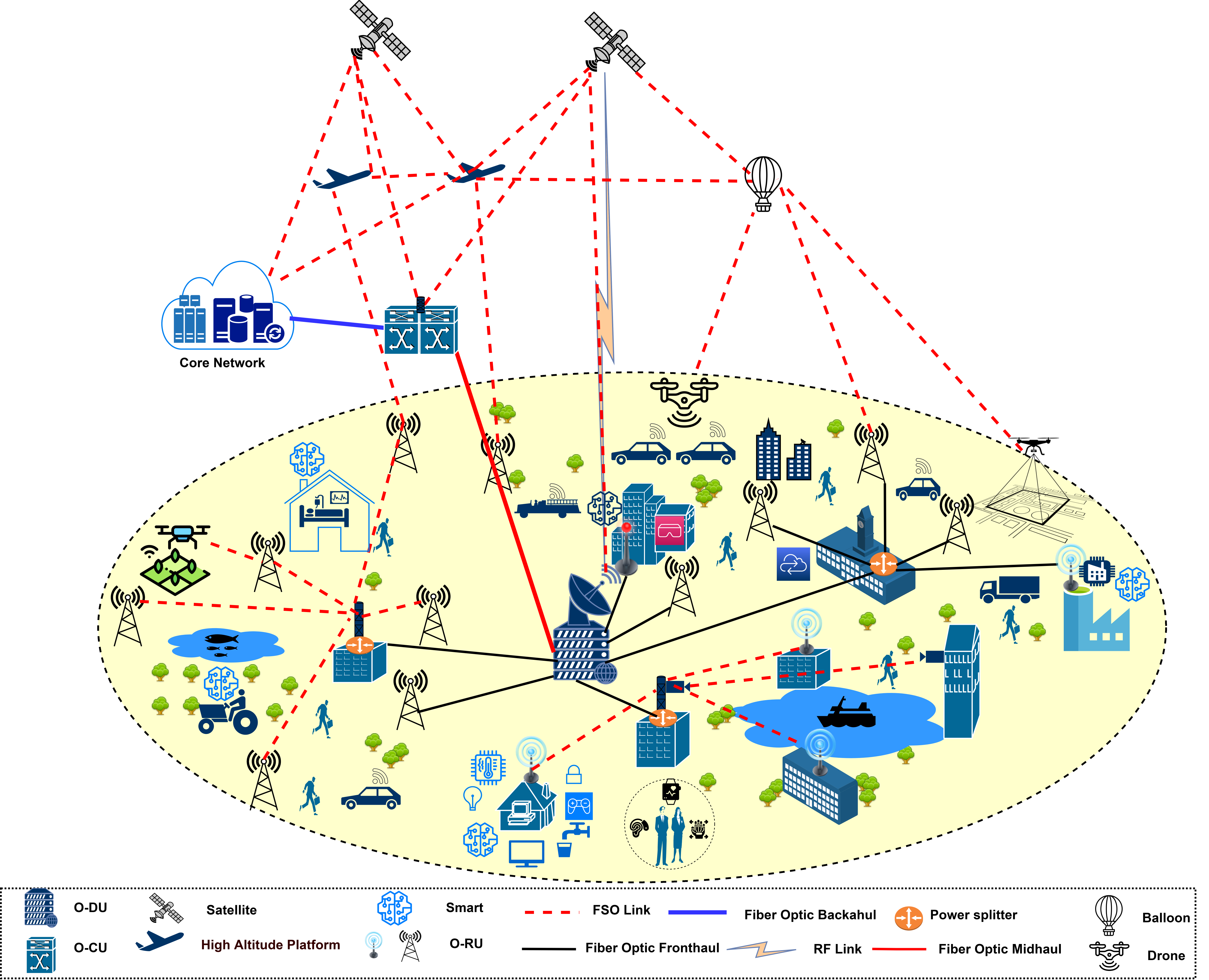}
	\caption{Envisioned 6G Network using various optical communications for fronthauling.}
	\label{fig:arch}
\end{figure*}

Furthermore, due to a growing interest in underwater activities, including military and scientific research, suitable communication technologies, including Optical Wireless Underwater Networks based on FSO links, are being investigated.
FSO can be utilized for short-range, high-speed underwater communication due to its directed beam and resistance to water turbulence. Nevertheless,  FSO systems have several limitations, including:
\begin{itemize}
\item  Their range is restricted by \emph{absorption and dispersion} caused by Water molecules or carbon dioxide, and it requires a clear line of sight between transmitter and receiver, which can be problematic in the ocean environment. 
\item  They are sensitive to temperature changes, leading to \emph{scintillation loss}.
 \item  Power attenuation occurs due to beam spreading, known as \emph{geometric loss}.
\item  At 1550 nm, haze attenuation is lower than other wavelengths, but fog causes wavelength-independent \emph{atmospheric attenuation}.
\item  They experience \emph{scattering} when the optical beam encounters a scatterer, reducing the beam's intensity over longer distances. \cite{fsowater, fsowater1}.
\end{itemize}
Table~\ref{tab:opticpons} represents a comparison of different optical fronthaul technologies mentioned up to this point.
	\begin{table*}[htbp]
	\centering
	\caption{Comparison of pros and cons of optical fronthaul technologies.}
	\begin{tabular}{|l|p{7 cm}|p{8 cm}|}
		\hline
		\textbf{Technology} & \textbf{Advantages}& \textbf{Disadvantages} \\
		\hline
		P2P  & 
			Deployment simplicity;  low latency; high capacity; longer transmission distances; low power consumption.
		& 
Higher deployment cost; low flexibility; limited sharing of resources; scalability problems.
		 \\\hline
	TDM-PON & Cost-effective; simple architecture; scalability.
	
	& Limited capacity; limited distance less than 20 km; bandwidth sharing  leading to security limitations.
	 \\\hline
	WDM-PON  & 
Low latency; high capacity; ease of implementation; security.
&Each ONU must be equipped with a wavelength filter; inefficient bandwidth utilization; limited wavelength resources; high cost.
	 \\\hline
	TWDM-PON  & High capacity more than 40 Gbps; cost reduction; compatibility with previous PON; combines advantages of TDM-PON and WDM-PON; flexible bandwidth allocation.
	& 
		Cross-talk issues, requires colorless ONUs.
	 \\\hline
	OFDM-PON & 
	High capacity more than 40 Gbps; efficient bandwidth utilization; high spectral efficiency; sharing OLT costs.
	& 
Complexity of transmitters and receivers; requires advanced signal processing; frequency offset caused by carrier frequency mismatch (sensitive to phase noise and frequency offset).
\\\hline
	OCDM-PON  &
Effective bandwidth usage; asynchronous transmission; user allocation flexibility; high security.
& 
Complexity of transmitters and receivers; limited code resources; performance relies on the address code.
\\\hline
	NOMA-PON & High spectral efficiency; user fairness and QoS; power budget; scalability and flexibility.

& Complex receiver design and coordination to implement SIC; sensitive to channel estimation errors and feedback delays; It suffers from SSII.

\\\hline
PDM-PON &High data rate ($\geq$ 100 Gbps per $\lambda$) and spectral efficiency; low-cost and low-complexity components.

& 
Requires complex polarization management; not compatible with legacy PON systems, as it uses different wavelengths and modulation formats.
\\\hline
	FSO  & 
high bandwidth; wireless alternative to optical fiber; unlicensed frequency; easy to deploy
	& 
Sensitivity to weather conditions, e.g., fog, rain, or snow; line-of-sight requirement; short transmission distances.
 \\\hline
	\end{tabular}
	\label{tab:opticpons}
\end{table*}
Figure \ref{fig:arch} shows the envisioned 6G network architecture with different use cases considering various optical fronthaul technologies (i.e., P2P, PON, FSO). As shown in the figure, we assume that the considered architecture for 6G is the O-RAN, considering different use cases highlighted in Fig.~\ref{fig:6g}.
\subsection{Lessons Learned}
Optical communications will be the main player in realizing the 6G fronthaul, and we can conclude the following lessons:
\begin{itemize}
	\item The motivation for optical fronthaul technologies for 5G and 6G is driven by the anticipated surge in data traffic, demanding higher data rates, lower latency, and increased reliability. The market growth projections highlight the critical role of optical fronthaul in meeting these requirements.
	\item There is no one-size-fits-all solution for 5G/6G fronthaul. Architectural approaches like P2P optical fiber, PONs, and FSO offer diverse benefits and drawbacks. The selection of fronthaul technologies should consider the specific requirements of the use case, including data rates, latency constraints, deployment scenarios, and cost considerations. Combining different technologies, hybrid solutions may offer optimal outcomes in specific scenarios.
	
	\item P2P optical fiber architecture provides dedicated high-capacity links, reducing latency and offering secure data transmission. However, its drawbacks include the need for extensive fiber resources and limited flexibility in resource allocation. P2P is suitable for high-capacity demand scenarios.
	
	\item PONs, including advancements like 50G-PON, and coherent PON offer cost-efficient solutions with fiber resource sharing. The evolution of PON standards demonstrates a progression towards higher capacities to support 5G and beyond fronthaul requirements.
	
	\item FSO emerges as a wireless alternative for 5G and beyond optical fiber fronthaul, offering high capacity, low latency, and immunity to electromagnetic interference. Challenges include susceptibility to weather conditions and potential security concerns.
	
	\item FSO finds applications in terrestrial and non-terrestrial settings, supporting point-to-point links in terrestrial deployments and facilitating communication between airborne entities in non-terrestrial scenarios.
	
	\item FSO faces challenges such as atmospheric interference, scintillation loss, and security concerns. Ongoing research focuses on addressing these limitations through adaptive optics, improved beam steering, and innovative solutions.
\end{itemize}

\section{Cutting-edge Research Related to 5G/6G Optical Fronthaul}\label{sec6}
In this section we present the  state-of-the-art research, and emerging trends that are likely to shape the future of 5G/6G optical fronthaul networks. The main surveyed studies are listed in Table~\ref{tab:comparison}. 
 
The section is organized into several subsections, each focusing on a specific aspect of optical fronthaul research. These include, deployment cost reduction, energy efficiency and sustainability, latency and jitter aspects, integration with other communication technologies, resource allocation, ML and AI for optical fronthaul, flexible and re-configurable fronthaul enabled by Software Defined Networking (SDN), high-capacity solutions through space division multiplexing, and security and privacy concerns.
\subsection{Optical Fronthaul Deployment Cost Reduction}
One of the biggest obstacles in deploying ultra-dense 5G/6G networks is the cost of optical fronthaul, which becomes increasingly expensive as the network density increases \cite{Integrated}. For that, numerous studies in the literature have explored cost-effective solutions for optical fronthaul, with the ultimate goal of facilitating the deployment of a cost-efficient optical fronthaul for future 6G systems. 
Several methods have been proposed to minimize the deployment costs of the network. 

For example, based on \cite{sharing1, sharing2}, network operators can minimize the cost of deploying new optical fronthaul networks by sharing network resources such as fiber infrastructure. Moreover, exploring new network planning techniques and algorithms, topology optimization, and the usage of PON can be seen as appealing solutions for the MNOs to reduce their optical fronthaul deployment for 5G and 6G networks in comparison to the P2P architecture\cite{cost8, cost10}. Additionally, the reuse of the existing optical infrastructure is also one of the key methods for reducing the cost of the optical fronthaul deployments \cite{brown}.
\subsection{Energy Efficiency and Sustainability in the Optical Fronthaul}
The rapid growth of data traffic in wireless networks has led to a significant increase in energy consumption. As a result, improving energy efficiency and sustainability in the optical fronthaul is essential for several reasons, such as environmental impact, cost-effectiveness, and scalability. 

The works \cite{fronthaulenergy1, cost61} provide models for assessing the energy consumption of different optical fronthaul architectures. Taking into account the C-RAN architecture and considering PON as a fronthaul, the work in \cite{fronthaulenergy3} highlights the main methods for reducing the energy consumption, including, \emph{power-saving modes} that are techniques that reduce the energy consumption of ONUs by switching off or adjusting the transmission rate of their transmitters and receivers during low traffic periods. \emph{Dynamic bandwidth allocation} that is a mechanism that allocates resources to ONUs based on their traffic demand and quality of service requirements and enables them to enter power-saving modes when idle, and \emph{radio-over-fiber} that integrates optical and wireless networks in the fronthaul, and offer different trade-offs between bandwidth, latency, complexity, and energy efficiency. 

Another method is by using a decomposed Arrayed Waveguide Grating Router (AWGR)-enabled passive WDM fronthaul \cite{fronthaulenergy2}. Additionally,  the work \cite{fronthaulenergy4} studied the energy efficiency in optical fronthaul considering the usage of FSO/mmWave for fronthauling.
\subsection{Latency and Jitter Aspects in Optical Fronthaul}
Latency refers to the time it takes for a signal to travel from one point to another. In optical fronthaul, low latency is crucial for ensuring seamless and efficient communication between the BBU and RRH. High latency can lead to:
\begin{itemize}
\item Degraded user experience: High latency can result in poor voice and video quality, thus affecting the overall user experience.
\item Increased handover failures: As latency increases, the probability of handover failures between adjacent cells also rises, leading to dropped connections and lower network reliability.
\end{itemize}

Moreover, jitter is the variation in latency over time. In optical fronthaul, jitter can cause:
\begin{itemize}
\item Packet loss: High jitter can result in the loss of data as packets may arrive out of order, or some of them may be discarded due to buffer overflows.
\item Reduced stability of network functioning: High jitter can lead to fluctuations in the network's performance, causing intermittent service disruptions.
\end{itemize}

Reducing latency and jitter over the optical fronthaul is crucial for a variety of time-sensitive applications in 5G and beyond networks, such as intelligent transportation systems and factory automation \cite{1914, latency8}. Therefore, there is a need for methods to limit the latency and jitter over the optical fronthaul, as these factors directly impact the overall performance and QoS in the network. For instance, based on \cite{latency8}, encapsulating CPRI over Ethernet (CoE) can meet the delay and jitter requirements for 5G applications. Another method to minimize the delay over the optical fronthaul is by topology optimizations applied to the optical fronthaul  \cite{latency2, latency10}. 
\subsection{Integration of Optical Fronthaul  with Other Communication Technologies}
There are two types of integration: 1) optical fiber communications (PON or P2P) with wireless technologies, such as microwave and mmWave technologies. 2) integration of FSO with wireless technologies.
\subsubsection{Integration of optical fiber with wireless technologies}
One promising approach to realizing the full potential of 6G is to integrate optical fiber with microwave or mmWave technologies in the fronthaul network. This combination offers several significant advantages, which we will discuss below.
\begin{itemize}
	\item \textit{Improved Flexibility and Scalability:} The hybrid nature of an integrated optical fiber and microwave or mmWave fronthaul network enables greater flexibility and scalability. Optical fiber provides a robust and reliable backbone for the network, while the wireless capabilities of microwave and mmWave technologies allow for dynamic and adaptable connections. This flexibility is particularly advantageous in scenarios where installing additional fiber infrastructure may be challenging, such as in dense urban environments or remote rural areas. Furthermore, the network can be easily scaled to accommodate the growing number of connected devices and increasing data traffic demands associated with 5G and future 6G \cite{fibermmwave}.
	\item \textit{Enhanced Resilience:} An integrated optical fiber and microwave or mmWave fronthaul network can offer enhanced resilience. By combining the strengths of both wired and wireless technologies, the network can better withstand potential failures or disruptions. For example, if a fiber link is damaged or disconnected, the microwave or mmWave link can act as a backup, ensuring uninterrupted communication  at least for critical services. This redundancy helps maintain a high level of reliability required for critical 6G applications and services \cite{survive1, survive2}.
	
	\item \textit{Cost-effectiveness:} Integrating optical fiber with microwave or mmWave technologies can lead to cost savings in the deployment and operation of the fronthaul network. By leveraging existing fiber infrastructure and supplementing it with wireless links, network operators can significantly lower the costs associated with deploying backup fiber-optic cables. Moreover, dynamic and flexible nature of a hybrid architecture can result in more efficient resource utilization, reducing operational costs in the long run \cite{fibermicrowave}.
\end{itemize}
\subsubsection{Integration of FSO with wireless technologies}

FSO is an attractive solution for future 6G cellular networks fronthaul. However, atmospheric conditions may significantly influence the performance of FSO, particularly for long-distance FSO links. The presence of dense clouds, intense fog, or dust storms can drastically degrade an FSO link performance. Adverse weather conditions may lead to suboptimal transmission performance in FSO systems. A hybrid FSO/mmWave approach is recommended to attain both high capacity and improved link availability, potentially offering carrier-grade link availability of 99.999 percent.  

Although mmWave performance may be compromised in the rain, it can penetrate fog. Conversely, FSO signals employing 800–1700 nm lasers cannot transmit through dense fog, yet rain minimally impacts the FSO system.
The hybrid FSO/RF networks can thus improve availability, reliability, and weather resistance \cite{fsommwave1, fsommwave2}.
\subsection{Resource Allocation in the Optical Fronthaul}
Efficient resource allocation in the optical fronthaul is crucial for ensuring high  performance of future 6G networks. 
The following advantages can be obtained from the efficient resource allocation in the Optical Fronthaul of 6G networks:
\begin{enumerate}
	\item \textit{Increased total network throughput:} Efficient resource allocation in the optical fronthaul may enable a more efficient use of available resources, leading to an increased number of flows served by the network. This is particularly important to handle a large volume of traffic in 5G and also future 6G networks, where the demand for data rates is expected to be much higher than in previous generations \cite{rofsurvey}.
	\item \textit{Improved spectral efficiency:} Allocating resources in a better way in the optical fronthaul may result in a better spectral efficiency, which can translate into higher data rates and lower latency for end users. This may be a significant advantage for future 6G networks, as they aim to deliver ultra-low latency and high data rates for various applications \cite{spectral1, Integrated}.
	\item \textit{Enhanced energy efficiency:} Optimized resource allocation in the optical fronthaul reduceing energy consumption by minimizing the total power required for signal transmission and processing \cite{improveenergy, improvenergy1}. This will be critical for 6G networks, as energy efficiency is essential for economic and environmental reasons.
	\item \textit{Enhancing the scalability and adaptability:} Efficient resource allocation in the optical fronthaul may allow for better scalability and adaptability, as it enables the network to accommodate a higher number of users and devices without compromising the network performance and adjusting to varying traffic demands and network conditions in real time. 
\end{enumerate}

Efficient resource allocation in the optical fronthaul can be achieved using \textit{Dynamic bandwidth allocation (DBA)}, which is a crucial technique that dynamically adjusts bandwidth allocation based on user requirements and network conditions. DBA optimizes resource utilization, improves network efficiency, and enhances user experience. It ensures optimal quality of service for various applications and supports the scalability of the network \cite{DBA1,DBA2, DBA41}. As technology advances, DBA is expected to evolve to incorporate ML and AI for further optimization \cite{DBA4, DBA5}.
	
\subsection{ML- and AI-empowered 6G Optical Fronthaul} 

Integrating ML and AI into 6G fronthaul technology heralds a new era of intelligent operations and resource allocation. The rationale for employing ML is twofold: (a) it becomes particularly compelling in scenarios where there is an absence of a physics-based mathematical model, attributed to a deficiency in domain-specific knowledge, or what is termed as a ``model deficit''; and (b) it is also justified when the existing algorithms, which execute established physics-based mathematical models, are overly intricate in terms of computation or require an excessive amount of time to process, a situation referred to as an ``algorithm deficit'' \cite{DBA5}. Implementing AI and ML within optical fronthaul networks presents a challenging research area. This is primarily due to the networks' sensitivity to costs, the constraints on computational resources, the dynamic nature of the network environment, and the absence of comprehensive data acquisition and monitoring systems, which results in a scarcity of representative data. Additionally, the imperative to facilitate real-time applications imposes extra requirements on ML models to be highly responsive and swiftly adapt to changes. However, they can help in optimize performance, mitigate congestion, and ensure a seamless user experience. Additionally, advanced optimization techniques can help balance values of such factors as latency, bandwidth, and energy efficiency. This approach may improve efficiency and connectivity in the dynamic 6G landscape \cite{DBA5, DBA6,DBA7}. Moreover, by leveraging AI algorithms, network management becomes intelligent, enabling dynamic resource allocation, predictive maintenance, and fault detection. For instance, AI can be employed to analyze real-time data from optical fronthaul networks, predicting potential failures and proactively addressing issues, thus enhancing reliability and minimizing downtime \cite{DBA7}. 

Additionally, ML-driven traffic prediction models can dynamically adjust the fronthaul interface splitting options based on varying demand, ensuring efficient resource utilization. This adaptive approach enhances network performance by aligning capacity with real-time requirements. Moreover, AI and ML can process vast amounts of data to make precise predictions about network behavior, leading to more accurate resource allocation. These models can adapt to new patterns in data, making them suitable for the dynamic nature of 6G networks. AI-driven systems reduce the need for manual intervention, allowing for automated network adjustments and maintenance \cite{MLfronthaul, AIoran}. Moreover, ML can play a crucial role in ensuring the high security of 5G/6G optical fronthaul, as will be shown in Section \ref{sec6}-(I).

Expanding on ML techniques, several methods can further enhance the capabilities of AI-driven 6G optical fronthaul networks as follows \cite{DBA5, ml6g}:
	
\begin{itemize}
\item \emph{Deep Learning:} utilized for complex pattern recognition in traffic flow, which is essential for predicting and managing network loads.
\item \emph{Supervised Learning: } By training on labeled datasets, it can accurately map network inputs to expected outputs, aiding in both regression tasks like predicting traffic load and classification tasks such as identifying types of service requests over the optical fronthual.
\item \emph{Unsupervised Learning:} It is utilized to discern underlying patterns or detect anomalies in the data traffic of 6G networks without the need for pre-labeled data. This is particularly useful for clustering similar types of traffic and spotting unusual patterns that may indicate network faults or security breaches.
\item \emph{Semi-Supervised Learning:} This approach leverages a combination of small amount of labeled data and a larger pool of unlabeled data, which is often the case in real-world network environments. It enhances the performance of supervised learning models, making it a cost-effective solution for improving network security and anomaly detection in 6G optical fronthaul systems where obtaining extensive labeled datasets can be challenging.
\item \emph{Reinforcement Learning algorithms:} optimize routing decisions in real-time, adapting to changing network conditions to improve latency and throughput.
\item \emph{Support Vector Machines:} classify and predict network anomalies, aiding in preventive maintenance and security.
\end{itemize}
\ However, it is important to highlight the limitations of these techniques:
\begin{itemize}
\item \emph{Data Dependency:} The performance of ML models is heavily dependent on the quality and quantity of the data available, which can be a limiting factor.
\item \emph{Complexity:} The complexity of ML models can make them difficult to implement and understand, requiring specialized knowledge.
\item \emph{Cost of Deployment:} Initial setup and maintenance of ML systems can be expensive, especially for large-scale networks
\end{itemize}

\subsection{Enhancement of Flexibility  and Reconfigurability  in Optical Fronthaul using SDN}
SDN enables dynamic provisioning and reconfiguration of the optical fronthaul network, adapting to varying traffic patterns and bandwidth needs while prioritizing critical services. Advanced traffic engineering techniques, such as dynamic bandwidth allocation and traffic grooming, optimize resource utilization and reduce latency \cite{SDNflex1, SDNflex2}. Furthermore, SDN's centralized control facilitates efficient network monitoring, troubleshooting, and real-time performance metric collection, allowing for proactive monitoring, fault detection, and quick response to incidents \cite{SDNflex3, SDNflex4, SDNflex5}. By leveraging SDN's programmability, operators can employ intelligent algorithms and analytics to optimize network performance, detect anomalies, and predict potential failures, ultimately enhancing the resilience of the fronthaul infrastructure.
\subsection{Capacity Enhancement of Optical Fronthaul using SDM}
The main idea behind Space Division Multiplexing (SDM) is to divide the cross-sectional area of an optical fiber into multiple separate channels, each of which can carry a different data stream.
 Various techniques, such as using numerous cores inside a multicore fiber or using multiple modes within a few-mode fiber, can be used to define these channels. By utilizing these spatial channels, SDM offers a significantly higher data capacity than conventional single-channel optical fibers \cite{capacity8, SDM1}. One of the primary advantages of SDM for high-capacity optical fronthaul is its ability to support the increasing data demands of modern communication networks. With the proliferation of bandwidth-intensive applications and the advent of technologies such as 5G and future 6G, there is a growing need for higher data rates and lower latency. SDM offers a scalable solution to address these requirements by effectively increasing the capacity of optical fibers \cite{SDM2, SDM3}.
\subsection{Security and Privacy Aspects in Optical Fronthaul}\label{sub1}

The advent of 5G/6G brings forth significant security and privacy challenges, particularly in the optical fronthaul \cite{furdek2016overview}.  Consequently, there is a need to explore innovative approaches to security provisioning, certificate enrollment, secure environments, and synchronization plane security \cite{security1}. Furthermore, the integration of ML for network security management has become a pivotal strategy in addressing these challenges. In fact, ML algorithms are instrumental in detecting and mitigating security breaches, ranging from physical-layer attacks to more sophisticated intrusions \cite{furdek2022machine}.

Moreover, the architecture of optical network security management systems is being redefined to incorporate efficient ML models that can detect a wide array of evolving threats \cite{furdek2021optical}. These systems are designed to be proactive, with the ability to learn from new attack patterns and enhance their detection capabilities over time. Additionally, the use of ML extends to enhancing optical network security by interpreting complex patterns within Optical Performance Monitoring (OPM) data, which is crucial for identifying subtle signs of security breaches \cite{furdek2020machine}. For instance, demonstrations of ML-assisted security monitoring in optical networks provide practical insights into the effectiveness of these approaches, showcasing their ability to detect, identify, and localize optical-layer attacks within real-life network environments \cite{furdek2020machine1}.

Also, Root Cause Analysis (RCA) plays a crucial role in autonomous optical network security management, enabling the identification of underlying causes of security incidents \cite{natalino2022root}. This is complemented by scalable physical layer security components designed for microservice-based optical SDN controllers, which enhance the network’s ability to adapt to and mitigate evolving threats \cite{natalino2021scalable}. As a result,  autonomous security management systems leverage these components to ensure robust security measures are in place, capable of evolving with the network itself \cite{natalino2021autonomous}.

As optical networks continue to evolve, so too must the security frameworks that protect them, ensuring that the privacy and integrity of data transmission remain in the face of ever-growing security challenges \cite{li2019optical}.
\begin{table*}[!ht]
		\centering
		\caption{State-of-the-art research in 5G/6G optical fronthaul.}
		\label{tab:comparison}
		\begin{tabular}{|c|c|c|c|c|c|c|c|c|c|c|c|c|c|c|c|c|c|c|c|c|c|}
			\hline
			& \multicolumn{8}{c|}{\textbf{Technology}} & \multicolumn{8}{c|}{\textbf{Focus}}  & \multicolumn{5}{c|}{\textbf{Architecture}} \\ \hline
			Reference  & \rotatebox[origin=c]{90}{P2P}  & \rotatebox[origin=c]{90}{TDM-PON} & \rotatebox[origin=c]{90}{WDM-PON}&  \rotatebox[origin=c]{90}{TWDM-PON} & \rotatebox[origin=c]{90}{OFDM-PON} & \rotatebox[origin=c]{90}{OCDM-PON} &\rotatebox[origin=c]{90}{Terrestrial FSO}&\rotatebox[origin=c]{90}{Non-terrestrial FSO}  & \rotatebox[origin=c]{90}{Cost}  & \rotatebox[origin=c]{90}{Latency/Jitter}  & \rotatebox[origin=c]{90}{SDN}  & \rotatebox[origin=c]{90}{Energy Efficiency} & \rotatebox[origin=c]{90}{Resource allocation} &\rotatebox[origin=c]{90}{Capacity} & \rotatebox[origin=c]{90}{AI/ML} & \rotatebox[origin=c]{90}{Security}  & \rotatebox[origin=c]{90}{C-RAN}  & \rotatebox[origin=c]{90}{H-RAN} & \rotatebox[origin=c]{90}{F-RAN}  & \rotatebox[origin=c]{90}{v-RAN} & \rotatebox[origin=c]{90}{O-RAN} \\ \hline
			\cite{skubic2017optical, bitrate,cost8} &$\checkmark$  &   & $\checkmark$ & $\checkmark$&  & & &  & $\checkmark$&  &  &   & && & &  &  &  &  &  \\ \hline
			\cite{tdm1}&&$\checkmark$&&&&&&&&$\checkmark$&&&&$\checkmark$&&&$\checkmark$&&&&\\ \hline
			\cite{fiber-fso} &$\checkmark$  &   &  & &    & & $\checkmark$ &  &$\checkmark$ &  &  &   & && & & $\checkmark$ &  &  &  &  \\ \hline
			\cite{fiber-fso1} &$\checkmark$  &   &  & &    & & $\checkmark$ &  &$\checkmark$ & $\checkmark$ &  &   & && & & $\checkmark$ &  &  &  &  \\ \hline
			
			\cite{pon-fso} &  &   & $\checkmark$ & &    & & $\checkmark$ &  &$\checkmark$ &  &  &   & && & & $\checkmark$ &  &  &  &  \\ \hline
			\cite{fso7} &$\checkmark$  &   &  & &    & & $\checkmark$ & $\checkmark$ &$\checkmark$ &  &  &   & && & & $\checkmark$ &  &  &  &  \\ \hline
			\cite{fso9} &$\checkmark$  &   &  & &    & & $\checkmark$ & $\checkmark$ &$\checkmark$ &  &  &   & && & & $\checkmark$ &  &  &  &  \\ \hline
			\cite{fso10} &  &   &  & &    & & $\checkmark$ & $\checkmark$ &$\checkmark$ &  &  &   & && & & $\checkmark$ &  &  &  &  \\ \hline
			\cite{cost10} &  &   & &  $\checkmark$&    & & &  & $\checkmark$ &  & &   & && & & $\checkmark$ &  &  &  &  \\ \hline
			\cite{fronthaulenergy2}&  &   & $\checkmark$ & &    & & &  & &  &  & $\checkmark$  &$\checkmark$ &&& &$\checkmark$  &  &  &  &  \\ \hline
			\cite{fronthaulenergy4}&  &   &  & & & & $\checkmark$&  & &  &  & $\checkmark$  &$\checkmark$ &&& &$\checkmark$  &  &  &  &  \\ \hline
			\cite{latency8} &$\checkmark$  &   &  & &  & & &  & & $\checkmark$ &  &   & && & & $\checkmark$ &  &  & $\checkmark$ &  \\ \hline
		
			\cite{latency8} &$\checkmark$  &   &  & &  & & &  & & $\checkmark$ &  &   & && & & $\checkmark$ &  &  & $\checkmark$ &  \\ \hline
			\cite{latency2, latency3} &  &  $\checkmark$ &  & &  & & &  & $\checkmark$& $\checkmark$ &  &   & && & & $\checkmark$ &  &  &  &  \\ \hline
			\cite{	latency10} &  &   & $\checkmark$ & &  & & &  & $\checkmark$& $\checkmark$ &  &   & && & & $\checkmark$ &  &  &  &  \\ \hline
			\cite{ofdm} &  &   &  & & $\checkmark$   & & &  &$\checkmark$ &  &  &   & &$\checkmark$&& & $\checkmark$ &  &  &  &  \\ \hline
			
			\cite{ocdm} &  &   &  & &    & $\checkmark$& &  & &  &  &   & &$\checkmark$&& & $\checkmark$ &  &  &  &  \\ \hline
			\cite{DBA4}&  &$\checkmark$&  & &    & & &  & & $\checkmark$ &  &   &$\checkmark$ &&& &$\checkmark$  &  &  &  &  \\ \hline
			\cite{capacity8} & $\checkmark$&   &   & &  & & &  & &  &  &   & &$\checkmark$& & &  & $\checkmark$ &  &  &  \\ \hline
			\cite{security1} &$\checkmark$  &   &  & &    & & &  & &  &  &   & && & $\checkmark$& $\checkmark$ &  &  &  &  \\ \hline
			\cite{cost6, cost61} &$\checkmark$  &   &  &$\checkmark$ &    & &$\checkmark$ &  & &  &  &   & & & & &  &  &  & $\checkmark$ &  \\ \hline
			\cite{cost1} &  &   &  &$\checkmark$ &   & &  &  &$\checkmark$ &  &  &   & && & & $\checkmark$ &  &  & $\checkmark$ &  \\ \hline
			
			\cite{cost2} & $\checkmark$ &   &  & &    & & $\checkmark$ & &$\checkmark$ &  &  &   & && & & $\checkmark$ &  &  &  &  \\ \hline
			
			\cite{cost3} &  &   & $\checkmark$ & &    & &   &$\checkmark$ & $\checkmark$ &  &   & && && & $\checkmark$ &  &  &  &  \\ \hline
			\cite{cost4} &  & $\checkmark$  &  & &    & & $\checkmark$&   &$\checkmark$ &  &  &   & && & & $\checkmark$ &  &  &  &  \\ \hline
			\cite{cost5} &$\checkmark$  &   & $\checkmark$ & &    & & &  &$\checkmark$ &  &  &   & &$\checkmark$& & & $\checkmark$ &  &  &  &  \\ \hline
			\cite{cost7} &  &   &$\checkmark$  & &    & &$\checkmark$ &  & &  &  &   & & & & & $\checkmark$ &  &  &  &  \\ \hline
			\cite{costoran} &  &   &  & $\checkmark$& & & &  & $\checkmark$&$\checkmark$  &  &   & & & & &  &  &  &  & $\checkmark$ \\ \hline
			\cite{cost9} & $\checkmark$ &   & $\checkmark$ & &    & & &  & $\checkmark$ &  & &   & && & & $\checkmark$ &  &  &  &  \\ \hline
			\cite{capacity1} &  &   &$\checkmark$  & &    & & &  & &$\checkmark$  &  &   & &$\checkmark$& & & $\checkmark$ &  &  &  &  \\ \hline
			\cite{capacity3} &  &   & $\checkmark$ & &    & & &  &$\checkmark$ &  &  &   & && & & $\checkmark$ &  &  &  &  \\ \hline
			\cite{capacity4} &$\checkmark$  &   &  & &    & & &  &$\checkmark$ &  &  &   & &$\checkmark$& & & $\checkmark$ &  &  &  &  \\ \hline
			\cite{capacity5} &  &   & $\checkmark$ & &    & &$\checkmark$ &  & &  &  &   & &$\checkmark$& & & $\checkmark$ &  &  &  &  \\ \hline
			
			\cite{capacity6} & &   &  $\checkmark$ & &  & & &  & &  &  &   & &$\checkmark$& & & $\checkmark$ &  &  &  &  \\ \hline
			\cite{capacity7} & &   &   & &  & &$\checkmark$ &  & &  &  &   & &$\checkmark$& & &  & $\checkmark$ &  &  &  \\ \hline
			\cite{resource1} &$\checkmark$  &   & $\checkmark$ & &    & & &  & &  &  &   &$\checkmark$ &$\checkmark$& & & $\checkmark$ &  &  &  &  \\ \hline
			\cite{resource2} &  &   &  &$\checkmark$ &    & & &  & &  & $\checkmark$ &   & $\checkmark$&&$\checkmark$ & & $\checkmark$ &  &  &  &  \\ \hline
			\cite{resource3} &$\checkmark$  &   & $\checkmark$ & &    & & &  & &  & $\checkmark$ &   & $\checkmark$&& & & $\checkmark$ &  &  &  &  \\ \hline
			\cite{SDN1, SDN2} &$\checkmark$  &   &  & &    & & &  &$\checkmark$ &  &  $\checkmark$&   & $\checkmark$&$\checkmark$& & & $\checkmark$ &  &  &  &  \\ \hline
			\cite{SDN3} &$\checkmark$  &   & $\checkmark$ & &    & & &  & &  &  $\checkmark$ &  & && & & $\checkmark$ &  &  &  &  \\ \hline
			\cite{ML1} &$\checkmark$  &   &  & &    & & &  &$\checkmark$ & $\checkmark$ &  &   & &$\checkmark$&$\checkmark$ & & $\checkmark$ &  &  &  &  \\ \hline
			\cite{ML2} &$\checkmark$  &   &  & &    & & &  & & $\checkmark$ &  &   & &$\checkmark$&$\checkmark$ & & $\checkmark$ &  &  &  &  \\ \hline
			\cite{energy1} &$\checkmark$  &   &  & &    & &$\checkmark$ &  &$\checkmark$ &  &  & $\checkmark$  & &$\checkmark$& & & $\checkmark$ &  &  &  &  \\ \hline
			\cite{energy2} &  & $\checkmark$  &  & &    & & &  & &  &  & $\checkmark$  & && & & $\checkmark$ &  &  &  &  \\ \hline
			\cite{energy3} &  &   &  & $\checkmark$&    & & &  & &  &  & $\checkmark$  & && & & $\checkmark$ &  &  &  &  \\ \hline
			\cite{energy4}&  &   &  &$\checkmark$ &    & & &  & &  &  & $\checkmark$  &$\checkmark$ &&& & $\checkmark$ &  &  & $\checkmark$ &  \\ \hline
			\cite{energy5}&  &   &  &$\checkmark$ &    & & &  & &  &  & $\checkmark$  &$\checkmark$ &&& &  &  &  $\checkmark$&  &  \\ \hline
			\cite{latency1, latency4} & $\checkmark$ &   &  & &  & & &  & & $\checkmark$ &  &   & &$\checkmark$& & & $\checkmark$ &  &  &  &  \\ \hline
		
			\cite{latency5, latency6} &$\checkmark$  &   &  & &  & & &  & & $\checkmark$ &  &   & && & & $\checkmark$ &  &  &  &  \\ \hline
			\cite{latency7} &$\checkmark$  &   &$\checkmark$  & &  & & &  &$\checkmark$ & $\checkmark$ &  &   & && & & $\checkmark$ &  &  &  &  \\ \hline
			\cite{latency9}&  &   &  &$\checkmark$ &    & & &  & &$\checkmark$  &  &   &$\checkmark$ &&& & $\checkmark$ &  &  &  &  \\ \hline

			\cite{FRANcost} & &   & $\checkmark$  & $\checkmark$&  & & &  & &  &  &   & &$\checkmark$& & & $\checkmark$ &  & $\checkmark$ &  &  \\ \hline
			\cite{FRANcost1, FRANcost2} & &   &   & $\checkmark$&  & & &  & &  &  &  $\checkmark$ &$\checkmark$ &$\checkmark$& & & $\checkmark$ &  & $\checkmark$ &  $\checkmark$ &  \\ \hline
			\cite{FRANcost3} & $\checkmark$&   &   & &  & & &  & & $\checkmark$ &$\checkmark$  &   & && & & $\checkmark$ &  &  &  $\checkmark$&  \\ \hline
			\cite{virtual} &$\checkmark$  &   & $\checkmark$ &$\checkmark$ &    & & &  & &$\checkmark$  &  $\checkmark$ &  & && & & $\checkmark$ & $\checkmark$ &$\checkmark$  & $\checkmark$ &  \\ \hline
			\cite{fsolatency} &  &   &  & &    & &$\checkmark$ &  & &  $\checkmark$ &  &   & && && $\checkmark$ &  &  &  &  \\ \hline
		
			\cite{colorwdm} &  &   & $\checkmark$ & &    & & &  & &   &  &   & &$\checkmark$& && $\checkmark$ &  &  &  &  \\ \hline	    
    \end{tabular}
	\end{table*}
\subsection{Lessons Learned}
From the surveyed papers, we can conclude the following learning points:
\begin{itemize}
	\item Addressing deployment cost challenges emphasizes collaborative solutions, such as resource sharing and infrastructure reuse, to facilitate cost-efficient optical fronthaul for future 6G networks.
	
	\item In response to escalating data traffic, a focus on energy efficiency underscores the pivotal role of optical fronthaul. Beyond environmental concerns, sustainable practices are deemed essential for the economic viability of optical fronthaul networks.
	
	\item The section on latency and jitter highlights their profound impact on user experience and network reliability. Innovative solutions, like CoE encapsulation and topology optimizations, underscore the imperative of minimizing latency for time-sensitive applications.
	
	\item Integration with other technologies emerges as a dual-fold strategy, marrying optical fiber with wireless technologies and FSO with wireless. The hybrid nature of integrated networks offers flexibility, scalability, and cost-effectiveness, while the fusion of FSO with wireless technologies enhances link availability in adverse weather conditions.
	
	\item Efficient resource allocation takes center stage, ensuring high 6G network performance. Dynamic bandwidth allocation proves pivotal, enhancing throughput, spectral efficiency, and energy efficiency. The adaptive nature of resource allocation aligns with the evolving demands of 6G networks.
	
	\item The infusion of ML and AI into 6G optical fronthaul networks transforms operational efficiency. Intelligent resource allocation, predictive maintenance, and fault detection become feasible, optimizing network performance and reliability in the dynamic 6G landscape.
	
	\item SDN introduces a dynamic paradigm, optimizing resource utilization and enhancing network monitoring. The programmability of SDN enables intelligent algorithms and proactive incident response, fortifying the resilience of fronthaul infrastructure.
	
	\item SDM emerges as a solution to modern communication networks' escalating data demands, offering scalable capacity for higher data rates and lower latency—a crucial lesson for future network scalability.
	
	\item In the security and privacy domain, fortifying fronthaul against emerging threats becomes imperative. Focus areas include security provisioning, certificate enrollment, and synchronization plane security. The overarching lesson is clear—a secure and reliable telecommunication network necessitates prioritizing improvements in fronthaul security.
\end{itemize}
\section{Research Projects Related to 5G/6G Fronthaul}\label{sec7}
Several projects have been launched in the past years focusing on the fronthaul and backhaul architectures for 5G and beyond networks. The different projects are summarized in Table.~\ref{tab:projects}. In this subsection, we highlight  the selected projects as follows:
\begin{itemize}
	\item \textbf{5G-PICTURE:} An EU-funded project and coordinated by the Centre Tecnològic de Telecommunications de Catalunya in Spain and has 18 partners from different countries. It started in 2017, and ended in 2020. Its objectives are as follows. Firstly, it developed and showcased a converged fronthaul and backhaul infrastructure, seamlessly integrating cutting-edge wireless technologies with innovative optical network solutions.
	Additionally, the project introduced the Dis-Aggregated-RAN (DA-RAN) concept, a novel approach that disentangles hardware and software components across wireless, optical, and compute/storage domains. A pivotal goal is the realization of ``resource disaggregation'' establishing a shared pool of resources that can be independently chosen and allocated on demand. The project undertakes demonstrations within a 5G railway experimental testbed and a stadium to validate these advancements, emphasizing ultra-high user density scenarios. Funded by the EU, the overarching ambition is to revolutionize conventional network infrastructures, fostering openness, scalability, and elasticity. The project placed a significant emphasis on optimizing resource and energy efficiency through dynamic functional splits and the software-driven evolution of the network \cite{project1}.
	\item \textbf{5G-XHaul:} An EU project funded and coordinated by IHP - Innovations for High-Performance Microelectronics in Germany, engaged 12 partners from diverse countries from 2015, 2018. This collaborative effort focused on developing a converged optical and wireless network solution capable of flexibly connecting Small Cells to the core network. The project's objectives included creating a dynamic resource allocation system leveraging user mobility, incorporating advanced technologies such as dynamically programmable mm-Wave transceivers and a Time Shared Optical Network, and implementing a software-defined cognitive control plane. Additionally, 5G-XHaul aimed to contribute to the establishment of international standards and integrate its technologies into a city-wide testbed in Bristol (UK) for comprehensive evaluation. These strategic goals were designed to meet the increasing demand for cost-effective and flexible broadband connectivity, harnessing key enablers like Small Cells, C-RAN, SDN, and NVF \cite{project2}.
	\item \textbf{6G Flagship:} It started in 2018 and will end in 2026. It is an eight-year research program that aims to envision and define 6G-enabled digital world towards 2030, aligned with the UN Sustainable Development Goals, where the physical and virtual worlds meet and wireless connectivity expands to all areas of life. It is funded by the Academy of Finland and the University of Oulu.  
	The project conducts high-quality research, creates impact through innovation and knowledge sharing, and supports the development of international standards and 6G trials to develop key technologies and concepts for 6G, including 6G fronthaul. 6G Flagship’s experts develop fundamental 6G technology components, novel wireless solutions, and business approaches in four research areas - wireless connectivity, device and circuit technologies, distributed intelligent computing, and sustainable human-centric services and applications \cite{project3, project31}.
	\item \textbf{5G-Crosshaul:}
	An EU-funded project involved 21 partners from 8 countries. The project started in 2015 and ended in 2017.
	The Xhaul project, situated within the 5G Infrastructure Public Private Partnership in Europe (5G PPP) framework, is a strategic initiative to address the intricate challenges of transport networks within the context of 5G. Coined from the fusion of "fronthaul" and "backhaul," the term "xhaul" encapsulates the project's focus on developing solutions that seamlessly integrate these vital segments of the transport network. With a vision of scalability and flexibility, the project endeavored to create transport solutions capable of accommodating the diverse and evolving demands of 5G networks. Prioritizing cost-efficiency, the Xhaul project explored approaches for economically viable deployment and operation of transport networks. Emphasizing interoperability and standardization, it contributed to developing harmonized solutions for transport networks, fostering a unified 5G ecosystem. Additionally, the project addressed network slicing requirements and placed a significant emphasis on enhancing the energy efficiency of transport networks, aligning with broader sustainability objectives \cite{project4}.
	\item \textbf{Metro-Haul:} An EU-funded project, and it has 20 partners. It is a forward-looking research initiative aimed at revolutionizing metro networks in the era of 5G and beyond. The project focuses on scalability and seeks to develop innovative solutions to accommodate the escalating demand for high-capacity, low-latency connectivity within urban areas. Emphasizing flexibility and programmability, METRO-HAUL aims to create metro networks that dynamically adapt to diverse traffic patterns and evolving service requirements. Integrating optical and wireless technologies seamlessly, the project envisions a cost-effective and efficient metro infrastructure. Addressing environmental concerns, METRO-HAUL is committed to designing energy-efficient solutions and incorporating end-to-end network slicing, which allows for virtualized network segments tailored to specific applications and services \cite{project5}.
	\item \textbf{Hexa-X:} An EU project has 23 partners from 9 countries and started in 2021. It aimed at developing key 6G technologies, including advancements in fronthaul and backhaul networks to shape the future of wireless technology. It aims to create a transformative fabric that connects the human, physical, and digital realms, addressing the immense challenges and opportunities for growth and sustainability beyond 2030. It envisions a future where wireless technology and architectural research are pivotal in achieving green deal efficiency, digital inclusion, and health and safety assurance. The project's ambition extends to developing key technology enablers such as novel radio access technologies at high frequencies, AI-driven air interfaces, and 6G architectural enablers that support network disaggregation and dynamic dependability, all contributing to a robust, connected intelligence framework \cite{project6}.
	
	\item \textbf{5G-COMPLETE:} An EU-funded project and has 13 partners from 7 countries. It is a collaborative effort to revolutionize the 5G architecture by creating a unified ultra-high capacity converged digital/analog FiberWireless (FiWi) RAN that combines key technologies such as analog modulation, packet-switched transport, uni-kernel technology, and post-Quantum cryptosystems. The project also aims to transform the RAN into a low-power distributed computer that integrates 5G New Radio fronthaul/midhaul/backhaul faculties into one Ethernet-based platform and to validate the results in the scalable lab- and field-trial demonstrators, as well as to introduce new business models and research opportunities \cite{project7}. 
	\item \textbf{TERAWAY:} An EU-funded project started in 2019 with 12 partners from 6 countries. It aimed to develop a new generation of photonics-enabled THz transceivers for high-capacity backhaul and fronthaul links in 5G networks. Additionally, it aimed to support ultra-broadband and ultra-wideband operation within carrier frequencies covering the W-band, D-band, and THz band (from 92 up to 322 GHz)and to introduce a software-defined networking controller for homogeneous management of network and radio resources. The project also seeks to provide the possibility to flexibly coordinate and use the spectral resources of a network within these bands as a shared pool of radio resources, enhancing network performance and energy efficiency \cite{project8}. 
	
	\item \textbf{Int5gent:} An EU-funded project started in 2020 and has 15 partners from 8 countries. It focuses on integrating advanced data plane technologies into a flexible 5G network infrastructure. The project aims to enhance 5G core technologies, creating an innovative 5G ecosystem. Key technologies of the project include multi-RAT baseband signal processing, beam steering, mmWave solutions, hardware-based edge processors, Graphics Processing Unit (GPU), 5G terminals, and SDN-based photonic data transport.
		
	Additionally, the project aims to showcase benefits such as increased bandwidth, low latency, and high reliability. It intends to create new market opportunities, especially for participating SMEs, through pilot validation of their solutions. Additionally, It focuses on developing advanced, flexible, and efficient fronthaul/backhaul connectivity for 5G networks fronthaul solutions, such as mmWave point-to-multipoint mesh nodes and D-band 5G terminal nodes with high transport capacity, in addition to supporting low-latency computing at the edge \cite{project9}.
	\item \textbf{MARSAL:} An EU-funded project started in 2021 and has 14 partners. It aimed to create a comprehensive framework for efficiently managing and orchestrating network resources in the 5G and 6G landscape. It strategically employs a converged optical-wireless network infrastructure in the access, fronthaul, and midhaul segments. The project's key objectives encompass the development of scalable cell-free networks, enabling a substantial increase in wireless access points through innovative distributed cell-free concepts and serial fronthaul approaches. Additionally, MARSAL focuses on enhancing the flexibility of optical access architectures for Beyond 5G Cell Site connectivity and implements a disaggregated SDN control plane for Fixed-Mobile Convergence. The initiative extends to deploying an Elastic Edge Computing paradigm using Cloud-Native technologies, incorporating ML-based algorithms in edge and midhaul Data Centers. Furthermore, MARSAL addresses network security concerns by introducing mechanisms for privacy and security, employing AI and Blockchain technologies to ensure a secure multi-tenant slicing environment. The ultimate aim is to deliver a self-driven infrastructure with pervasive, ML-driven control, showcasing a proof-of-concept for the developed solutions and advancing the landscape of network resource management in the evolving 5G and beyond ecosystem \cite{project10}. 
	\item \textbf{FLEX-SCALE:} An EU project funded under the Horizon Europe Program. It is coordinated by the University of Patras in Greece and has 11 participants from different countries. It started in 2023 and will end in 2025. It represents groundbreaking research on optical x-haul (x = front/mid/back) network technologies for 6G with a focus on several key aspects. First, it aims to achieve massive capacity scaling, boasting a flexible capacity of 10 Tb/s rate per interface, 1 Pb/s capacity per link, and 10 Pb/s throughput per optical node. Furthermore, the project emphasizes the utilization of ultra-high bandwidth photonic/plasmonic technologies, tapping into optical spatial and spectral switching for ultra-high. Energy efficiency takes center stage as the project strives for record-breaking efficiency, targeting sub-pJ per switched/transmitted bit, all while maintaining cost-effectiveness through photonic integration and optical transparency. Additionally, the initiative delves into innovative technologies such as autonomous SDN control, streaming telemetry, ML-enabled data analytics architectures, and energy-efficient routing algorithms, all aimed at minimizing energy consumption without compromising the QoS \cite{project11, project112}. 
	\item \textbf{EMPOWER-6G}: It is a Marie Skłodowska-Curie Innovative Doctoral Network funded by the European Commission under the Horizon 2020 program. It started in 2023 and will end 2027. It aims to revolutionize 6G cellular networks through a converged optical-wireless architecture, focusing on efficient Cell-Free access networks for high-density and high-coverage scenarios. Aligned with the O-RAN Alliance, the project leverages distributed processing CF concepts and wireless mmWave solutions. The vision encompasses the development of novel radio access solutions, innovations in the optical transport domain, and the evolution of Multi-access Edge Computing (MEC) towards fully elastic Edge Computing. The network configuration involves a two-tiered distributed Edge infrastructure with Data Centres (DCs), addressing both real-time and non-real-time network functions. Challenges at the radio-edge, regional-edge, and network-management domains are tackled, focusing on scalable CF networking mechanisms, mmWave Hybrid MIMO solutions, optical regional edge networks, and innovative control-plane protocols \cite{EMPOWER6G}.
	\item \textbf{PROTEUS-6G:} An EU-funded project. It starts in 2024 and aims to design and develop a dynamic, flexible, scalable, cost-effective, high-bandwidth, and low-latency packet-optical fronthaul and midhaul for 6G networks. This initiative is geared towards enabling the dynamic management of radio functional splits, facilitating the adaptation of 6G networks to varying environments, including changes in services and traffic. In this novel scenario, the 6G network adaptation involves both selecting the most appropriate functional split and reconfiguring the underlying packet-optical fronthaul and midhaul network. This ensures the provision of the required transport capacity and latency demanded by the selected centralization level \cite{6Gprotus}.
\end{itemize}
\begin{table*}
	\caption{Summary of research projects of 5G and beyond fronthaul.}
	\label{tab:projects}
	\begin{tabular}{|p{2cm}|p{1.3cm}|p{6cm}|p{7cm}|}
		\hline
		\textbf{Project} & \textbf{Time Frame} &\textbf{Title} & \textbf{Research Topics} \\ \hline
		5G-PICTURE & 2017-2020 & 5G Programmable Infrastructure Converging disaggregated neTwork and compUte REsources.&  DA-RAN, Network softwarization, flexible functional splits, converged fronthaul and backhaul services.\\ \hline
		
		 5G-XHaul&2015-2018&Dynamically Reconfigurable Optical-Wireless Backhaul/Fronthaul with Cognitive Control Plane for Small Cells and Cloud-RANs.& High-capacity and low-latency transport solutions, Network slicing, Fronthaul and backhaul solutions, Integration of optical and wireless technologies,  Energy efficiency, Network management and orchestration.\\ \hline
		 
		6G Flagship&2018-2026&6G-Enabled Wireless Smart Society \& Ecosystem.&New 6G Enablers, Scalable and 3GPP Compliant Federation Solution, Federated AI Plane, 6G architectures.\\ \hline
		
		5G-Crosshaul&2015-2017&The 5G Integrated fronthaul/backhaul transport network.&High-capacity and low-latency transport solutions, Network slicing, Fronthaul and backhaul solutions, Integration of optical and wireless technologies,  Energy efficiency, Network management and orchestration. \\ \hline
		
		Metro-Haul&2017-2020&METRO High bandwidth, 5G Application-aware optical network, with edge storage, compUte and low Latency.&Disaggregated optical metro network nodes, Intelligent control and orchestration, Multi-layer network slicing, Edge computing and network slicing.\\ \hline
		
		Hexa-X&2021-2024& A flagship for B5G/6G vision and intelligent fabric of technology enablers connecting human, physical, and digital worlds.&New radio access technologies, AI-driven air interface and governance for future networks, 6G architectural enablers, high capacity fronthaul/backhaul.\\ \hline
		
		5G-COMPLETE&2019-2022&A unified network, Computational and stOrage resource Management framework targeting end-to-end Performance optimization for secure 5G muLti-tEchnology and multi-Tenancy Environments.&Fiber wireless fronthaul, Ethernet fronthauling and eCPRI standard, Analog modulation and coding schemes, Unikernel technology and post-Quantum cryptosystems.\\ \hline
		
		TERAWAY&2019-2023&Terahertz technology for ultra-broadband and ultra-wideband operation of backhaul and fronthaul links in systems with SDN management of network and radio resources.&THz transceivers, Optical concepts and photonic integration, management of the network and radio resources,  backhaul and fronthaul solutions.\\ \hline
		
		Int5gent&2020-2024&Integrating 5G enabling technologies in a holistic service to physical layer 5G system platform.&Integration of innovative data plane technology blocks, 
		5G architecture that promotes edge processing
		Validation and showcasing of advanced 5G services and IoT solutions.
		\\ \hline
		
		MARSAL&2021-2024&ML-based, networking and computing infrastructure resource management of 5G and beyond intelligent networks.&Optical-wireless convergence, fixed-mobile convergence, distributed cell-free O-RAN for B5G, AI, elastic edge computing, self-driven infrastructure.\\ \hline
		
		FLEX-SCALE&2023-2025&Flexible Scalable Energy Efficient Networking.& Optical x-haul for 6G network, Transformational transceivers and optical switches, x-haul planning and operation algorithms, ML-enabled SDN control plane, reduced cost and power consumption.\\ \hline
		EMPOWER-6G&2023-2027&Empower converged optical wireless configurations with cell-free technologies for high-density 6G networks. & Converged optical-wireless architecture,  Cell-free access networks, Radio Access, Wireless mmWave, Optical transport, Multi-Access Edge Computing (MEC) Evolution, NFV, and Control-plane protocols.\\
		\hline
		PROTEUS-6G&2024-2027&Programmable Reconfigurable Optical Transport for Efficiently offering Unconstrained Services in 6G.&  Packet-optical fronthaul and midhaul,
		Dynamic management of radio functional splits. \\
		\hline
	\end{tabular}
\end{table*}

\section{Future Research Directions}\label{sec8}
Extensive research endeavors have been dedicated to 6G optical fronthaul, aiming to overcome challenges, enhance data transmission capacity, reduce latency, improve energy efficiency, and increase network flexibility. However, despite significant progress, numerous challenges and new research directions exist. This section highlights the key challenges and outlines essential research directions for realizing an efficient optical fronthaul in future 6G networks. These include:
	\subsection{Capacity Enhancement} Undoubtedly, the relentless pursuit of  Tbps capacities in the imminent 6G networks mandates a strategic and comprehensive approach, compelling the amalgamation of cutting-edge modulation techniques with the seamless integration of coherent PON and FSO. coherent optical communications have been widely recognized as a promising technology for enhancing the capacity, performance, and efficiency of optical fronthaul networks for 5G and beyond.
		
		Intriguingly, the incorporation of coherent PON-based 6G fronthaul emerges as a paramount avenue, introducing a new layer of complexity and, simultaneously, unlocking unparalleled potential for capacity enhancement. Coherent PON harnesses the power of advanced modulation formats and digital signal processing to catapult the data-carrying capabilities of optical fronthaul networks to unprecedented heights. By capitalizing on the coherence properties of light, coherent PON seamlessly orchestrates the simultaneous transmission of multiple wavelengths over a single optical fiber, thereby enhancing overall network throughput up to 100 Gbps/$\lambda$ or higher \cite{coherentpon}.

	Additionally, the recent work \cite{coherentfso} highlights the ability of coherent FSO transmission link to deliver more than 800 Gbps over a 42 m distance in an outdoor deployment exposed to time-varying turbulence and meteorological conditions, which meet the high capacity fronthaul links required by 6G networks.
	While substantial progress has been made, there exists a need for a more profound \emph{investigation} into sophisticated modulation methods, such as higher-order modulation and multi-dimensional signaling. This research direction not only addresses current limitations but also positions optical fronthauls as robust infrastructures capable of supporting the diverse and data-intensive requirements anticipated in future 6G scenarios \cite{capacityenhance}.

	\subsection{Latency Reduction} In the era of 6G networks, the pivotal role of mobile networks extends far beyond mere connectivity, influencing critical facets of daily life such as telemedicine and automated vehicles. Unlike previous generations, where long delays merely affected connections, in 6G, extended latency could potentially translate to life-threatening consequences. This paradigm shift necessitates an urgent focus on latency reduction, aiming to achieve levels as low as 100  $\mu$s or even less than 10  $\mu$s, as in Table.~\ref{tab:6glatency} that shows the different latency requirements of various 6G use cases based on \cite{6gfrontlatency}. 
	The primary sources of latency in the physical layer stem from three key components: optical fronthaul (assuming the use of optical fiber), Electrical/Optical (E/O) devices, and signal processing. Consider a coherent system covering a distance of 20 km as an illustration; in this scenario, the three components typically contribute approximately 98  $\mu$s, 10  $\mu$s, and 5 $\mu$s of latency, respectively. This results in a cumulative latency of approximately 113  $\mu$s, meeting the existing 5G fronthaul requirements. Nevertheless, it falls short of the requirements for 6G \cite{spectral1}. 
	
	In the mentioned fronthaul technologies, the only architecture that can meet such latencies is P2P.
	For that, it is essential to have a unified management layer for different network domains, which can control the QoS parameters from end to end in real-time \cite{6Glatency}. 
	 The optical fronthaul network needs methods for latency planning that can satisfy the strict requirements of 6G services. Also, it is important to deal with the challenges of fronthaul traffic, which requires precise latency engineering because of its large bandwidth usage and tight delay and jitter limits. Real-time monitoring mechanisms that can split end-to-end latency into layers and domains are necessary to automatically find the network elements that cause latency violations.
	\begin{table}[h]
		\centering
		\caption{Latency requirements for various 6G use cases.}
		\label{tab:6glatency}
		\begin{tabular}{|c|c|}
			\hline
			\textbf{6G Use Case} & \textbf{Latency Requirements} \\
			\hline
			Remote health monitoring & 1 ms \\ \hline
			Telemedicine & 0.1 ms \\ \hline
			Servo motors & 0.1 $\mu$s \\ \hline
			Motion control & 1 $\mu$s \\ \hline
			Delivery drones & 1 ms \\ \hline
			V2X & 0.1 ms \\ 
			\hline
		\end{tabular}
	\end{table}
	\subsection{Improving Energy Efficiency}
	The current energy crisis and growing concerns about climate change highlight the urgency of adopting sustainable and energy-efficient solutions. In the context of 6G optical fronthaul, future research should prioritize innovative \emph{transceiver architectures}, advanced materials, and energy-efficient modulation schemes to minimize power consumption while maintaining performance standards. Furthermore, \emph{dynamic resource allocation algorithms}, incorporating ML and AI, can adapt to varying traffic loads and channel conditions, enhancing predictive and optimized resource utilization for energy-efficient operations. 
	
	Additionally, research should delve into \emph{routing algorithms} that consider both energy efficiency and traditional performance metrics, exploring network topologies like hierarchical or mesh structures that inherently support energy efficiency. Moreover, integrating \emph{renewable energy sources} such as solar or wind power into the optical fronthaul infrastructure aligns with sustainability goals, significantly reducing the environmental impact of 6G networks. Emphasizing \emph{cross-layer optimization} is crucial, as understanding the interplay between physical layer parameters, data link layer protocols, and network layer operations is essential for achieving optimal energy efficiency across the entire fronthaul network.
	\subsection{Cost-efficient Deployment} One critical aspect that demands attention is the cost of deploying and maintaining 6G optical fronthaul networks \cite{cost8, cost61, Integrated}.
	To address this concern, an effective avenue for achieving cost-efficient deployment involves maximizing the utilization of existing infrastructure. Operators can achieve this by repurposing and upgrading current optical fiber networks, minimizing the need for new installations, and significantly reducing deployment costs. Furthermore, integrating advanced technologies in P2P links, such as WDM and SDN, facilitates efficient sharing of optical resources, optimizing the overall cost structure.
	PON architectures also present a promising opportunity for cost-efficient deployments through \emph{network sharing and virtualization}. By allowing multiple users or service providers to share the same optical infrastructure, the initial capital and operational expenditures can be distributed among stakeholders. This approach effectively reduces the financial burden on individual entities and enhances the overall cost-effectiveness of PON-based fronthaul networks.

	Additionally, FSO technology offers a wireless alternative for optical fronthaul, creating opportunities for cost-efficient deployment, especially in scenarios where physical fiber installations are challenging. To enhance the feasibility of FSO links, future research directions should focus on improving the robustness and reliability of these links, particularly in adverse weather conditions. This includes the development of adaptive modulation and coding techniques, which can maintain link performance under varying atmospheric conditions \cite{cost61}. 
	\subsection{Network Resilience} To fortify the reliability of 6G networks, researchers are increasingly focusing on network resilience, which involves the ability of a system to maintain its functionality in the face of various challenges, such as hardware failures and environmental disruptions \cite{resilliency, resilliency1}.
	One direction for future research in 6G optical fronthaul is the exploration of advanced hardware redundancy and fault-tolerant mechanisms. This includes investigating novel architectures that seamlessly reroute traffic during component failures or malfunctions. 
	Moreover, developing intelligent algorithms to dynamically adapt to changing network conditions and quickly identify and isolate faults will be crucial for maintaining continuous connectivity. 
	
	The ability to dynamically reconfigure network resources in response to evolving demands and disruptions is essential to network resilience. Additionally, future research should explore intelligent algorithms and protocols that enable dynamic resource allocation, rerouting, and optimization. This includes developing self-healing mechanisms that can autonomously detect and mitigate issues without human intervention, contributing to a more resilient and self-sustaining network.
	Optical communication is susceptible to environmental factors such as adverse weather conditions and physical obstructions. Investigating ways to enhance the resilience of optical fronthaul networks against these factors is a promising research direction. This may involve developing adaptive FSO systems that dynamically adjust to environmental changes.
	
\subsection{Security and Privacy Aspects} Numerous studies have delved into the security implications of 5G and beyond fronthaul, as evidenced by the works cited \cite{security1, security3, security31, furdek2021optical}. However, the security dimensions of  6G fronthaul remain an active area of research. Here, we mention critical security and privacy research directions in 6G optical fronthaul. 
Primarily, there is a need for in-depth exploration of robust encryption mechanisms to fortify communication channels between fronthaul elements. This initiative aims to secure the fronthaul against potential threats. Simultaneously, the development of efficient authentication protocols is imperative to thwart unauthorized access, thereby reinforcing the security of optical fronthaul networks.
Furthermore, the deployment of PONs introduces privacy concerns, particularly in densely populated areas. Consequently, it is essential to scrutinize privacy-preserving mechanisms such as secure user authentication and data anonymization. Implementing these measures becomes crucial to safeguard end-user privacy while upholding the efficacy of PON-based optical fronthaul networks.

Additionally, FSO-based fronthaul, despite its advantages in data rates and latency, is susceptible to eavesdropping due to its broadcast nature. To counter potential security threats, future research efforts should concentrate on devising physical layer security solutions for FSO links. This includes exploring technologies like beamforming, quantum key distribution, and advanced modulation schemes.

Lastly, the increasing interconnectedness and software-defined nature of optical fronthaul expose it to cyber-physical attacks. To fortify against potential vulnerabilities in both the optical and digital domains, future research should focus on enhancing the resilience of optical fronthaul. This involves the development of intrusion detection systems and secure software-defined networking frameworks to mitigate cyber-physical threats effectively.

\subsection{Standardization and Interoperability} One of the pivotal challenges in developing the optical fronthaul of 6G O-RAN architecture is the establishment of standardized interfaces, protocols, and interoperability requirements. This imperative arises from the need to seamlessly integrate and interconnect heterogeneous network components from various vendors, enabling the deployment of multi-vendor fronthaul solutions. Standardization ensures uniformity, provides a common language for diverse elements, and enhances overall network reliability. 

Efforts should focus on defining protocols tailored to the unique characteristics of 6G, addressing factors such as high data rates, low latency, and dynamic resource allocation. With an ecosystem comprising diverse elements like P2P links, PON architectures, and FSO communication systems, the challenge is to develop standards that accommodate this diversity. The ultimate goal is to facilitate the deployment of multi-vendor solutions, allowing network operators to select the best components from different vendors, fostering innovation, and avoiding vendor lock-in. While initiatives by organizations like the O-RAN Alliance \cite{oran1} and related standards bodies \cite{oran2} are laying the groundwork, ongoing and future research is vital to refine and expand these standards, ensuring they evolve to meet the dynamic needs of 6G networks.
\subsection{Accurate Time and Frequency Synchronization} Accurate time and frequency synchronization is crucial for the evolution of 6G optical fronthaul networks, especially in the context of coordinated multi-point transmission and reception techniques within wireless networks. As 5G and beyond scenarios require ultra-low-latency communication, synchronization methodologies become more essential, as stated in \cite{sync, sync1}. Therefore, novel synchronization approaches are needed to ensure precise alignment of time and frequency across the optical fronthaul network. Synchronization enables coherent collaboration among distributed network elements and helps reduce latency and optimize network performance. Thus, pursuing such synchronization advancements can unlock the full potential of coordinated multi-point transmission and reception, propelling 6G optical fronthaul networks towards unprecedented levels of reliability and performance.

\subsection{Caching and Edge Computing} Integrating intelligent caching and edge computing technologies emerges as a pivotal research direction in 6G optical fronthaul networks, offering substantial improvements in latency reduction, core network offloading, and real-time data processing \cite{cach1, cach2}.

By strategically deploying caches at the fronthaul layer, end-user latency is minimized, benefiting XR applications. This approach also enables offloading data from the core network to the edge, optimizing resource utilization and enhancing overall bandwidth management. The real-time processing capabilities at the edge empower the network to efficiently handle applications like video analytics and industrial automation, unlocking new possibilities. However, challenges related to resource management, security, and coordination among edge nodes underscore the need for future research to develop dynamic caching algorithms, address security concerns, and explore collaborative edge computing strategies. In navigating these challenges, researchers aim to harness the full potential of caching and edge computing, making significant strides toward meeting the stringent requirements of 6G communication systems.
\subsection{Advanced Monitoring and Diagnostics Techniques} In the context of 5G and beyond optical fronthaul networks, the imperative need for advanced monitoring and diagnostics techniques is evident \cite{monitor1}. The development of sophisticated real-time monitoring tools is crucial to meet the stringent requirements of 6G optical fronthaul. These tools should provide comprehensive insights into various network parameters, including latency, bandwidth utilization, signal quality, and overall network health. Leveraging AI and ML algorithms can enhance the capabilities of these tools, allowing for predictive analytics and anomaly detection. By continuously analyzing data patterns, these tools can anticipate potential issues and trigger preemptive measures before they escalate into critical problems.

Moreover, efficient maintenance is essential for the sustained performance of 6G optical fronthaul networks. Future research directions should explore innovative maintenance strategies that leverage automation and robotics. Autonomous robotic systems equipped with optical inspection capabilities can conduct routine inspections, identify physical layer impairments, and even perform minor repairs without human intervention. This approach not only reduces operational costs but also enhances the overall reliability of the network.

\subsection{Digital Twins (DTs) Empowered Optical Fronthaul}

In the rapidly evolving landscape of 6G networks, the integration of cutting-edge technologies such as DTs becomes imperative to address unprecedented demands for connectivity, reliability, and efficiency \cite{twin6g}.

DTs, serving as virtual replicas of physical systems or processes, play a pivotal role in reshaping the design, operation, and optimization of 6G optical networks \cite{twinoptic}. It offers a comprehensive virtual representation of the entire optical fronthaul infrastructure, encompassing elements such as fiber optic cables, network components, and connected devices. The resulting virtual environment provides operators with real-time insights into the network's status and performance, enabling informed decision-making \cite{twinoptic1, twinoptic, twinoptic2, twinoptic3, twinoptic4}. As a future research direction in this field, we highlight the following:

One promising avenue for future research involves developing methods for utilizing DTs in proactive decision-making regarding resource allocation, bandwidth management, and capacity planning. This can be achieved by analyzing historical data and simulating various scenarios, providing a predictive capability that is indispensable in optimizing optical fronthauls' performance within 6G networks. This approach ensures efficient resource utilization to meet dynamically changing demands.

Furthermore, the dynamic configuration and adaptive management of optical fronthaul networks are made possible through DTs. The virtual representation allows operators to simulate different configurations and scenarios, assessing the impact of changes before implementation. This capability enhances network operations' efficiency and enables adaptive management strategies, where the digital twin continuously learns from real-time data to adjust parameters for optimized performance under changing conditions autonomously.

Real-time monitoring capabilities provided by DTs contribute significantly to early fault detection and rapid response. Identifying anomalies and potential issues within the optical fronthaul network in the digital twin environment allows operators to take preventive measures before they impact the network. This proactive approach is crucial in improving reliability and reducing downtime, ensuring a seamless user experience in 6G networks.

The synergy between DTs and AI/ML algorithms further amplifies the capabilities of optical fronthaul networks. Leveraging data generated by DTs, AI algorithms optimize resource allocation, predict traffic patterns, and enhance overall network performance. This intelligent integration contributes to the self-optimizing nature of 6G networks, where DTs serve as dynamic models providing real-time insights to AI/ML algorithms for informed decision-making. As part of future research endeavors, exploring the full potential of this synergy in the context of 6G fronthaul applications is an intriguing avenue to pursue.

\subsection{Spatial Modulation for FSO-based 6G Fronthaul}
The exploration of Sparse Massive-Multiple Input Multiple Output (SM-MIMO) technologies within the realm of 6G networks offers promising advancements, particularly concerning FSO-based optical fronthauls. SM-MIMO, representing an evolution from traditional MIMO technology, distinguishes itself through the deployment of a substantial number of antennas with a sparse configuration, leading to notable improvements in spectral efficiency and communication capabilities.

A prospective direction for future research involves delving deeper into the implications of SM-MIMO on FSO-based optical fronthauls. The substantial increase in data throughput and capacity, facilitated by the adept use of spatial diversity within SM-MIMO, presents an intriguing area for further investigation. This exploration may involve optimizing the sparse arrangement of antennas to maximize the benefits of spatial diversity, thereby addressing the escalating demands for high-bandwidth applications and services in the evolving landscape of 6G networks.

Additionally, future research efforts could focus on enhancing the adaptability of SM-MIMO to atmospheric effects on FSO links. Investigating how the sparse antenna arrangement aids in effectively mitigating challenges such as turbulence and beam wander could lead to the development of more robust and weather-resilient FSO-based optical fronthauls. Understanding the intricate interplay between SM-MIMO and atmospheric conditions will be crucial for ensuring consistent performance in adverse weather scenarios.

Furthermore, there is an opportunity for research to explore the role of SM-MIMO in achieving ultra-low latency, a critical requirement for emerging applications like XR, and autonomous systems in 6G networks. Investigating the potential optimizations and configurations that can further reduce communication latency through SM-MIMO's parallel transmission and reception capabilities could pave the way for improved QoS in time-sensitive applications.

\section{Conclusions}\label{sec9}
The advent of 6G technology introduces novel applications that cater to various societal requirements, offering enhanced performance and capacity beyond current capabilities. The fronthaul segment is an essential part of 6G networks, providing high capacity, low latency connections to the end user. However, the optimization of fronthaul infrastructure for 6G presents numerous challenges, demanding concerted efforts from academia and industry to realize the full potential of 6G networks.
Optical technologies, including P2P, PON, and FSO are essential for achieving massive capacity, ultra-low latency, ultra-reliability, and security required by different splitting options of 5G/6G fronthaul (1-7.3, 7.2, 7.1, and 8).
P2P is suitable for high-capacity and low-latency
demand scenarios. However, deploying P2P architecture requires massive investments. PON architecture can be considered a cost-efficient solution to overcome the high deployment cost of P2P. However, there is a need for the development of sophisticated coherent optical transmitters and receivers in order to increase the capacity of PON to the level of $\geq$ 1 Tbps per wavelength to meet the massive capacity required by 6G. Additionally, there is a need for a new latency management mechanism that splits and prioritizes different services based on their urgency to meet various latency requirements of different emergent use cases. FSO is considered a viable alternative for optical fiber-based fronthaul solutions in scenarios where deploying optical cables is unenviable due to physical or cost constraints. However, there is a need for new routing and modulation schemes to overcome the sensitivity of FSO systems for different weather conditions. 
Recognizing the multifaceted demands of 6G, it becomes evident that no single optical technology can meet all the diverse requirements of 6G. Combining and integrating various optical technologies will likely be the key to handling these diverse needs.

In this paper, we provided a comprehensive survey on the enabling technologies and research perspectives toward efficient 6G optical fronthaul. We started by discussing the evolution of communication networks toward 6G, highlighting the new challenges facing the advent of 6G. The evolving landscape of radio access networks was introduced, emphasizing the need for efficient fronthaul interfaces. 
Meanwhile, we discussed the fronthaul interface and delved into the various optical technologies available for 6G fronthaul and their implications, including P2P, PON, and FSO. Each approach was analyzed regarding its advantages, limitations, and applicability in the context of 5G and beyond.
Furthermore, we have briefly discussed cutting-edge research on 5G/6G optical fronthaul, highlighting recent advancements and research projects. 
Based on this survey, we identified several future research directions to enhance further the development of 6G optical fronthaul systems. 

\bibliographystyle{IEEEtran}
\bibliography{refs}

\begin{thebibliography}{100}
\providecommand{\url}[1]{#1}
\csname url@samestyle\endcsname
\providecommand{\newblock}{\relax}
\providecommand{\bibinfo}[2]{#2}
\providecommand{\BIBentrySTDinterwordspacing}{\spaceskip=0pt\relax}
\providecommand{\BIBentryALTinterwordstretchfactor}{4}
\providecommand{\BIBentryALTinterwordspacing}{\spaceskip=\fontdimen2\font plus
\BIBentryALTinterwordstretchfactor\fontdimen3\font minus
  \fontdimen4\font\relax}
\providecommand{\BIBforeignlanguage}[2]{{%
\expandafter\ifx\csname l@#1\endcsname\relax
\typeout{** WARNING: IEEEtran.bst: No hyphenation pattern has been}%
\typeout{** loaded for the language `#1'. Using the pattern for}%
\typeout{** the default language instead.}%
\else
\language=\csname l@#1\endcsname
\fi
#2}}
\providecommand{\BIBdecl}{\relax}
\BIBdecl

\bibitem{itu}
ITU-R, ``{IMT} traffic estimates for the years 2020 to 2030,'' ITU, available
  online: \url{www.itu.int/pub/R-REP-M.2370-2015}.

\bibitem{5g1}
M.~Agiwal, A.~Roy, and N.~Saxena, ``Next generation {5G} wireless networks: a
  comprehensive survey,'' \emph{IEEE Communications Surveys \& Tutorials},
  vol.~18, no.~3, pp. 1617--1655, 2016.

\bibitem{6g1}
W.~Jiang, B.~Han, M.~A. Habibi, and H.~D. Schotten, ``The road towards {6G}: a
  comprehensive survey,'' \emph{IEEE Open Journal of the Communications
  Society}, vol.~2, pp. 334--366, 2021.

\bibitem{6g2}
M.~Boldi \emph{et~al.}, ``{{6G} architecture landscape – European
  perspective},'' White Paper, 2022,
  \url{https://5g-ppp.eu/6g-architecture-landscape-european-perspective-white-paper}.

\bibitem{ericsson}
\BIBentryALTinterwordspacing
Ericsson, ``Ever-present intelligent communication: a research outlook towards
  {6G},'' 2020. [Online]. Available:
  \url{https://www.ericsson.com/en/reports-and-papers/white-papers/a-research-outlook-towards-6g}
\BIBentrySTDinterwordspacing

\bibitem{nokia}
\BIBentryALTinterwordspacing
Nokia, ``Communications in the {6G} era,'' 2020. [Online]. Available:
  \url{https://www.bell-labs.com/institute/white-papers/communications-6g-era-white-paper/}
\BIBentrySTDinterwordspacing

\bibitem{samsung}
\BIBentryALTinterwordspacing
Samsung, ``The next hyper-connected experience for all,'' 2020. [Online].
  Available: \url{https://cdn.codeground.org/nsr/downloads/researchareas/2020
  12 01-6G-Vision-web.pdf}
\BIBentrySTDinterwordspacing

\bibitem{huawei}
\BIBentryALTinterwordspacing
Huawei, ``{6G}: the next horizon white paper,'' 2022, [Accessed: 03-Feb-2023].
  [Online]. Available:
  \url{https://www.huawei.com/en/huaweitech/future-technologies/6g-white-paper}
\BIBentrySTDinterwordspacing

\bibitem{projects}
S.~A.~A. Hakeem, H.~H. Hussein, and H.~Kim, ``Vision and research directions of
  {6G} technologies and applications,'' \emph{Journal of King Saud University -
  Computer and Information Sciences}, vol.~34, no.~6, pp. 2419--2442, 2022.

\bibitem{oran1}
\BIBentryALTinterwordspacing
O.~Alliance, ``{{O-RAN}: towards an open and smart RAN},'' 2018. [Online].
  Available: \url{https://www.o-ran.org/resources}
\BIBentrySTDinterwordspacing

\bibitem{oran2}
P.~Wireless, ``Everything you need to know about open {RAN},'' 2020.

\bibitem{oran22}
M.~Polese, L.~Bonati, S.~D’Oro, S.~Basagni, and T.~Melodia, ``Understanding
  {O-RAN}: architecture, interfaces, algorithms, security, and research
  challenges,'' \emph{IEEE Communications Surveys \& Tutorials}, vol.~25,
  no.~2, pp. 1376--1411, 2023.

\bibitem{wireless}
M.~Z. Chowdhury, M.~Shahjalal, M.~K. Hasan, and Y.~M. Jang, ``The role of
  optical wireless communication technologies in {5G}/{6G} and {I}o{T}
  solutions: prospects, directions, and challenges,'' \emph{Appl. Sciences},
  vol.~9, no.~20, pp. 1--20, 2019.

\bibitem{optical}
S.~Miladić-Tešić, G.~Marković, D.~Peraković, and I.~Cvitić, ``A review of
  optical networking technologies supporting {5G} communication
  infrastructure,'' \emph{Wireless Networks}, vol.~28, no.~1, pp. 459--467,
  2021.

\bibitem{pon2}
S.~Aleksic, ``A survey on optical technologies for {I}o{T}, smart industry, and
  smart infrastructures,'' \emph{Journal of Sensor and Actuator Networks},
  vol.~8, no.~3, pp. 1--18, 2019.

\bibitem{related10}
M.~Jaber, M.~A. Imran, R.~Tafazolli, and A.~Tukmanov, ``{5G} backhaul
  challenges and emerging research directions: a survey,'' \emph{IEEE Access},
  vol.~4, pp. 1743--1766, 2016.

\bibitem{related1}
D.~H. Hailu, B.~G. Gebrehaweria, S.~H. Kebede, G.~G. Lema, and G.~T.
  Tesfamariam, ``Mobile fronthaul transport options in {C-RAN} and emerging
  research directions: A comprehensive study,'' \emph{Optical Switching and
  Networking}, vol.~30, pp. 40--52, 2018.

\bibitem{related2}
I.~A. Alimi, A.~L. Teixeira, and P.~P. Monteiro, ``Toward an efficient {C-RAN}
  optical fronthaul for the future networks: a tutorial on technologies,
  requirements, challenges, and solutions,'' \emph{IEEE Communications Surveys
  \& Tutorials}, vol.~20, no.~1, pp. 708--769, 2018.

\bibitem{related7}
E.~Yaacoub and M.-S. Alouini, ``A key {6G} challenge and
  opportunity—connecting the base of the pyramid: a survey on rural
  connectivity,'' \emph{Proceedings of the IEEE}, vol. 108, no.~4, pp.
  533--582, 2020.

\bibitem{related8}
S.~S. Jaffer, A.~Hussain, M.~A. Qureshi, and W.~S. Khawaja, ``Towards the
  shifting of {5G} front haul traffic on passive optical network,''
  \emph{Wireless Personal Communications}, vol. 112, no.~3, pp. 1549--1568,
  2020.

\bibitem{related9}
T.~Sharma, A.~Chehri, and P.~Fortier, ``{Review of optical and wireless
  backhaul networks and emerging trends of next generation 5G and 6G
  technologies},'' \emph{Transactions on Emerging Telecommunications
  Technologies}, vol.~32, no.~3, pp. 1--16, 2021.

\bibitem{related3}
H.~Guo, Y.~Wang, J.~Liu, and N.~Kato, ``Super-broadband optical access networks
  in {6G}: vision, architecture, and key technologies,'' \emph{IEEE Wireless
  Communications}, vol.~29, no.~6, pp. 152--159, 2022.

\bibitem{related5}
M.~Jiang, J.~Cezanne, A.~Sampath, O.~Shental, Q.~Wu, O.~Koymen, A.~Bedewy, and
  J.~Li, ``Wireless fronthaul for {5G} and future radio access networks:
  challenges and enabling technologies,'' \emph{IEEE Wireless Communications},
  vol.~29, no.~2, pp. 108--114, 2022.

\bibitem{related6}
B.~Tezergil and E.~Onur, ``Wireless backhaul in {5G} and beyond: issues,
  challenges and opportunities,'' \emph{IEEE Communications Surveys \&
  Tutorials}, vol.~24, no.~4, pp. 2579--2632, 2022.

\bibitem{related4}
H.~R. D.~F. et~al., ``Wireless and optical convergent access technologies
  toward {6G},'' \emph{IEEE Access}, vol.~11, pp. 9232--9259, 2023.

\bibitem{xhaul}
C.~Ranaweera, C.~Lim, Y.~Tao, S.~Edirisinghe, T.~Song, L.~Wosinska, and
  A.~Nirmalathas, ``Design and deployment of optical x-haul for {5G}, {6G}, and
  beyond: progress and challenges [invited],'' \emph{Journal of Optical
  Communications and Networking}, vol.~15, no.~9, pp. D56--D66, 2023.

\bibitem{rofsurvey}
B.~Bismi and S.~Azeem, ``A survey on increasing the capacity of {5G} fronthaul
  systems using {RoF},'' \emph{Optical Fiber Technology}, vol.~74, pp. 1--15,
  2022.

\bibitem{1g1}
P.~Sharma, ``Evolution of mobile wireless communication networks {1G} to {5G}
  as well as future prospective of next generation communication network,''
  \emph{International Journal of Computer Science and Mobile Computing},
  vol.~2, no.~8, pp. 47--53, 2013.

\bibitem{1g2}
L.~Afolabi, E.~Olawole, F.~Taofeek-Ibrahim, T.~Mohammed, and O.~Shogo,
  ``Evolution of wireless networks technologies, history and emerging
  technology of {5G} wireless network: a review,'' \emph{Journal of Telecommun
  System Management}, vol.~7, pp. 2167--0919, 2018.

\bibitem{3g1}
R.~Baldemair, E.~Dahlman, G.~Fodor, G.~Mildh, S.~Parkvall, Y.~Selen,
  H.~Tullberg, and K.~Balachandran, ``Evolving wireless communications:
  addressing the challenges and expectations of the future,'' \emph{IEEE
  Vehicular Technology Magazine}, vol.~8, no.~1, pp. 24--30, 2013.

\bibitem{3g2}
T.~Mshvidobadze, ``Evolution mobile wireless communication and {LTE}
  networks,'' in \emph{6th International Conference on Application of
  information and communication Technologies (AICT)}.\hskip 1em plus 0.5em
  minus 0.4em\relax IEEE, 2012, pp. 1--7.

\bibitem{4g1}
S.~Won and S.~W. Choi, ``Three decades of {3GPP} target cell search through
  {3G}, {4G}, and {5G},'' \emph{IEEE Access}, vol.~8, pp. 116\,914--116\,960,
  2020.

\bibitem{5g2}
A.~Gupta and R.~K. Jha, ``A survey of {5G} network: architecture and emerging
  technologies,'' \emph{IEEE Access}, vol.~3, pp. 1206--1232, 2015.

\bibitem{6g3}
N.~Chen and M.~Okada, ``{Toward 6G internet of things and the convergence with
  RoF system},'' \emph{IEEE Internet of Things Journal}, vol.~8, no.~11, pp.
  8719--8733, 2021.

\bibitem{6g4}
M.~H. Alsharif, A.~H. Kelechi, M.~A. Albreem, S.~A. Chaudhry, M.~S. Zia, and
  S.~Kim, ``Sixth generation ({6G}) wireless networks: vision, research
  activities, challenges and potential solutions,'' \emph{Symmetry}, vol.~12,
  no.~4, pp. 1--21, 2020.

\bibitem{6g6}
W.~Jiang, B.~Han, M.~A. Habibi, and H.~D. Schotten, ``The road towards {6G}: A
  comprehensive survey,'' \emph{IEEE Open Journal of the Communications
  Society}, vol.~2, pp. 334--366, 2021.

\bibitem{hexa}
\BIBentryALTinterwordspacing
K.~Roth and et~al., ``Towards {TBPS} communication in {6G}: Use cases and gap
  analysis,'' 2021. [Online]. Available:
  \url{https://hexa-x.eu/wpcontent/uploads/2021/06/Hexa-X-D2.1.pdf}
\BIBentrySTDinterwordspacing

\bibitem{thz}
S.~Tripathi, N.~Sabu, A.~Gupta, and H.~Dhillon, ``Millimeter-wave and terahertz
  spectrum for {6G} wireless,'' in \emph{6G Mobile Wireless Networks}.\hskip
  1em plus 0.5em minus 0.4em\relax Springer International Publishing, 2021, pp.
  83--121.

\bibitem{thz2}
C.-X. Wang, J.~Wang, S.~Hu, Z.~H. Jiang, J.~Tao, and F.~Yan, ``Key technologies
  in {6G} terahertz wireless communication systems: A survey,'' \emph{IEEE
  Vehicular Technology Magazine}, vol.~16, no.~4, pp. 27--37, 2021.

\bibitem{thz3}
A.~Fayad and T.~Cinkler, ``Energy-efficient joint user and power allocation in
  {5G} millimeter wave networks: A genetic algorithm-based approach,''
  \emph{IEEE Access}, vol.~12, pp. 20\,019--20\,030, 2024.

\bibitem{thz4}
A.~Fayad, T.~Cinkler, and J.~Rak, ``{5G} millimeter wave network optimization:
  Dual connectivity and power allocation strategy,'' \emph{IEEE Access},
  vol.~11, pp. 82\,079--82\,094, 2023.

\bibitem{capacity}
F.~Tang, X.~Chen, M.~Zhao, and N.~Kato, ``The roadmap of communication and
  networking in {6G} for the metaverse,'' \emph{IEEE Wireless Communications},
  pp. 1--15, 2022.

\bibitem{latency}
M.~Adhikari and A.~Hazra, ``{6G}-enabled ultra-reliable low-latency
  communication in edge networks,'' \emph{IEEE Communications Standards
  Magazine}, vol.~6, no.~1, pp. 67--74, 2022.

\bibitem{Integrated}
T.~Sizer, D.~Samardzija, H.~Viswanathan, S.~T. Le, S.~Bidkar, P.~Dom,
  E.~Harstead, and T.~Pfeiffer, ``Integrated solutions for deployment of {6G}
  mobile networks,'' \emph{Journal of Lightwave Technology}, vol.~40, no.~2,
  pp. 346--357, 2022.

\bibitem{ozger20236g}
M.~Ozger, I.~Godor, A.~Nordlow, T.~Heyn, S.~Pandi, I.~Peterson, A.~Viseras,
  J.~Holis, C.~Raffelsberger, A.~Kercek \emph{et~al.}, ``{6G for Connected Sky:
  A Vision for Integrating Terrestrial and Non-Terrestrial Networks},'' in
  \emph{2023 Joint European Conference on Networks and Communications \& 6G
  Summit (EuCNC/6G Summit)}.\hskip 1em plus 0.5em minus 0.4em\relax IEEE, 2023,
  pp. 711--716.

\bibitem{sky}
M.~Mozaffari, X.~Lin, and S.~Hayes, ``Toward {6G} with connected sky: {UAV}s
  and beyond,'' \emph{IEEE Communications Magazine}, vol.~59, no.~12, pp.
  74--80, 2021.

\bibitem{AI}
K.~B. Letaief, W.~Chen, Y.~Shi, J.~Zhang, and Y.-J.~A. Zhang, ``The roadmap to
  {6G}: {AI} empowered wireless networks,'' \emph{IEEE Communications
  Magazine}, vol.~57, no.~8, pp. 84--90, 2019.

\bibitem{ran1}
M.~A. Habibi, M.~Nasimi, B.~Han, and H.~D. Schotten, ``A comprehensive survey
  of {RAN} architectures toward {5G} mobile communication system,'' \emph{IEEE
  Access}, vol.~7, pp. 70\,371--70\,421, 2019.

\bibitem{ran2}
Z.~Guizani and N.~Hamdi, ``{CRAN, H-CRAN,} and {F-RAN} for {5G} systems: Key
  capabilities and recent advances,'' \emph{International Journal of Network
  Management}, vol.~27, no.~5, pp. 1--22, 2017.

\bibitem{ran3}
S.~Singh, R.~Singh, and B.~Kumbhani, ``The evolution of radio access network
  towards open-{RAN}: Challenges and opportunities,'' in \emph{Proceedings of
  the 2020 IEEE Wireless Communications and Networking Conference Workshops
  (WCNCW)}, 2020, pp. 1--6.

\bibitem{ran4}
V.~S. Pana, O.~P. Babalola, and V.~Balyan, ``{5G} radio access networks: A
  survey,'' \emph{Array}, vol.~14, pp. 1--10, 2022.

\bibitem{cran}
\BIBentryALTinterwordspacing
K.~Chen and R.~Duan, ``{C-RAN the road towards green RAN. White Paper},''
  \emph{China Mobile Research Institute, Tech. Rep}, 2011. [Online]. Available:
  \url{http://labs.chinamobile.com/cran/}
\BIBentrySTDinterwordspacing

\bibitem{ran5}
M.~Peng, Y.~Sun, X.~Li, Z.~Mao, and C.~Wang, ``Recent advances in cloud radio
  access networks: system architectures, key techniques, and open issues,''
  \emph{IEEE Communications Surveys \& Tutorials}, vol.~18, no.~3, pp.
  2282--2308, 2016.

\bibitem{ran6}
M.~F. Hossain, A.~U. Mahin, T.~Debnath, F.~B. Mosharrof, and K.~Z. Islam,
  ``Recent research in cloud radio access network ({C-RAN}) for {5G} cellular
  systems - a survey,'' \emph{Journal of Network and Computer Applications},
  vol. 139, pp. 31--48, 2019.

\bibitem{hran}
M.~Peng, Y.~Li, J.~Jiang, J.~Li, and C.~Wang, ``Heterogeneous cloud radio
  access networks: a new perspective for enhancing spectral and energy
  efficiencies,'' \emph{IEEE Wireless Communications}, vol.~21, no.~6, pp.
  126--135, 2014.

\bibitem{hcran6g}
I.~Al-Samman, R.~Almesaeed, A.~Doufexi, and M.~Beach, ``Heterogeneous cloud
  radio access networks: enhanced time allocation for interference
  mitigation,'' \emph{Wireless Communications and Mobile Computing}, vol. 2018,
  pp. 1--17, 2018.

\bibitem{hcran6g1}
M.~Peng, Y.~Li, J.~Jiang, J.~Li, and C.~Wang, ``Heterogeneous cloud radio
  access networks: a new perspective for enhancing spectral and energy
  efficiencies,'' \emph{IEEE Wireless Communications}, vol.~21, no.~6, pp.
  126--135, 2014.

\bibitem{fran}
M.~Peng, S.~Yan, K.~Zhang, and C.~Wang, ``Fog-computing-based radio access
  networks: issues and challenges,'' \emph{IEEE Network}, vol.~30, no.~4, pp.
  46--53, 2016.

\bibitem{fogcompution}
C.~Mouradian, D.~Naboulsi, S.~Yangui, R.~H. Glitho, M.~J. Morrow, and P.~A.
  Polakos, ``A comprehensive survey on {Fog} computing: State-of-the-art and
  research challenges,'' \emph{IEEE communications surveys \& tutorials},
  vol.~20, no.~1, pp. 416--464, 2017.

\bibitem{frantrail}
X.~Zhang and M.~Peng, ``Testbed design and performance emulation in {F}og radio
  access networks,'' \emph{IEEE Network}, vol.~33, no.~3, pp. 49--57, 2019.

\bibitem{vran}
M.~A. Habibi, B.~Han, M.~Nasimi, N.~P. Kuruvatti, A.~Fellan, and H.~D.
  Schotten, \emph{Towards a fully virtualized, cloudified, and slicing-aware
  {RAN} for {6G} mobile networks}.\hskip 1em plus 0.5em minus 0.4em\relax
  Springer International Publishing, 2021, pp. 327--358.

\bibitem{vRANValue}
\BIBentryALTinterwordspacing
J.-J.~L. Young~Lee, Hyunjeong~Lee, ``{vRAN} value proposition and cost
  modeling,'' 2021. [Online]. Available:
  \url{https://www.samsung.com/global/business/networks/insights/white-papers/vran-value-proposition-and-cost-modeling/}
\BIBentrySTDinterwordspacing

\bibitem{vran1}
X.~W. et~al., ``Virtualized cloud radio access network for {5G} transport,''
  \emph{IEEE Communications Magazine}, vol.~55, no.~9, pp. 202--209, 2017.

\bibitem{oran3}
D.~Wypiór, M.~Klinkowski, and I.~Michalski, ``Open {RAN}—radio access
  network evolution, benefits and market trends,'' \emph{Applied Sciences},
  vol.~12, no.~1, pp. 1--18, 2022.

\bibitem{fronthaul1}
{ITU-T}, ``Transport network support of {IMT-2020/5G},'' ITU-T, Technical
  Report, 2018.

\bibitem{fronthaul2}
\BIBentryALTinterwordspacing
{Ericsson AB and Huawei Technologies Co. Ltd and NEC Corporation and Nokia},
  ``{eCPRI Interface Specification V2.0},'' {Common Public Radio Interface
  (CPRI)}, {eCPRI Specification}, May 2019. [Online]. Available:
  \url{http://www.cpri.info/downloads/eCPRI_v_2.0_2019_05_10c.pdf}
\BIBentrySTDinterwordspacing

\bibitem{fronthaul3}
\BIBentryALTinterwordspacing
{5G PPP Architecture Working Group}, ``View on {5G} architecture,'' {5G Public
  Private Partnership (5G PPP)}, White Paper, 2018. [Online]. Available:
  \url{https://www.5g-ppp.eu/wp-content/uploads/2018/01/5G-PPP-5G-Architecture-White-Paper-Jan-2018-v2.0.pdf}
\BIBentrySTDinterwordspacing

\bibitem{split}
\BIBentryALTinterwordspacing
{3rd Generation Partnership Project (3GPP)}, ``Study on new radio access
  technology radio access architecture and interfaces,'' 3GPP, Technical Report
  TR 38.801, 2017. [Online]. Available:
  \url{https://www.3gpp.org/ftp/Specs/archive/38_series/38.801/}
\BIBentrySTDinterwordspacing

\bibitem{sp1}
L.~M.~P. Larsen, A.~Checko, and H.~L. Christiansen, ``A survey of the
  functional splits proposed for {5G} mobile crosshaul networks,'' \emph{IEEE
  Communications Surveys \& Tutorials}, vol.~21, no.~1, pp. 146--172, 2019.

\bibitem{7x}
V.~Q. Rodriguez, F.~Guillemin, A.~Ferrieux, and L.~Thomas, ``Cloud-{RAN}
  functional split for an efficient fronthaul network,'' in \emph{2020
  International Wireless Communications and Mobile Computing (IWCMC)}, 2020,
  pp. 245--250.

\bibitem{bitrate}
C.~Ranaweera, P.~Monti, B.~Skubic, E.~Wong, M.~Furdek, L.~Wosinska, C.~M.
  Machuca, A.~Nirmalathas, and C.~Lim, ``Optical transport network design for
  {5G} fixed wireless access,'' \emph{Journal of Lightwave Technology},
  vol.~37, no.~16, pp. 3893--3901, 2019.

\bibitem{split1}
F.~Ponzini, K.~Kondepu, F.~Giannone, P.~Castoldi, and L.~Valcarenghi, ``Optical
  access network solutions for {5G} fronthaul,'' in \emph{2018 20th
  International Conference on Transparent Optical Networks (ICTON)}.\hskip 1em
  plus 0.5em minus 0.4em\relax IEEE, 2018, pp. 1--5.

\bibitem{skubic2017optical}
B.~Skubic, M.~Fiorani, S.~Tombaz, A.~Furuskär, J.~Mårtensson, and P.~Monti,
  ``Optical transport solutions for {5G} fixed wireless access [invited],''
  \emph{Journal of Optical Communications and Networking}, vol.~9, no.~9, pp.
  D10--D18, 2017.

\bibitem{forcast2030}
M.~El-Moghazi and J.~Whalley, ``The {ITU IMT-2020} standardization: Lessons
  from {5G} and future perspectives for {6G},'' \emph{Journal of Information
  Policy}, 2022.

\bibitem{opticalforcast}
{Benzinga}, ``Mobile backhaul \& fronthaul market business strategies
  2023-2030: Latest trends, future demand with predictions,'' 2023, [Online].
  Available:
  \url{https://www.benzinga.com/pressreleases/23/08/34148634/mobile-backhaul-fronthaul-market-business-strategies-2023-2030}.

\bibitem{p2p2}
{ITU Telecommunication Standardization Sector}, ``{G.9806 : Higher-speed
  bidirectional, single fibre, point-to-point optical access system
  (HS-PtP)},'' International Telecommunication Union, Geneva, Switzerland,
  {ITU-T Recommendation} G.9806, 2020.

\bibitem{p2p3}
F.~Effenberger, ``Call for interest: bidirectional 10 {G}b/s and 25 {G}b/s
  optical access {PHY}s,'' CFI\_01\_0318, IEEE 802.3 Meeting, 2018.

\bibitem{p2p}
P.~Chanclou, H.~Suzuki, J.~Wang, Y.~Ma, M.~R. Boldi, K.~Tanaka, S.~Hong,
  C.~Rodrigues, L.~A. Neto, and J.~Ming, ``How does passive optical network
  tackle radio access network evolution?'' \emph{Journal of Optical
  Communications and Networking}, vol.~9, no.~11, pp. 1030--1040, 2017.

\bibitem{Infinera2022XROptics}
Infinera, ``{XR Optics: Innovation for Next-generation Networks},''
  \url{https://www.infinera.com/innovation/xr-optics/}, 2022, accessed:
  2024-04-24.

\bibitem{p2p1}
P.~J. Winzer and D.~T. Neilson, ``From scaling disparities to integrated
  parallelism: A decathlon for a decade,'' \emph{Journal of Lightwave
  Technology}, vol.~35, no.~5, pp. 1099--1115, 2017.

\bibitem{fiberchannel}
Broadcom, ``Fibre channel standards,''
  \url{https://www.broadcom.com/support/fibre-channel-networking/san-standards/fc-standards},
  accessed: 2024-04-20.

\bibitem{ethernet}
{IEEE Standards Association}, ``{IEEE Standard for Ethernet},'' IEEE, Tech.
  Rep. 802.3-2018, 2018, accessed: 2024-04-20.

\bibitem{edr}
\BIBentryALTinterwordspacing
{InfiniBand Trade Association}, ``Introduction to infiniband,'' White Paper,
  2003, accessed: accessed Jan. 03, 2024. [Online]. Available:
  \url{https://network.nvidia.com/pdf/whitepapers/IB_Intro_WP_190.pdf}
\BIBentrySTDinterwordspacing

\bibitem{p2p4}
\BIBentryALTinterwordspacing
Optcore, ``{The Big Differences Between SFP, SFP+, SFP28, QSFP+, QSFP28,
  QSFP-DD, and OSFP},'' 2022, [Online; accessed 24-Jul-2023]. [Online].
  Available: \url{https://www.optcore.net/article018/}
\BIBentrySTDinterwordspacing

\bibitem{fronthaul4}
F.~Saliou, P.~Chanclou, L.~A. Neto, G.~Simon, J.~Potet, M.~Gay, L.~Bramerie,
  and H.~Debregeas, ``Optical access network interfaces for {5G} and beyond
  [invited],'' \emph{Journal of Optical Communications and Networking},
  vol.~13, no.~8, pp. D32--D42, 2021.

\bibitem{p2p5}
Cisco, ``Cisco 400g qsfp-dd cable and transceiver modules data sheet,''
  \url{https://www.cisco.com/c/en/us/products/collateral/interfaces-modules/transceiver-modules/datasheet-c78-743172.html},
  2022, [Online; accessed 24-Jul-2023].

\bibitem{p2p6}
{Cisco Blogs}, ``{QSFP-DD800 Archives},''
  \url{https://blogs.cisco.com/tag/qsfp-dd800}, 2021, accessed: 2024-04-24.

\bibitem{pon1}
{ITU-T}, ``{5G wireless fronthaul requirements in a PON context},'' ITU
  Telecommunication Standardization Sector, Geneva, Switzerland, G.sup.5GP,
  2018.

\bibitem{pon3}
E.~Wong, ``Towards {6G}: The evolution of passive optical networks,'' in
  \emph{IEEE Photonics Conference (IPC)}.\hskip 1em plus 0.5em minus
  0.4em\relax IEEE, 2022, pp. 1--2.

\bibitem{ocdm2}
T.~Muciaccia, F.~Gargano, and V.~Passaro, ``Passive optical access networks:
  State of the art and future evolution,'' \emph{Photonics}, vol.~1, no.~4, pp.
  323--346, 2014.

\bibitem{pon}
D.~Zhang, D.~Liu, X.~Wu, and D.~Nesset, ``Progress of {ITU-T} higher speed
  passive optical network ({50G-PON}) standardization,'' \emph{Journal of
  Optical Communications and Networking}, vol.~12, no.~10, pp. D99--D108, 2020.

\bibitem{pon0}
H.~S. Abbas and M.~A. Gregory, ``The next generation of passive optical
  networks: A review,'' \emph{Journal of Network and Computer Applications},
  vol.~67, pp. 53--74, 2016.

\bibitem{ponsplit}
S.~Das, F.~Slyne, and M.~Ruffini, ``Optimal slicing of virtualized passive
  optical networks to support dense deployment of cloud-{RAN} and multi-access
  edge computing,'' \emph{IEEE Network}, vol.~36, no.~2, pp. 131--138, 2022.

\bibitem{xgpon}
A.~E. Ankouri, L.~A. Neto, G.~Simon, H.~L. Bras, A.~Sanhaji, and P.~Chanclou,
  ``High-speed train cell-less network enabled by {XGS-PON} and impacts on
  {vRAN} split interface transmission,'' in \emph{Optical Fiber Communication
  Conference (OFC)}, 2019, paper W4J.5.

\bibitem{xgpon1}
S.~Bidkar, J.~Galaro, and T.~Pfeiffer, ``First demonstration of an
  ultra-low-latency fronthaul transport over a commercial {TDM-PON} platform,''
  in \emph{Optical Fiber Communication Conference (OFC)}, 2018, paper Tu2K.3.

\bibitem{25pon}
``{IEEE Standard for Ethernet Amendment 9: Physical Layer Specifications and
  Management Parameters for 25 Gb/s and 50 Gb/s Passive Optical Networks},''
  IEEE Std. 802.3ca-2020, 2020, available: IEEE Xplore.

\bibitem{itu25pon}
\emph{{ITU-T Recommendation G.9804.3, "50-Gigabit-capable Passive Optical
  Networks ({50G-PON}): Physical Media Dependent ({PMD}) Layer
  specification."}}, ITU-T Std., 2021.

\bibitem{transport}
{IHS Markit}, ``{5G transport strategies \& vendor leadership, service provider
  survey},'' IHS Markit, Tech. Rep., 2020, available: IHS Markit.

\bibitem{InfineraPON2024}
Infinera, ``{PON Overlay Networks: Unlocking Revenue Potential},''
  \url{https://www.infinera.com/innovation/pon/}, 2024, accessed: 2024-04-24.

\bibitem{InfineraBlog2024}
J.~Baldry, ``{Turbocharge Your Single Fiber Access PON Network},''
  \url{https://www.infinera.com/blog/turbocharge-your-single-fiber-access-pon-network/tag/access-and-aggregation/},
  2024, accessed: 2024-04-24.

\bibitem{coherentpon}
J.~Zhang and Z.~Jia, ``{Coherent Passive Optical Networks for
  100G/$\lambda$-and-Beyond Fiber Access: Recent Progress and Outlook},''
  \emph{IEEE Network}, vol.~36, no.~2, pp. 116--123, 2022.

\bibitem{coherentpon1}
N.~Suzuki, H.~Miura, K.~Mochizuki, and K.~Matsuda, ``{Beyond-100G {PON} Systems
  for Integrated Access and Metro Networks in the {B5G}/{6G} Era},'' in
  \emph{Advanced Photonics Congress 2023, Technical Digest Series (Optica
  Publishing Group)}, 2023, p. SpM3D.4.

\bibitem{coherent2}
N.~Suzuki, H.~Miura, K.~Matsuda, R.~Matsumoto, and K.~Motoshima, ``{100 Gb/s to
  1 Tb/s Based Coherent Passive Optical Network Technology},'' \emph{Journal of
  Lightwave Technology}, vol.~36, no.~8, pp. 1485--1491, 2018.

\bibitem{coherent3}
G.~Simon, F.~Saliou, A.~Afonso, C.~Castro, A.~Napoli, M.~Bouzayani,
  F.~Bonnafous, J.~Potet, and P.~Chanclou, ``{200 Gb/s Coherent
  Point-to-Multipoint Coexistence with 50G-PON for Next-Generation Optical
  Access},'' \emph{IEEE Photonics Technology Letters}, 2024.

\bibitem{coherent4}
T.~Duthel, C.~R. Fludger, B.~Liu, S.~Ranzini, A.~Napoli, N.~S{\"o}lch, S.~M.
  Bilal, S.~B. Amado, S.~Alreesh, J.~Sime \emph{et~al.}, ``{DSP Design for
  Coherent Optical Point-to-Multipoint Transmission},'' \emph{Journal of
  Lightwave Technology}, 2023.

\bibitem{coherent5}
C.~Castro, A.~Tartaglia, R.~Magri, B.~Spinnler, J.~Pedro, and A.~Napoli,
  ``Point-to-multipoint coherent transceivers for next-generation mobile
  transport,'' in \emph{2023 23rd International Conference on Transparent
  Optical Networks (ICTON)}.\hskip 1em plus 0.5em minus 0.4em\relax IEEE, 2023,
  pp. 1--6.

\bibitem{B-PON}
\BIBentryALTinterwordspacing
``{ITU-T "G.983.1: Broadband optical access systems based on Passive Optical
  Networks ({PON})"},'' 2020. [Online]. Available:
  \url{https://www.itu.int/rec/T-REC-G.983.1/en}
\BIBentrySTDinterwordspacing

\bibitem{E-PON}
{IEEE Standards Association}, ``{CSMA/CD access method and physical layer
  specifications amendment: Media access control parameters, physical layers,
  and management parameters for subscriber access networks},'' 2020, [Online].
  Available: \url{https://standards.ieee.org/ieee/802.3ah/3179/}.

\bibitem{G-PON}
{ITU-T}, ``{G.984.1: Gigabit-capable passive optical networks (GPON): General
  characteristics},'' International Telecommunication Union, Tech. Rep., 2008,
  [Online]. Available: \url{https://www.itu.int/rec/T-REC-G.984.1/en}.

\bibitem{10G-EPON}
``{IEEE Standard for Information technology--Telecommunications and information
  exchange between systems--Local and metropolitan area networks--Specific
  requirements Part 3: Carrier Sense Multiple Access with Collision Detection
  (CSMA/CD) Access Method and Physical Layer Specifications Amendment: Physical
  Layer Specifications and Management Parameters for 10 Gb/s Passive Optical
  Networks},'' IEEE Standards Association, 2008, [Online]. Available:
  https://standards.ieee.org/ieee/802.3av/4060/.

\bibitem{XG-PON}
{ITU-T}, ``{G.987: 10-Gigabit-capable passive optical network (XG-PON) systems:
  Definitions, abbreviations and acronyms},'' International Telecommunication
  Union, Tech. Rep., 2012, [Online]. Available:
  \url{https://www.itu.int/rec/T-REC-G.987/en}.

\bibitem{NG-PON2}
ITU-T, ``{G.989: 40-Gigabit-capable passive optical networks (NG-PON2):
  Definitions, abbreviations and acronyms},'' International Telecommunication
  Union, Tech. Rep., 2015, [Online]. Available:
  \url{https://www.itu.int/rec/T-REC-G.989/en}.

\bibitem{XGS-PON}
{ITU-T}, ``{G.9807.1: 10-Gigabit-capable symmetric passive optical network
  (XGS-PON)},'' International Telecommunication Union, Tech. Rep., 2023,
  [Online]. Available: \url{https://www.itu.int/rec/T-REC-G.9807.1/en}.

\bibitem{NG-EPON}
{IEEE Standards Association}, ``{IEEE 802.3ca-2020: IEEE Standard for Ethernet
  Amendment 9: Physical Layer Specifications and Management Parameters for 25
  Gb/s and 50 Gb/s Passive Optical Networks},'' IEEE, Tech. Rep., 2020,
  [Online]. Available:
  \url{https://standards.ieee.org/standard/802_3ca-2020.html}.

\bibitem{hpon1}
{ITU-T}, ``{G.9804.2: Higher speed passive optical networks - Common
  transmission convergence layer specification},'' International
  Telecommunication Union, Tech. Rep., 2021, [Online]. Available:
  \url{https://www.itu.int/rec/T-REC-G.9804.2/en}.

\bibitem{hpon2}
ITU-T, ``{G.9804.3: 50-Gigabit-capable passive optical networks (50G-PON) -
  Physical media dependent (PMD) layer specification},'' International
  Telecommunication Union, Tech. Rep., 2021, [Online]. Available:
  \url{https://www.itu.int/rec/T-REC-G.9804.3/en}.

\bibitem{tdm}
E.~Harstead, D.~van Veen, V.~Houtsma, and P.~Dom, ``{Technology roadmap for
  time-division multiplexed passive optical networks ({TDM PONs})},''
  \emph{Journal of Lightwave Technology}, vol.~37, no.~2, pp. 657--664, 2019.

\bibitem{tdm1}
H.~Chung, H.~H. Lee, K.~O. Kim, K.-H. Doo, Y.~Ra, and C.~Park,
  ``{TDM-PON}-based optical access network for tactile internet, {5G}, and
  beyond,'' \emph{IEEE Network}, vol.~36, no.~2, pp. 76--–81, 2022.

\bibitem{tdm3}
H.~Uzawa, K.~Honda, H.~Nakamura, Y.~Hirano, K.-I. Nakura, S.~Kozaki, and
  J.~Terada, ``{Dynamic bandwidth allocation scheme for network-slicing-based
  {TDM-PON} toward the beyond-{5G} era},'' \emph{Journal of Optical
  Communications and Networking}, vol.~12, pp. A135--A143, 2020.

\bibitem{tdm2}
S.~Hatta, N.~Tanaka, and T.~Sakamoto, ``{Low latency dynamic bandwidth
  allocation method with high bandwidth efficiency for {TDM-PON}},'' \emph{NTT
  Technical Review}, vol.~15, no.~4, pp. 50--56, 2017.

\bibitem{wdm}
J.~S. Wey and J.~Zhang, ``Passive optical networks for {5G} transport:
  technology and standards,'' \emph{Journal of Lightwave Technology}, vol.~37,
  no.~12, pp. 2830--2837, 2019.

\bibitem{wdmfh}
{ITU-T}, ``G.9802.1: Wavelength division multiplexed passive optical networks
  (wdm pon): General requirements,'' International Telecommunication Union,
  Tech. Rep., 2021, [Online]. Available:
  \url{https://www.itu.int/rec/T-REC-G.9802.1/en}.

\bibitem{wdmfh1}
B.~Yang, ``Key {WDM-PON} technologies for {5G} fronthaul,'' \emph{ZTE
  Technol.}, [Accessed: 13-Feb-2023].

\bibitem{capacity6}
F.~J. Effenberger and D.~Zhang, ``{WDM-PON} for{5G} wireless fronthaul,''
  \emph{IEEE Wireless Communications}, vol.~29, no.~2, pp. 94--99, 2022.

\bibitem{twdm}
Y.~Luo, X.~Zhou, F.~Effenberger, X.~Yan, G.~Peng, Y.~Qian, and Y.~Ma, ``Time-
  and wavelength-division multiplexed passive optical network ({TWDM-PON}) for
  next-generation {PON} stage 2 ({NG-PON2}),'' \emph{Journal of Lightwave
  Technology}, vol.~31, no.~4, pp. 587--593, 2013.

\bibitem{twdm1}
D.~Nesset, ``{NG-PON2} technology and standards,'' \emph{Journal of Lightwave
  Technology}, vol.~33, no.~5, pp. 1136--1143, 2015.

\bibitem{twdmpon}
P.~Torres-Ferrera, G.~Rizzelli, H.~Wang, V.~Ferrero, and R.~Gaudino,
  ``{Experimental Demonstration of 100 Gbps/$\lambda$ C-Band Direct-Detection
  Downstream {PON} Using Non-Linear and CD Compensation with 29 dB+ OPL Over 0
  Km–100 Km},'' \emph{Journal of Lightwave Technology}, vol.~40, no.~2, pp.
  547--556, 2022.

\bibitem{ofdm0}
K.~Qiu, X.~Yi, J.~Zhang, H.~Zhang, M.~Deng, and C.~Zhang, ``{OFDM-PON} optical
  fiber access technologies,'' in \emph{Asia Communications and Photonics
  Conference and Exhibition}, 2011, pp. 1--6.

\bibitem{ofdm1}
R.~Q. Shaddad, A.~B. Mohammad, S.~A. Al-Gailani, A.~M. Al-hetar, and M.~A.
  Elmagzoub, ``A survey on access technologies for broadband optical and
  wireless networks,'' \emph{Journal of Network and Computer Applications},
  vol.~41, pp. 459--472, 2014.

\bibitem{ofdm2}
M.~Riva, H.~Donâncio, F.~Almeida, G.~Figueiredo, R.~Tinini, R.~C. Jr, and
  D.~Batista, ``An elastic optical network-based architecture for the {5G}
  fronthaul,'' in \emph{Anais do XXXVI Simpósio Brasileiro de Redes de
  Computadores e Sistemas Distribuídos}.\hskip 1em plus 0.5em minus
  0.4em\relax SBC, 2018, pp. 631--641.

\bibitem{ocdm1}
S.~Yoshima, Y.~Tanaka, N.~Kataoka, N.~Wada, J.~Nakagawa, and K.-I. Kitayama,
  ``{Full-duplex, extended-reach {10G-TDM-OCDM-PON} system without En/Decoder
  at ONU},'' \emph{Journal of Lightwave Technology}, vol.~31, no.~1, pp.
  43--49, 2013.

\bibitem{ocdm0}
H.~S. Abbas and M.~A. Gregory, ``The next generation of passive optical
  networks: A review,'' \emph{Journal of Network and Computer Applications},
  vol.~67, p. 53––74, 2016.

\bibitem{noma-pon1}
G.~Wang, Z.~Fan, and J.~Zhao, ``{Experimental Demonstration of Adaptive Bit and
  Power Loading Algorithm for OFDM-NOMA PON},'' in \emph{2021 Optical Fiber
  Communications Conference and Exhibition (OFC)}, 2021, pp. 1--3.

\bibitem{noma-pon2}
\BIBentryALTinterwordspacing
B.~Lin, H.~Yang, R.~Wang, Z.~Ghassemlooy, and X.~Tang, ``{Convolutional neural
  network-based signal demodulation method for NOMA-PON},'' \emph{Opt.
  Express}, vol.~28, no.~10, pp. 14\,357--14\,365, 2020. [Online]. Available:
  \url{https://opg.optica.org/oe/abstract.cfm?URI=oe-28-10-14357}
\BIBentrySTDinterwordspacing

\bibitem{noma-pon3}
Y.~Wan, B.~Liu, J.~Ren, R.~Ullah, Y.~Mao, S.~Zhu, S.~Chen, X.~Wu, F.~Wang,
  T.~Sun, Y.~Wu, and L.~Zhao, ``Performance-enhanced optical non-orthogonal
  multiple access enabled by orthogonal chirp division multiplexing,''
  \emph{Journal of Lightwave Technology}, vol.~40, no.~16, pp. 5440--5449,
  2022.

\bibitem{pdm-pon1}
J.~Zhang, J.~Yu, X.~Li, K.~Wang, W.~Zhou, J.~Xiao, L.~Zhao, X.~Pan, B.~Liu, and
  X.~Xin, ``200 {G}bit/s/$\lambda$ {PDM-PAM-4 PON} system based on intensity
  modulation and coherent detection,'' \emph{Journal of Optical Communications
  and Networking}, vol.~12, no.~1, pp. A1--A8, 2019.

\bibitem{pdm-pon2}
H.~Xin, D.~Kong, K.~Zhang, S.~Jia, Y.~Fu, W.~Hu, and H.~Hu, ``100 {G}bps
  simplified coherent {PON} using carrier-suppressed {PDM-PAM-4} and
  phsae-recovery-free {KK} detection,'' in \emph{45th European Conference on
  Optical Communication (ECOC 2019)}, 2019, pp. 1--4.

\bibitem{pdm-pon4}
J.~Zhang, J.~Yu, K.~Wang, W.~Zhou, X.~Xiao, J.~Xiao, L.~Zhao, X.~Pan, B.~Liu,
  and X.~Xin, ``{200-Gb/s/$\lambda$ PDM-PAM-4 PON with 29-dB power budget based
  on heterodyne coherent detection},'' in \emph{2019 Optical Fiber
  Communications Conference and Exhibition (OFC)}.\hskip 1em plus 0.5em minus
  0.4em\relax IEEE, 2019, pp. 1--3.

\bibitem{pdm-pon3}
Y.~Fan, M.~Fu, H.~Jiang, X.~Liu, Q.~Liu, Y.~Xu, L.~Yi, W.~Hu, and Q.~Zhuge,
  ``{Point-to-Multipoint Coherent Architecture with Joint Resource Allocation
  for B5G/6G Fronthaul},'' \emph{IEEE Wireless Communications}, vol.~29, no.~2,
  pp. 100--106, 2022.

\bibitem{fso0}
H.~Henniger and O.~Wilfert, ``An introduction to free-space optical
  communications,'' in \emph{Radio engineering}, vol.~19, no.~2, 2010, p.
  1––10.

\bibitem{fso}
A.~Bekkali, H.~Fujita, and M.~Hattori, ``Free-space optical communication
  systems for {B5G}/{6G} networks,'' in \emph{Optoelectronics and
  Communications Conference}.\hskip 1em plus 0.5em minus 0.4em\relax Optical
  Society of America, 2021, pp. W1A--1.

\bibitem{fso1}
F.~Corral, C.~Cuenca, and I.~Soto, ``{Design of an optical wireless network
  using Free Space Optics technology (FSO), in {5G}/{6G} networks
  environment},'' in \emph{2021 IEEE International Conference on
  Automation/XXIV Congress of the Chilean Association of Automatic Control
  (ICA-ACCA)}.\hskip 1em plus 0.5em minus 0.4em\relax IEEE, 2021, pp. 1--5.

\bibitem{fso4}
T.~Brasini, M.~Domecq, T.~Iliev, I.~Stoyanov, G.~Mihaylov, and I.~Beloev, ``On
  free-space optical communication as backhauls applications for {5G},'' in
  \emph{2022 International Conference on Communications, Information,
  Electronic and Energy Systems (CIEES)}.\hskip 1em plus 0.5em minus
  0.4em\relax IEEE, 2022, pp. 1--4.

\bibitem{fso3}
F.~Guiomar, M.~Fernandes, J.~Nascimento, and P.~Monteiro, ``{400G+ wireless
  transmission via free-space optics},'' in \emph{2021 European Conference on
  Optical Communication (ECOC)}.\hskip 1em plus 0.5em minus 0.4em\relax IEEE,
  2021, pp. 1--4.

\bibitem{fso2}
V.~W.~S. Chan, ``Free-space optical communications,'' \emph{Journal of
  Lightwave Technology}, vol.~24, no.~12, pp. 4750--4762, 2006.

\bibitem{fso5}
A.~Trichili, M.~A. Cox, B.~S. Ooi, and M.-S. Alouini, ``Roadmap to free space
  optics,'' \emph{Journal of the Optical Society of America B}, vol.~37,
  no.~11, pp. A184--A201, 2020.

\bibitem{fso6}
A.~Jahid, M.~H. Alsharif, and T.~J. Hall, ``A contemporary survey on free space
  optical communication: Potentials, technical challenges, recent advances and
  research direction,'' \emph{Journal of Network and Computer Applications},
  vol. 200, pp. 1--35, 2022.

\bibitem{fso55}
T.~McKenna, J.~Juarez, J.~Nanzer, and T.~Clark, ``Hybrid
  millimeter-wave/free-space optical system for high data rate
  communications,'' in \emph{2013 IEEE Photonics Conference}.\hskip 1em plus
  0.5em minus 0.4em\relax IEEE, 2013, pp. 203--204.

\bibitem{fiber-fso1}
A.~Fayad and T.~Cinkler, ``Cost-effective delay-constrained optical fronthaul
  design for {5G} and beyond,'' \emph{Infocommunications journal}, vol.~14,
  no.~2, pp. 19--27, 2022.

\bibitem{fiber-fso}
F.~Tonini, C.~Raffaelli, L.~Wosinska, and P.~Monti, ``{Cost-Optimal Deployment
  of a {C-RAN} With Hybrid Fiber/FSO Fronthaul},'' \emph{Journal of Optical
  Communications and Networking}, vol.~11, no.~7, pp. 397--408, 2019.

\bibitem{pon-fso}
S.~S. Jaffer, A.~Hussain, M.~A. Qureshi, J.~Mirza, and K.~K. Qureshi, ``A low
  cost {PON-FSO} based fronthaul solution for {5G} cran architecture,''
  \emph{Optical Fiber Technology}, vol.~63, pp. 1--15, 2021.

\bibitem{pon-fso1}
T.~Nguyen, H.~Nguyen, H.~Le, and N.~Nguyen, ``Performance analysis of
  gigabit-capable mobile backhaul networks exploiting {TWDM-PON} and {FSO}
  technologies,'' in \emph{2016 International Conference on Advanced
  Technologies for Communications (ATC)}.\hskip 1em plus 0.5em minus
  0.4em\relax IEEE, 2016, pp. 180--185.

\bibitem{fso7}
E.~Yaacoub and M.-S. Alouini, ``{Efficient fronthaul and backhaul connectivity
  for IoT traffic in rural areas},'' \emph{IEEE Internet of Things Magazine},
  vol.~4, no.~1, pp. 60–--66, 2021.

\bibitem{fso9}
M.~Alzenad, M.~Z. Shakir, H.~Yanikomeroglu, and M.-S. Alouini, ``{FSO-based
  vertical backhaul/fronthaul framework for {5G+} wireless networks},''
  \emph{IEEE Communications Magazine}, vol.~56, no.~1, pp. 218–--224, 2018.

\bibitem{fso10}
Y.~Zhou, Z.~Gu, J.~Zhang, and Y.~Ji, ``{Efficient deployment of aerial relays
  in FSO-based backhaul networks},'' \emph{Journal of Optical Communications
  and Networking}, vol.~15, no.~1, pp. 29--42, 2022.

\bibitem{fsowater}
S.~Zafar and H.~Khalid, ``Free space optical networks: Applications, challenges
  and research directions,'' \emph{Wireless Personal Communications}, vol. 121,
  no.~1, pp. 429--457, 2021.

\bibitem{fsowater1}
J.~A. Simpson, B.~L. Hughes, and J.~F. Muth, ``Smart transmitters and receivers
  for underwater free-space optical communication,'' \emph{IEEE Journal on
  Selected Areas in Communications}, vol.~30, no.~5, pp. 964–--974, 2012.

\bibitem{sharing1}
S.~K.~A. Kumar and E.~J. Oughton, ``Infrastructure sharing strategies for
  wireless broadband,'' \emph{IEEE Communications Magazine}, 2023.

\bibitem{sharing2}
P.~Chanclou, L.~A. Neto, K.~Grzybowski, Z.~Tayq, F.~Saliou, and N.~Genay,
  ``{Mobile fronthaul architecture and technologies: A RAN equipment assessment
  [Invited]},'' \emph{Journal of Optical Communications and Networking},
  vol.~10, no.~1, pp. A1--A7, 2017.

\bibitem{cost8}
C.~Ranaweera, A.~Nirmalathas, E.~Wong, C.~Lim, P.~Monti, M.~Furdek,
  L.~Wosinska, B.~Skubic, and C.~Machuca, ``Rethinking of optical transport
  network design for {5G}/{6G} mobile communication,'' \emph{IEEE Future
  Networks Tech Focus}, vol.~12, 2021.

\bibitem{cost10}
A.~Fayad, T.~Cinkler, J.~Rak, and M.~Jha, ``Design of cost-efficient optical
  fronthaul for {5G}/{6G} networks: An optimization perspective,''
  \emph{Sensors}, vol.~22, no.~23, pp. 1--22, 2022.

\bibitem{brown}
C.~Ranaweera, M.~G.~C. Resende, K.~Reichmann, P.~Iannone, P.~Henry, B.-J. Kim,
  P.~Magill, K.~N. Oikonomou, R.~K. Sinha, and S.~Woodward, ``Design and
  optimization of fiber optic small-cell backhaul based on an existing
  fiber-to-the-node residential access network,'' \emph{IEEE Communications
  Magazine}, vol.~51, no.~9, pp. 62–--69, 2013.

\bibitem{fronthaulenergy1}
M.~Fiorani, S.~Tombaz, J.~Mårtensson, B.~Skubic, L.~Wosinska, and P.~Monti,
  ``Modeling energy performance of {C-RAN} with optical transport in {5G}
  network scenarios,'' \emph{Journal of Optical Communications and Networking},
  vol.~8, no.~11, pp. B21--B34, 2016.

\bibitem{cost61}
A.~Fayad, T.~Cinkler, and J.~Rak, ``{5G}/{6G} optical fronthaul modelling: Cost
  and energy consumption assessment,'' \emph{Journal of Optical Communications
  and Networking}, vol.~15, no.~9, pp. D33--D46, 2023.

\bibitem{fronthaulenergy3}
J.~Lorincz, Z.~Klarin, and D.~Begusic, ``Advances in improving energy
  efficiency of fiber–wireless access networks: A comprehensive overview,''
  \emph{Sensors}, vol.~23, no.~4, pp. 1--37, 2023.

\bibitem{fronthaulenergy2}
H.~Yu, J.~Zhang, Y.~Ji, and M.~Tornatore, ``{Energy-efficient dynamic lightpath
  adjustment in a decomposed AWGR-based passive WDM fronthaul},'' \emph{Journal
  of Optical Communications and Networking}, vol.~10, no.~9, pp. 749–--759,
  2018.

\bibitem{fronthaulenergy4}
P.~Zhu, J.~Zhang, and Y.~Ji, ``{Resource Allocation in Energy Efficient Hybrid
  FSO/mmW Fronthaul: A Differential Evolution Approach},'' in \emph{ICC 2019 -
  2019 IEEE International Conference on Communications (ICC)}, Shanghai, China,
  2019, pp. 1--6.

\bibitem{1914}
A.~Lometti and V.~Sestito, ``{Fronthaul in {5G} transport networks: IEEE1914.1
  architecture and requirements},'' in \emph{2020 22nd International Conference
  on Transparent Optical Networks (ICTON)}, Bari, Italy, 2020, pp. 1--4.

\bibitem{latency8}
D.~Chitimalla, K.~Kondepu, L.~Valcarenghi, M.~Tornatore, and B.~Mukherjee,
  ``{{5G} fronthaul--latency and jitter studies of CPRI over Ethernet},''
  \emph{Journal of Optical Communications and Networking}, vol.~9, no.~2, pp.
  172--182, 2017.

\bibitem{latency2}
X.~Wang, Y.~Ji, J.~Zhang, L.~Bai, and M.~Zhang, ``{Joint optimization of
  latency and deployment cost over {TDM-PON} based {MEC}-enabled cloud radio
  access networks},'' \emph{IEEE Access}, vol.~8, pp. 681--696, 2020.

\bibitem{latency10}
X.~Wang, L.~B. Y.~Ji, J.~Zhang, and M.~Zhang, ``Low-latency oriented network
  planning for {MEC}-enabled {WDM-PON} based fiber-wireless access networks,''
  \emph{IEEE Access}, vol.~7, pp. 183\,383--183\,395, 2019.

\bibitem{fibermmwave}
G.~Kalfas, C.~Vagionas, A.~Antonopoulos, E.~Kartsakli, A.~Mesodiakaki,
  S.~Papaioannou, P.~Maniotis, J.~S. Vardakas, C.~Verikoukis, and N.~Pleros,
  ``Next generation fiber-wireless fronthaul for {5G} mmwave networks,''
  \emph{IEEE Communications Magazine}, vol.~57, no.~3, pp. 138–--144, 2019.

\bibitem{survive1}
Y.~Yang, K.~W. Sung, L.~Wosinska, and J.~Chen, ``Hybrid fiber and microwave
  protection for mobile backhauling,'' \emph{Journal of Optical Communications
  and Networking}, vol.~6, no.~10, pp. 869--878, 2014.

\bibitem{survive2}
N.~Chouhan, U.~R. Bhatt, and R.~Upadhyay, ``{An optimization framework for FiWi
  access network: Comprehensive solution for green and survivable
  deployment},'' \emph{Optical Fiber Technology}, vol.~53, pp. 1--16, 2019.

\bibitem{fibermicrowave}
F.~Farias, M.~Fiorani, S.~Tombaz, M.~Mahloo, L.~Wosinska, J.~C. W.~A. Costa,
  and P.~Monti, ``Cost- and energy-efficient backhaul options for heterogeneous
  mobile network deployments,'' \emph{Photonic Network Communications},
  vol.~32, no.~3, pp. 422–--437, 2016.

\bibitem{fsommwave1}
H.~Zhang, Y.~Dong, J.~Cheng, M.~J. Hossain, and V.~C.~M. Leung, ``{Fronthauling
  for {5G} LTE-U ultra dense cloud small cell networks},'' \emph{IEEE Wireless
  Communications}, vol.~23, no.~6, pp. 48–--53, 2016.

\bibitem{fsommwave2}
S.~A.~H. Mohsan, M.~A. Khan, and H.~Amjad, ``{Hybrid FSO/RF networks: A review
  of practical constraints, applications and challenges},'' \emph{Optical
  Switching and Networking}, vol.~47, pp. 1--17, 2023.

\bibitem{spectral1}
Y.~Fan, M.~Fu, H.~Jiang, X.~Liu, Q.~Liu, Y.~Xu, L.~Yi, W.~Hu, and Q.~Zhuge,
  ``Point-to-multipoint coherent architecture with joint resource allocation
  for {B5G}/{6G} fronthaul,'' \emph{IEEE Wireless Communications}, vol.~29,
  no.~2, pp. 100–--106, 2022.

\bibitem{improveenergy}
P.~Georgiadis, M.~Anastasopoulos, A.~Manolopoulos, V.~Alevizaki, N.~Nikaein,
  and A.~Tzanakaki, ``Demonstration of energy efficient optimization in beyond
  {5G} systems supported by optical transport networks,'' pp. W4F--4, 2023.

\bibitem{improvenergy1}
N.~Rajatheva, I.~Atzeni, E.~Bjornson, A.~Bourdoux, S.~Buzzi, J.~Dore,
  S.~Erkucuk, M.~Fuentes, K.~Guan, Y.~Hu, and X.~Huang, ``White paper on
  broadband connectivity in {6G},'' \emph{arXiv preprint arXiv:2004.14247},
  2020.

\bibitem{DBA1}
Garima, V.~Jha, and R.~K. Singh, ``A novel dynamic bandwidth allocation scheme
  for {XGPON} based mobile fronthaul for small cell {CRAN},'' \emph{Optical
  Switching and Networking}, vol.~45, pp. 1--8, 2022.

\bibitem{DBA2}
A.~Zaouga, A.~de~Sousa, M.~Najja, and P.~Monteiro, ``Low latency dynamic
  bandwidth allocation algorithms for {NG-PON2} to support {5G} fronthaul and
  data services,'' in \emph{2019 21st International Conference on Transparent
  Optical Networks (ICTON)}, 2019, pp. 1--4.

\bibitem{DBA41}
E.~Wong and L.~Ruan, ``Towards {6G}: fast and self-adaptive dynamic bandwidth
  allocation for next-generation mobile fronthaul [invited],'' \emph{Journal of
  Optical Communications and Networking}, vol.~15, no.~8, pp. C203--C211, 2023.

\bibitem{DBA4}
Garima, V.~Jha, and R.~K. Singh, ``A novel dynamic bandwidth allocation scheme
  towards improving the performance of {XG-PON} system,'' \emph{Optical
  Switching and Networking}, vol.~47, pp. 1--14, 2023.

\bibitem{DBA5}
E.~Wong, S.~Mondal, and L.~Ruan, ``Machine learning enhanced next-generation
  optical access networks—challenges and emerging solutions [invited
  tutorial],'' \emph{Journal of Optical Communications and Networking},
  vol.~15, no.~2, pp. A49--A62, 2023.

\bibitem{DBA6}
A.~Quran, S.~Troia, O.~Ayoub, N.~D. Cicco, and M.~Tornatore, ``A reinforcement
  learning-based dynamic bandwidth allocation for {XGS-PON} networks,'' in
  \emph{26th International Conference on Optical Network Design and Modeling},
  2022, pp. 1--3.

\bibitem{DBA7}
E.~Wong and L.~Ruan, ``Towards {6G}: machine learning driven resource
  allocation in next generation optical access networks (invited),'' in
  \emph{European Conference on Optical Communication (ECOC) 2022}, 2022, pp.
  Mo3C--5.

\bibitem{MLfronthaul}
E.~Wong, S.~Mondal, and L.~Ruan, ``Machine learning enhanced next-generation
  optical access networks—challenges and emerging solutions [invited
  tutorial],'' \emph{Journal of Optical Communications and Networking},
  vol.~15, no.~2, pp. A49--A62, 2023.

\bibitem{AIoran}
J.~S. Vardakas, K.~Ramantas, E.~Vinogradov, M.~A. Rahman, A.~Girycki,
  S.~Pollin, S.~Pryor, P.~Chanclou, and C.~Verikoukis, ``Machine learning-based
  cell-free support in the {O-RAN} architecture: an innovative converged
  optical-wireless solution toward {6G} networks,'' \emph{IEEE Wireless
  Communications}, vol.~29, pp. 20--26, 2022.

\bibitem{ml6g}
M.~K. Shehzad, L.~Rose, M.~M. Butt, I.~Z. Kovacs, M.~Assaad, and M.~Guizani,
  ``{Artificial intelligence for 6G networks: Technology advancement and
  standardization},'' \emph{IEEE Vehicular Technology Magazine}, vol.~17,
  no.~3, pp. 16--25, 2022.

\bibitem{SDNflex1}
G.~Talli, F.~Slyne, S.~Porto, D.~Carey, N.~Brandonisio, A.~Naughton,
  P.~Ossieur, S.~McGettrick, C.~Bl{\"u}mm, M.~Ruffini \emph{et~al.}, ``{SDN}
  enabled dynamically reconfigurable high capacity optical access architecture
  for converged services,'' \emph{Journal of Lightwave Technology}, vol.~35,
  no.~3, pp. 550--560, 2017.

\bibitem{SDNflex2}
N.~Psaromanolakis, A.~Ropodi, P.~Fragkogiannis, K.~Tsagkaris, L.~A. Neto,
  A.~El~Ankouri, M.~Wang, G.~Simon, and P.~Chanclou, ``Software defined
  networking in a converged {5G} fiber-wireless network,'' in \emph{2020
  European Conference on Networks and Communications (EuCNC)}, 2020, pp.
  225--230.

\bibitem{SDNflex3}
K.~Kondepu, A.~Sgambelluri, F.~Cugini, P.~Castoldi, R.~A. Morenilla,
  D.~Larrabeiti, B.~Vermeulen, and L.~Valcarenghi, ``{Performance evaluation of
  SDN-controlled green mobile fronthaul using a federation of experimental
  network testbeds},'' \emph{Photonic Network Communications}, vol.~37, no.~3,
  pp. 399--408, 2019.

\bibitem{SDNflex4}
D.~Camps-Mur, J.~Gutierrez, E.~Grass, A.~Tzanakaki, P.~Flegkas, K.~Choumas,
  D.~Giatsios, A.~F. Beldachi, T.~Diallo, J.~Zou \emph{et~al.}, ``{{5G-XHaul}:
  A novel wireless-optical SDN transport network to support joint {5G} backhaul
  and fronthaul services},'' \emph{IEEE Communications Magazine}, vol.~57,
  no.~7, pp. 99--105, 2019.

\bibitem{SDNflex5}
E.~Datsika, J.~S. Vardakas, K.~Ramantas, P.-V. Mekikis, I.~T. Monroy, L.~A.
  Neto, and C.~Verikoukis, ``{SDN-enabled resource management for converged
  Fi-Wi {5G} fronthaul},'' \emph{IEEE Journal on Selected Areas in
  Communications}, vol.~39, no.~9, pp. 2772--2788, 2021.

\bibitem{capacity8}
S.~Rommel, D.~Perez-Galacho, J.~M. Fabrega, R.~Munoz, S.~Sales, and I.~T.
  Monroy, ``High-capacity {5G} fronthaul networks based on optical space
  division multiplexing,'' \emph{IEEE Transactions on Broadcasting}, vol.~65,
  no.~2, pp. 434--443, 2019.

\bibitem{SDM1}
B.~J. Puttnam, G.~Rademacher, and R.~S. Luís, ``Space-division multiplexing
  for optical fiber communications,'' \emph{Optica}, vol.~8, pp. 1186--1203,
  2021.

\bibitem{SDM2}
I.~Gasulla and J.~Capmany, ``Multicore fibres for 5g fronthaul evolution,''
  \emph{Optical and Wireless Convergence for 5G Networks}, pp. 79--100, 2019.

\bibitem{SDM3}
S.~Wang, H.~Yang, Y.~Qin, D.~Peng, and S.~Fu, ``Power-over-fiber in support of
  {5G NR} fronthaul: Space division multiplexing versus wavelength division
  multiplexing,'' \emph{Journal of Lightwave Technology}, vol.~40, no.~13, pp.
  4169--4177, 2022.

\bibitem{furdek2016overview}
M.~Furdek, L.~Wosinska, R.~Go{\'s}cie{\'n}, K.~Manousakis, M.~Aibin,
  K.~Walkowiak, S.~Ristov, M.~Gushev, and J.~L. Marzo, ``An overview of
  security challenges in communication networks,'' in \emph{2016 8th
  International Workshop on Resilient Networks Design and Modeling
  (RNDM)}.\hskip 1em plus 0.5em minus 0.4em\relax IEEE, 2016, pp. 43--50.

\bibitem{security1}
M.~Wong, A.~Prasad, and A.~C.~K. Soong, ``The security aspect of {5G}
  fronthaul,'' \emph{IEEE Wireless Communications}, vol.~29, no.~2, pp.
  116--122, 2022.

\bibitem{furdek2022machine}
M.~Furdek and C.~Natalino, ``Machine learning for network security management,
  attacks, and intrusions detection,'' in \emph{Machine Learning for Future
  Fiber-Optic Communication Systems}.\hskip 1em plus 0.5em minus 0.4em\relax
  Elsevier, 2022, pp. 317--336.

\bibitem{furdek2021optical}
M.~Furdek, C.~Natalino, A.~Di~Giglio, and M.~Schiano, ``Optical network
  security management: requirements, architecture, and efficient machine
  learning models for detection of evolving threats,'' \emph{Journal of Optical
  Communications and Networking}, vol.~13, no.~2, pp. A144--A155, 2021.

\bibitem{furdek2020machine}
M.~Furdek and C.~Natalino, ``Machine learning for optical network security
  management,'' in \emph{2020 Optical Fiber Communications Conference and
  Exhibition (OFC)}.\hskip 1em plus 0.5em minus 0.4em\relax IEEE, 2020, pp.
  1--3.

\bibitem{furdek2020machine1}
M.~Furdek, C.~Natalino, F.~Lipp, D.~Hock, A.~Di~Giglio, and M.~Schiano,
  ``Machine learning for optical network security monitoring: A practical
  perspective,'' \emph{Journal of Lightwave Technology}, vol.~38, no.~11, pp.
  2860--2871, 2020.

\bibitem{natalino2022root}
C.~Natalino, M.~Schiano, A.~Di~Giglio, and M.~Furdek, ``Root cause analysis for
  autonomous optical network security management,'' \emph{IEEE Transactions on
  Network and Service Management}, vol.~19, no.~3, pp. 2702--2713, 2022.

\bibitem{natalino2021scalable}
C.~Natalino, C.~Manso, R.~Vilalta, P.~Monti, R.~Mun{\~o}z, and M.~Furdek,
  ``Scalable physical layer security components for microservice-based optical
  {SDN} controllers,'' in \emph{2021 European Conference on Optical
  Communication (ECOC)}.\hskip 1em plus 0.5em minus 0.4em\relax IEEE, 2021, pp.
  1--4.

\bibitem{natalino2021autonomous}
C.~Natalino, A.~Di~Giglio, M.~Schiano, and M.~Furdek, ``Autonomous security
  management in optical networks,'' in \emph{Optical Fiber Communication
  Conference}.\hskip 1em plus 0.5em minus 0.4em\relax Optica Publishing Group,
  2021, pp. Tu1I--1.

\bibitem{li2019optical}
Y.~Li, N.~Hua, J.~Li, Z.~Zhong, S.~Li, C.~Zhao, X.~Xue, and X.~Zheng, ``Optical
  spectrum feature analysis and recognition for optical network security with
  machine learning,'' \emph{Optics express}, vol.~27, no.~17, pp.
  24\,808--24\,827, 2019.

\bibitem{latency3}
S.~Bidkar, R.~Bonk, and T.~Pfeiffer, ``Low-latency {TDM-PON} for {5G} xhaul,''
  in \emph{2020 22nd International Conference on Transparent Optical Networks
  (ICTON)}, 2020, pp. 1--4.

\bibitem{ofdm}
P.~M.~A. Shah, S.~S. Qureshi, R.~A. Butt, S.~M. Idrus, and J.~Mirza, ``Design
  and analysis of {5G} network architecture with orthogonal frequency division
  multiple access based passive optical network,'' \emph{Optical Fiber
  Technology}, vol.~67, pp. 1--10, 2021.

\bibitem{ocdm}
P.~Rajasekaran, G.~Ganesan, and M.~Murugappan, ``{Performance analysis of
  hybrid Fi-Wi network employing {OCDMA} based {NG-PON}},'' \emph{Frequenz},
  vol.~76, no. 1-2, pp. 97--108, 2022.

\bibitem{cost6}
A.~Fayad, T.~Cinkler, J.~Rak, and B.~Sonkoly, ``Cost-efficient optical
  fronthaul architectures for {5G} and future {6G} networks,'' in \emph{2022
  IEEE Future Networks World Forum (FNWF)}, 2022, pp. 249--254.

\bibitem{cost1}
M.~Masoudi, S.~S. Lisi, and C.~Cavdar, ``Cost-effective migration toward
  virtualized {C-RAN} with scalable fronthaul design,'' \emph{IEEE Systems
  Journal}, vol.~14, no.~4, pp. 5100--5110, 2020.

\bibitem{cost2}
I.~Sousa, N.~Sousa, M.~P. Queluz, and A.~Rodrigues, ``Fronthaul design for
  wireless networks,'' \emph{Applied Sciences}, vol.~10, no.~14, p. 4754, 2020.

\bibitem{cost3}
J.~Kim, S.~H. Chang, and J.~K. Lee, ``Comparative study for evaluating the cost
  efficiency of {5G} ethernet mobile fronthaul networks,'' \emph{Journal of
  Optical Communications and Networking}, vol.~14, no.~12, pp. 960--969, 2022.

\bibitem{cost4}
A.~Mukhopadhyay and M.~Ruffini, ``Design methodology for wireless
  backhaul/fronthaul using free space optics and fibers,'' \emph{Journal of
  Lightwave Technology}, vol.~41, no.~1, pp. 17--30, 2023.

\bibitem{cost5}
G.~Pandey, A.~Choudhary, and A.~Dixit, ``Wavelength division multiplexed radio
  over fiber links for {5G} fronthaul networks,'' \emph{IEEE Journal on
  Selected Areas in Communications}, vol.~39, no.~9, pp. 2789--2803, 2021.

\bibitem{cost7}
J.~Han, P.~Han, and Y.~Liu, ``Survivable wavelength-division-multiplexed
  passive optical network for fronthaul in 5g and beyond,'' in \emph{2021 9th
  International Conference on Intelligent Computing and Wireless Optical
  Communications (ICWOC)}.\hskip 1em plus 0.5em minus 0.4em\relax IEEE, 2021,
  pp. 5--10.

\bibitem{costoran}
S.~Mondal and M.~Ruffini, ``Optical front/mid-haul with open access-edge server
  deployment framework for sliced {O-RAN},'' \emph{IEEE Transactions on Network
  and Service Management}, vol.~19, no.~3, pp. 3202--3219, 2022.

\bibitem{cost9}
C.~Ranaweera, E.~Wong, A.~Nirmalathas, C.~Jayasundara, and C.~Lim, ``{5G}
  {C-RAN} with optical fronthaul: An analysis from a deployment perspective,''
  \emph{Journal of Lightwave Technology}, vol.~36, no.~11, pp. 2059--2068,
  2018.

\bibitem{capacity1}
K.~Zeb, X.~Zhang, and Z.~Lu, ``{High Capacity Mode Division Multiplexing Based
  MIMO Enabled All-Optical Analog Millimeter-Wave Over Fiber Fronthaul
  Architecture for {5G} and Beyond},'' \emph{IEEE Access}, vol.~7, pp.
  89\,522--89\,533, 2019.

\bibitem{capacity3}
J.~Chaudhary, J.~Zou, and G.~Fettweis, ``Cost saving analysis in
  capacity-constrained {C-RAN} fronthaul,'' in \emph{2018 IEEE Globecom
  Workshops (GC Wkshps)}, 2018, pp. 1--7.

\bibitem{capacity4}
D.~Zhang, D.~Zhe, M.~Jiang, and J.~Zhang, ``High speed {WDM-PON} technology for
  {5G} fronthaul network,'' in \emph{Asia Communications and Photonics
  Conference (ACP) 2018, OSA Technical Digest}, 2018, pp. S3K--8.

\bibitem{capacity5}
F.~El-Nahal, T.~Xu, D.~AlQahtani, and M.~Leeson, ``A bidirectional {WDM-PON}
  free space optical ({FSO}) system for fronthaul {5G} {C-RAN} networks,''
  \emph{IEEE Photonics Journal}, vol.~15, no.~1, pp. 1--10, 2023.

\bibitem{capacity7}
M.~Elhattab and W.~Hamouda, ``{Performance Analysis for H-CRANs Under
  Constrained Capacity Fronthaul},'' \emph{IEEE Networking Letters}, vol.~2,
  no.~2, pp. 62--66, 2020.

\bibitem{resource1}
Y.~Fan, M.~Fu, H.~Jiang, X.~Liu, Q.~Liu, Y.~Xu, L.~Yi, W.~Hu, and Q.~Zhuge,
  ``Point-to-multipoint coherent architecture with joint resource allocation
  for {B5G}/{6G} fronthaul,'' \emph{IEEE Wireless Communications}, vol.~29,
  no.~2, pp. 100--106, 2022.

\bibitem{resource2}
A.~M. Mikaeil, W.~Hu, and L.~Li, ``{Joint allocation of radio and fronthaul
  resources in multi-wavelength-enabled C-RAN based on reinforcement
  learning},'' \emph{Journal of Lightwave Technology}, vol.~37, no.~23, pp.
  5780--5789, 2019.

\bibitem{resource3}
T.~Lagkas, D.~Klonidis, P.~Sarigiannidis, and I.~Tomkos, ``{Optimized Joint
  Allocation of Radio, Optical, and MEC Resources for the {5G} and Beyond
  Fronthaul},'' \emph{IEEE Transactions on Network and Service Management},
  vol.~18, no.~4, pp. 4639--4653, 2021.

\bibitem{SDN1}
K.~Ramantas, A.~Antonopoulos, E.~Kartsakli, P.-V. Mekikis, J.~Vardakas, and
  C.~Verikoukis, ``{A {C-RAN} Based {5G} Platform With a Fully Virtualized, SDN
  Controlled Optical/Wireless Fronthaul},'' in \emph{2018 20th International
  Conference on Transparent Optical Networks (ICTON)}, 2018, pp. 1--4.

\bibitem{SDN2}
V.~Alevizaki, M.~Anastasopoulos, A.~Tzanakaki, and D.~Simeonidou, ``{Joint
  Fronthaul Optimization and SDN Controller Placement in Dynamic {5G}
  Networks},'' in \emph{Optical Network Design and Modeling: 23rd IFIP WG 6.10
  International Conference, ONDM 2019}, 2020, pp. 181--192.

\bibitem{SDN3}
D.~Camps-Mur, J.~Gutierrez, E.~Grass, A.~Tzanakaki, P.~Flegkas, K.~Choumas,
  D.~Giatsios, A.~F. Beldachi, T.~Diallo, J.~Zou, P.~Legg, J.~Bartelt, J.~K.
  Chaudhary, A.~Betzler, J.~J. Aleixendri, R.~Gonzalez, and D.~Simeonidou,
  ``{5G-XHaul}: A novel wireless-optical sdn transport network to support joint
  {5G} backhaul and fronthaul services,'' \emph{IEEE Communications Magazine},
  vol.~57, no.~7, pp. 99--105, 2019.

\bibitem{ML1}
I.~Nascimento, R.~Souza, S.~Lins, A.~Silva, and A.~Klautau, ``Deep
  reinforcement learning applied to congestion control in fronthaul networks,''
  in \emph{2019 IEEE Latin-American Conference on Communications (LATINCOM)},
  2019, pp. 1--6.

\bibitem{ML2}
A.~Mikaeil, W.~Hu, S.~Hussain, and A.~Sultan, ``Traffic-estimation-based
  low-latency {XGS-PON} mobile fronthaul for small-cell {C-RAN} based on an
  adaptive learning neural network,'' \emph{Applied Sciences}, vol.~8, no.~7,
  pp. 1--15, 2018.

\bibitem{energy1}
P.~Agheli, M.~J. Emadi, and H.~Beyranvand, ``Designing cost-and
  energy-efficient cell-free massive mimo network with fiber and fso fronthaul
  links,'' \emph{arXiv preprint arXiv:2011.08511}, 2020.

\bibitem{energy2}
Z.~Tan, C.~Yang, and Z.~Wang, ``Energy evaluation for cloud {RAN} employing
  {TDM-PON} as front-haul based on a new network traffic modeling,''
  \emph{Journal of Lightwave Technology}, vol.~35, no.~13, pp. 2669--2677,
  2017.

\bibitem{energy3}
C.~Y. Z.~Tan and Z.~Wang, ``{Energy consume analysis for ring-topology
  {TWDM-PON} front-haul enabled cloud RAN},'' \emph{Journal of Lightwave
  Technology}, vol.~35, no.~20, pp. 4526--4534, 2017.

\bibitem{energy4}
X.~Wang, S.~Thota, M.~Tornatore, H.~S. Chung, H.~H. Lee, S.~Park, and
  B.~Mukherjee, ``{Energy-Efficient Virtual Base Station Formation in
  Optical-Access-Enabled Cloud-RAN},'' \emph{IEEE Journal on Selected Areas in
  Communications}, vol.~34, no.~5, pp. 1130--1139, 2016.

\bibitem{energy5}
R.~I. Tinini, D.~M. Batista, G.~B. Figueiredo, M.~Tornatore, and B.~Mukherjee,
  ``{Energy-Efficient vBBU Migration and Wavelength Reassignment in Cloud-Fog
  RAN},'' \emph{IEEE Transactions on Green Communications and Networking},
  vol.~5, no.~1, pp. 18--28, 2021.

\bibitem{latency1}
G.~O. Perez, D.~L. Lopez, and J.~A. Hernandez, ``{5G New Radio Fronthaul
  Network Design for eCPRI-IEEE 802.1CM and Extreme Latency Percentiles},''
  \emph{IEEE Access}, vol.~7, pp. 82\,218--82\,230, 2019.

\bibitem{latency4}
G.~O. Pérez, ``Design and analysis of ultra-low latency fronthaul and backhaul
  networks for {5G},'' 2020, arXiv:1607.01942.

\bibitem{latency5}
M.~Klinkowski, ``{Latency-Aware DU/CU Placement in Convergent Packet-Based {5G}
  Fronthaul Transport Networks},'' \emph{Applied Sciences}, vol.~10, no.~21,
  pp. 1--21, 2020.

\bibitem{latency6}
G.~O. Pérez, J.~A. Hernández, and D.~Larrabeiti, ``Fronthaul network modeling
  and dimensioning meeting ultra-low latency requirements for {5G},''
  \emph{Journal of Optical Communications and Networking}, vol.~10, no.~6, pp.
  573--581, 2018.

\bibitem{latency7}
F.~Musumeci, C.~Bellanzon, N.~Carapellese, M.~Tornatore, A.~Pattavina, and
  S.~Gosselin, ``{Optimal BBU Placement for {5G} {C-RAN} Deployment Over WDM
  Aggregation Networks},'' \emph{Journal of Lightwave Technology}, vol.~34,
  no.~8, pp. 1963--1970, 2016.

\bibitem{latency9}
Y.~Nakayama and D.~Hisano, ``Wavelength and bandwidth allocation for mobile
  fronthaul in {TWDM-PON},'' \emph{IEEE Transactions on Communications},
  vol.~67, no.~11, pp. 7642--7655, 2019.

\bibitem{FRANcost}
M.~R.~P. dos Santos, R.~I. Tinini, T.~O. Januario, and G.~B. Figueiredo,
  ``{Deep Recurrent Neural Network for Optical Fronthaul Dimensioning and
  Proactive vBBU Placement in CF-RAN},'' \emph{Photonic Network
  Communications}, vol.~43, no.~1, pp. 59--73, 2022.

\bibitem{FRANcost1}
R.~I. Tinini, D.~M. Batista, G.~B. Figueiredo, M.~Tornatore, and B.~Mukherjee,
  ``{Low-Latency and Energy-Efficient BBU Placement and {VPON} Formation in
  Virtualized Cloud-Fog RAN},'' \emph{Journal of Optical Communications and
  Networking}, vol.~11, no.~4, pp. B37--B48, 2019.

\bibitem{FRANcost2}
R.~Tinini, D.~Batista, and G.~Figueiredo, ``{Energy-Efficient {VPON} Formation
  and Wavelength Dimensioning in Cloud-Fog RAN over {TWDM-PON}},'' in
  \emph{2018 IEEE Symposium on Computers and Communications (ISCC)}, 2018, pp.
  521--526.

\bibitem{FRANcost3}
S.~Su, X.~Xu, Z.~Tian, M.~Zhao, and W.~Wang, ``{5G} fronthaul design based on
  software-defined and virtualized radio access network,'' in \emph{2019 28th
  Wireless and Optical Communications Conference (WOCC)}, 2019, pp. 1--5.

\bibitem{virtual}
M.~Habibi, B.~Han, M.~Nasimi, N.~Kuruvatti, A.~Fellan, and H.~Schotten,
  ``{Towards a Fully Virtualized, Cloudified, and Slicing-Aware RAN for {6G}
  Mobile Networks},'' in \emph{6G Mobile Wireless Networks}.\hskip 1em plus
  0.5em minus 0.4em\relax Springer International Publishing, 2021, pp.
  327--358.

\bibitem{fsolatency}
X.~Sun, L.~Yu, and T.~Zhang, ``Latency aware transmission scheduling for
  steerable free space optics,'' \emph{IEEE Transactions on Mobile Computing},
  vol.~22, no.~4, pp. 2221--2232, 2023.

\bibitem{colorwdm}
C.~L. M.~P. Plazas, A.~M. de~Souza, D.~R. Celino, and M.~A. Romero, ``Colorless
  {WDM-PON} fronthaul topology for beyond {5G C-RAN} architectures,''
  \emph{Optical Fiber Technology}, vol.~76, pp. 1--10, 2023.

\bibitem{project1}
{I. GmbH}, ``{5G-PICTURE: Home},'' \url{https://www.5g-picture-project.eu/}.

\bibitem{project2}
I.~GmbH, ``{5G-XHaul: Home},'' \url{https://www.5g-xhaul-project.eu/}.

\bibitem{project3}
``{Main page - {6G} Flagship},'' \url{https://www.6gflagship.com/}.

\bibitem{project31}
M.~Katz, M.~Matinmikko-Blue, and M.~Latva-Aho, ``6genesis flagship program:
  Building the bridges towards 6g-enabled wireless smart society and
  ecosystem,'' in \emph{2018 IEEE 10th Latin-American Conference on
  Communications (LATINCOM)}, 2018, pp. 1--9.

\bibitem{project4}
``{5G-Crosshaul: Home},'' \url{https://www.5g-crosshaul.eu/}.

\bibitem{project5}
``{METRO-HAUL – METRO-HAUL 5G Project},'' \url{https://metro-haul.eu/}.

\bibitem{project6}
``{Main - Hexa-X},'' \url{https://hexa-x.eu/}.

\bibitem{project7}
``{Concept – 5G Complete | EU Project},''
  \url{https://5gcomplete.eu/concept/}.

\bibitem{project8}
``{TERAWAY – Terahertz technology for ultra-broadband and ultra-wideband
  operation of backhaul and fronthaul links in systems with SDN management of
  network and radio resources},'' \url{https://ict-teraway.eu/}.

\bibitem{project9}
``{Int5Gent Worldsensing},''
  \url{https://www.worldsensing.com/project/int5gent/}.

\bibitem{project10}
``{Home - MARSAL},'' \url{https://www.marsalproject.eu/}.

\bibitem{project11}
``{FLEX-SCALE - Home},'' \url{https://6g-flexscale.eu/en/}.

\bibitem{project112}
``{Flexibly Scalable Energy Efficient Networking | FLEX-SCALE Project | Fact
  Sheet | HORIZON},'' \url{https://cordis.europa.eu/project/id/101096909}.

\bibitem{EMPOWER6G}
``{Empower Converged Optical Wireless Configurations with Cell-Free
  Technologies for High-Density 6G Networks},''
  \url{https://cordis.europa.eu/project/id/101120332}.

\bibitem{6Gprotus}
\BIBentryALTinterwordspacing
A.~Bienvenu, ``{6G research gets a 130 million EUR EU funding boost in Europe -
  SNS JU},'' Oct 2023. [Online]. Available:
  \url{https://smart-networks.europa.eu/6g-research-gets-a-130-million-eur-eu-funding-boost-in-europe/}
\BIBentrySTDinterwordspacing

\bibitem{coherentfso}
F.~P. Guiomar, M.~A. Fernandes, J.~L. Nascimento, V.~Rodrigues, and P.~P.
  Monteiro, ``Coherent free-space optical communications: Opportunities and
  challenges,'' \emph{Journal of Lightwave Technology}, vol.~40, no.~10, pp.
  3173--3186, 2022.

\bibitem{capacityenhance}
X.~Liu, ``Enabling optical network technologies for {5G} and beyond,''
  \emph{Journal of Lightwave Technology}, vol.~40, no.~2, pp. 358--367, 2022.

\bibitem{6gfrontlatency}
Y.~Liu, Y.~Deng, A.~Nallanathan, and J.~Yuan, ``Machine learning for {6G}
  enhanced ultra-reliable and low-latency services,'' \emph{IEEE Wireless
  Communications}, vol.~30, no.~2, pp. 48--54, 2023.

\bibitem{6Glatency}
D.~Larrabeiti, L.~M. Contreras, G.~Otero, J.~A. Hernández, and J.~P.
  Fernandez-Palacios, ``Toward end-to-end latency management of {5G} network
  slicing and fronthaul traffic (invited paper),'' \emph{Optical Fiber
  Technology}, vol.~76, pp. 1--10, 2023.

\bibitem{resilliency}
M.~Chiesa, A.~Kamisinski, J.~Rak, G.~Retvari, and S.~Schmid, ``A survey of
  fast-recovery mechanisms in packet-switched networks,'' \emph{IEEE
  Communications Surveys \& Tutorials}, vol.~23, no.~2, pp. 1253--1301, 2021.

\bibitem{resilliency1}
J.~Rak, R.~Girão-Silva, T.~Gomes, G.~Ellinas, B.~Kantarci, and M.~Tornatore,
  ``Disaster resilience of optical networks: State of the art, challenges, and
  opportunities,'' \emph{Optical Switching and Networking}, vol.~42, pp. 1--28,
  2021.

\bibitem{security3}
A.~S. Abdalla and V.~Marojevic, ``End-to-end {O-RAN} security architecture,
  threat surface, coverage, and the case of the open fronthaul,'' \emph{arXiv
  preprint arXiv:2304.05513}, 2023.

\bibitem{security31}
D.~Dik and M.~S. Berger, ``{Open-RAN Fronthaul Transport Security Architecture
  and Implementation},'' \emph{IEEE Access}, vol.~11, pp. 46\,185--46\,203,
  2023.

\bibitem{sync}
H.~Li, L.~Han, R.~Duan, and G.~M. Garner, ``Analysis of the synchronization
  requirements of {5G} and corresponding solutions,'' \emph{IEEE Communications
  Standards Magazine}, vol.~1, no.~1, pp. 52--58, 2017.

\bibitem{sync1}
S.~Bhattacharjee, R.~Schmidt, K.~Katsalis, C.~Y. Chang, T.~Bauschert, and
  N.~Nikaein, ``Time-sensitive networking for 5g fronthaul networks,'' in
  \emph{ICC 2020-2020 IEEE International Conference on Communications (ICC)},
  2020, pp. 1--7.

\bibitem{cach1}
I.~Dias, L.~Ruan, C.~Ranaweera, and E.~Wong, ``From {5G} to beyond: Passive
  optical network and multi-access edge computing integration for
  latency-sensitive applications,'' \emph{Optical Fiber Technology}, vol.~75,
  pp. 1--12, 2023.

\bibitem{cach2}
R.~Aghazadeh, A.~Shahidinejad, and M.~Ghobaei-Arani, ``Proactive content
  caching in edge computing environment: A review,'' \emph{Software: Practice
  and Experience}, vol.~53, no.~3, pp. 811--855, 2023.

\bibitem{monitor1}
P.~J. Urban, G.~C. Amaral, G.~Zeglinski, E.~Weinert-Raczka, and J.~P. von~der
  Weid, ``A tutorial on fiber monitoring for applications in analogue mobile
  fronthaul,'' \emph{IEEE Communications Surveys \& Tutorials}, vol.~20, no.~4,
  pp. 2742--2757, 2018.

\bibitem{twin6g}
N.~P. Kuruvatti, M.~A. Habibi, S.~Partani, B.~Han, A.~Fellan, and H.~D.
  Schotten, ``Empowering {6G} communication systems with digital twin
  technology: A comprehensive survey,'' \emph{IEEE Access}, vol.~10, pp.
  112\,158--112\,186, 2022.

\bibitem{twinoptic}
D.~Wang, Z.~Zhang, M.~Zhang, M.~Fu, J.~Li, S.~Cai, C.~Zhang, and X.~Chen, ``The
  role of digital twin in optical communication: fault management, hardware
  configuration, and transmission simulation,'' \emph{IEEE Communications
  Magazine}, vol.~59, no.~1, pp. 133--139, 2021.

\bibitem{twinoptic1}
Y.~Wu, M.~Zhang, L.~Zhang, J.~Li, X.~Chen, and D.~Wang, ``Dynamic network
  topology portrait for digital twin optical network,'' \emph{Journal of
  Lightwave Technology}, vol.~41, no.~10, pp. 2953--2968, 2023.

\bibitem{twinoptic2}
C.~Janz, Y.~You, M.~Hemmati, Z.~Jiang, A.~Javadtalab, and J.~Mitra, ``Digital
  twin for the optical network: Key technologies and enabled automation
  applications,'' in \emph{NOMS 2022-2022 IEEE/IFIP Network Operations and
  Management Symposium}, 2022, pp. 1--6.

\bibitem{twinoptic3}
M.~S. Faruk and S.~J. Savory, ``Measurement informed models and digital twins
  for optical fiber communication systems,'' \emph{Journal of Lightwave
  Technology}, pp. 1016--1030, 2023.

\bibitem{twinoptic4}
N.~Morette, H.~Hafermann, Y.~Frignac, and Y.~Pointurier, ``Machine learning
  enhancement of a digital twin for wavelength division multiplexing network
  performance prediction leveraging quality of transmission parameter
  refinement,'' \emph{Journal of Optical Communications and Networking},
  vol.~15, pp. 333--343, 2023.

\end{thebibliography}
\vspace{-1.5 cm}
\begin{IEEEbiography}[{\includegraphics[width=1in,height=1.25in,clip,keepaspectratio]{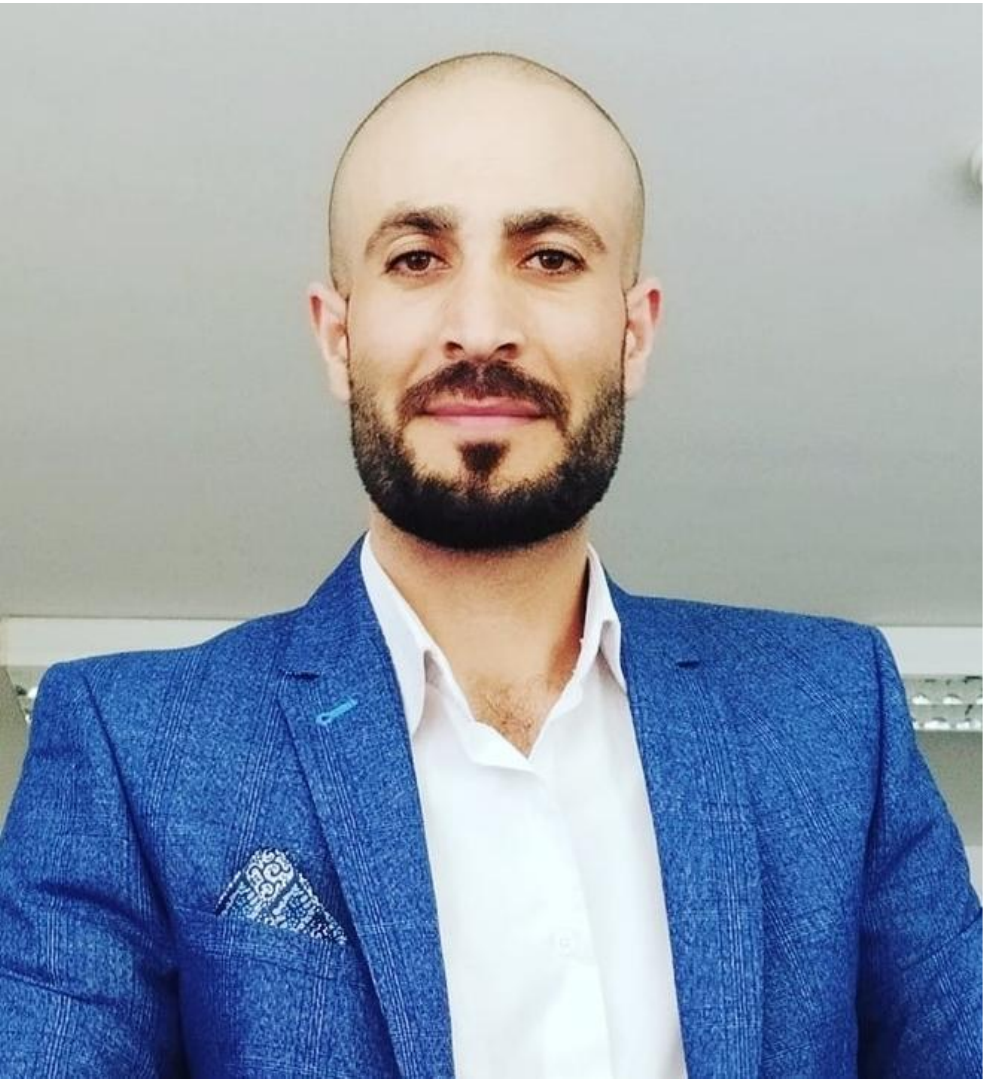}}]{Abdulhalim Fayad} received his B.Sc. and M.Sc. degrees in Electronics and Communication Engineering from Damascus University, Syria, in 2014 and 2019, respectively. He is currently pursuing a Ph.D. at the High Speed Networks Laboratory (HSN Lab) in the Department of Telecommunications and Media Informatics at the Budapest University of Technology and Economics (BME), Hungary. His research interests include optical access networks, 5G/6G fronthaul/backhaul design, communication network optimization, and developing cost- and energy-efficient architectures for 5G/6G networks.
\end{IEEEbiography}
\vspace{-1 cm}
\begin{IEEEbiography}[{\includegraphics[width=1in,height=1.25in,clip,keepaspectratio]{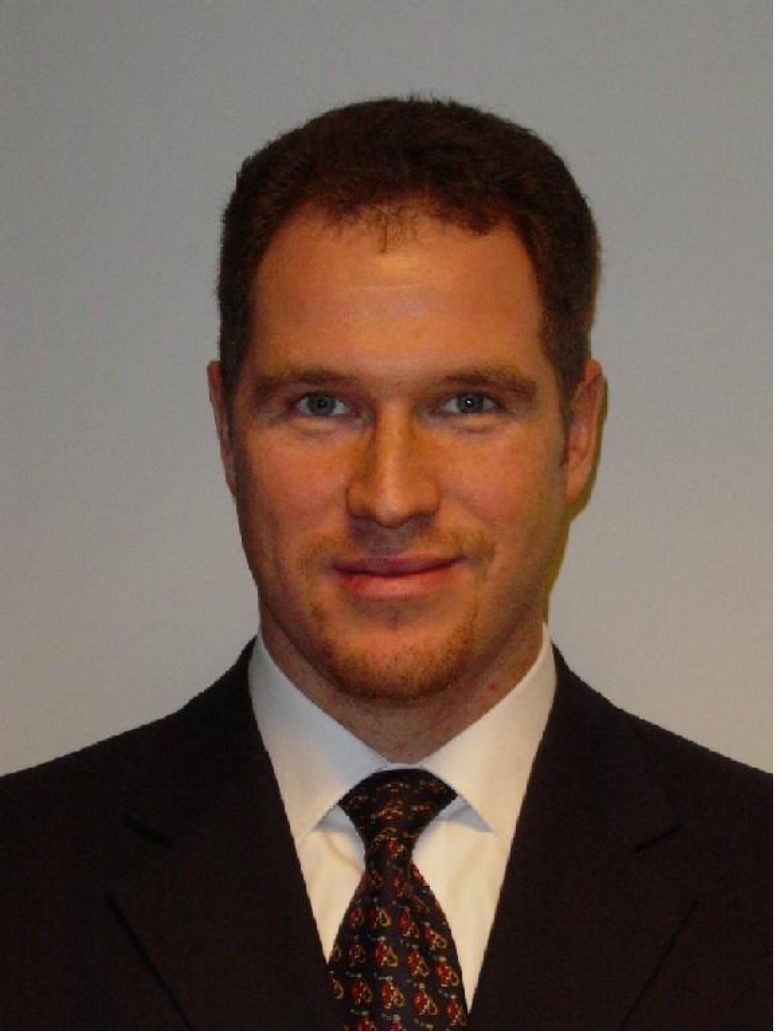}}]{Tibor Cinkler} received the M.Sc. and Ph.D. degrees from the Budapest University of Technology and Economics (BME), Hungary, in 1994 and 1999, respectively. He received the D.Sc. degree from the Hungarian Academy of Sciences in 2013, in the same year he habilitated.  He is currently a Full Professor at the Department of Telecommunications and Media Informatics (TMIT), BME. He is author of over 300 refereed scientific publications, including four patents, with over 2700 citations. His research interests include the optimization of communications networks, including optical networks, 5G/6G cellular mobile networks, IoT and IIoT, all with particular emphasis on enhanced energy efficiency and availability.
\end{IEEEbiography}
\vspace{-1 cm}
\begin{IEEEbiography}[{\includegraphics[width=1in,height=1.25in,clip,keepaspectratio]{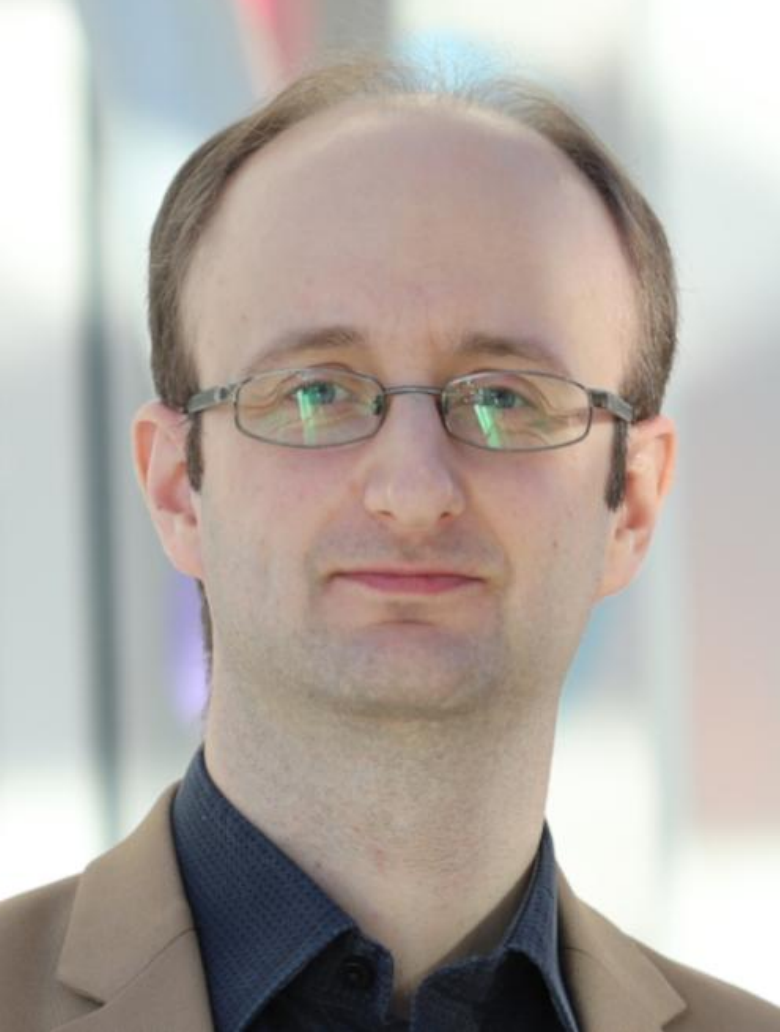}}]{Jacek Rak} received the M.Sc., Ph.D., and D.Sc. (Habilitation) degrees from Gda\'nsk University of Technology, Gda\'nsk, Poland, in 2003, 2009, and 2016, respectively.
He is currently the Head of the Department of Computer Communications at Gda\'nsk University of Technology, Poland, as well as the PICAIS Visiting Research Fellow at the Chair of Computer Networks and Computer Communications at the University of Passau, Germany.
	He has authored over 100 publications, including the book \textit{Resilient Routing in Communication
		Networks} (Springer, 2015). From 2016 and 2020, he was leading the COST CA15127 Action
	\textit{Resilient Communication Services Protecting End-User Applications From Disaster-Based Failures} (\textit{RECODIS}) involving over 170 members from 31 countries. His main research interests include the
	resilience of communication networks and networked systems. Recently, he has been the TPC Chair of ONDM 2017, and the TPC Co-Chair of IFIP Networking 2019. He is the Member of the Editorial Board of Optical Switching and Networking (Elsevier), Networks (Wiley) and the Founder of the International Workshop on Resilient Networks Design and Modeling (RNDM).
\end{IEEEbiography}
\end{document}